\documentclass[%
twocolumn,
physplasmas,
aip,
 amsmath,amssymb,
reprint,%
]{revtex4-2}

\usepackage{graphicx}% Include figure files

\usepackage{dcolumn}% Align table columns on decimal point
\usepackage{bm}% bold math
\usepackage{color}
\definecolor{RED}{rgb}{1,0,0}

\allowdisplaybreaks

\begin{document}

\preprint{AIP/123-QED}

%\title[Sample title]{Sample Title:\\with Forced Linebreak\footnote{Error!}}
% Force line breaks with \\
%\thanks{Footnote to title of article.}

\title{
Schr\"{o}dinger equations and fluctuation theorems for collisionless plasma systems
}

\author{H. Sugama}
\email{sugama.hideo@nifs.ac.jp}
\affiliation{
National Institute for Fusion Science, 
Toki 509-5292, Japan
}

\date{\today}% It is always \today, today,
             %  but any date may be explicitly specified

\begin{abstract}
The fluctuation theorem and detailed fluctuation theorem are formulated for classical systems whose governing equations can be written in Schr\"{o}dinger-type equations and which possess either a unitary or an antiunitary time-reversal operator. The initial state vector is treated as a random variable drawn from a time-reversal-symmetric probability distribution, and a stochastic relative entropy defined from its probability density is used to formulate these theorems. 
The framework is applied to two collisionless plasma systems: the linear Vlasov-Poisson and linear gyrokinetic systems. For the linear Vlasov-Poisson system, the governing equations are recast into Schr\"{o}dinger form, and Hamiltonian eigenvectors corresponding to Case-Van Kampen modes are derived to construct explicit solutions. The stochastic relative entropy is interpreted as entropy generation associated with Landau damping, in which electric-field energy is transferred from the lowest Hermite state to higher-order Hermite states acting as thermal reservoirs. For a specific class of initial distributions, a new analytical expression for the probability density function of the stochastic relative entropy is derived and validated numerically.
For the linear gyrokinetic system in a uniform magnetic field, the governing equations are likewise transformed into Schr\"{o}dinger form, and the corresponding time-reversal operators are identified. The state-vector space is constructed as a tensor product of species, perpendicular-velocity, and parallel-velocity spaces. The resulting state vectors decompose into two orthogonal components: one coupled to electromagnetic fluctuations and the other corresponding to ballistic modes. These results establish a nonequilibrium statistical-mechanical framework for collisionless plasma dynamics and provide useful examples for future quantum-computing applications to plasma simulations.
\end{abstract}

%\keywords{
%Landau damping, Vlasov-Poisson system, Schr\"{o}dinger equation, fluctuation theorem
 %}%Use showkeys class option if keyword
                              %display desired
\maketitle

\section{Introduction}
Collisionless plasmas support a wide range of collective phenomena through the long-range electromagnetic interaction of charged particles.
Among these phenomena, Landau damping, driven by resonant wave-particle interactions, has been extensively studied as a fundamental mechanism for stabilizing instabilities and heating particles in space and fusion plasmas.~\cite{Landau,Case,VK,VK&F,Nicholson,H&P,Sugama1999,Sugama2006,Biancalani,Schekochihin,Zocco2011,Loureiro,Plunk,Villani,Zocco,Maekaku,Sugama2025}
Although Landau damping is governed by the time-reversal-symmetric Vlasov equation, it appears macroscopically irreversible.
This apparent irreversibility arises because phase mixing transfers information from low-order velocity moments to increasingly fine velocity-space structures, while the Gibbs entropy defined from the velocity distribution function remains conserved.

In Ref.~\cite{Sugama2025}, Landau damping was examined from the perspective of nonequilibrium statistical mechanics 
by applying the fluctuation theorem.~\cite{FTES,Jarzynski,Shiraishi}
The fluctuation theorem, derived from reversible microscopic dynamics, relates the probabilities of entropy production and entropy reduction and thereby provides a microscopic basis for the second law of thermodynamics.
In Ref.~\cite{Sugama2025}, the linearized Vlasov-Poisson system is reformulated as a Schr\"{o}dinger equation, 
so that conservation laws and time-reversal symmetry can be expressed compactly.
The corresponding state vector is defined such that its squared norm equals an invariant of the linear Vlasov-Poisson system, 
given by the perturbed energy normalized by the equilibrium temperature minus the perturbed Gibbs entropy.
With a stochastic relative entropy defined from the probability distribution of this state vector, the fluctuation theorem can be applied directly to Landau damping. 
The same Schr\"{o}dinger form of the linear Vlasov-Poisson system was also derived by Ameri \textit{et al.}~\cite{Ameri}, who used it to construct a quantum algorithm for the system. 
Dodin and Startsev~\cite{Dodin} also investigated the quantum mechanical formulation and quantum computing 
for plasma simulations.

In this paper, we formulate the fluctuation theorem and the detailed fluctuation theorem for classical systems described by time-reversal-symmetric Schr\"{o}dinger equations.
A simple example of such classical systems is the harmonic oscillator with mass $m$ and angular frequency $\omega$.
Introducing the complex amplitude
$\psi = q + i \, p/(m\omega)$
from the coordinate $q$ and momentum $p=m\,dq/dt$, one obtains
\begin{equation}
i \, \frac{d\psi }{ dt } = \omega \, \psi
, 
\end{equation}
which has the form of a Schr\"{o}dinger equation with $\hbar=1$ and Hamiltonian operator $\widehat{H}=\omega$.
Here, however, $\psi$ is a classical variable, not a quantum-mechanical probability amplitude associated with measurement.
More generally, linear Hamiltonian canonical equations can be written in Schr\"{o}dinger form.~\cite{Heslot}
It is therefore natural, following Jarzynski's derivation for Hamiltonian systems,~\cite{Jarzynski} 
to prove fluctuation theorems for Schr\"{o}dinger-form systems with time-reversal symmetry.

We apply this formulation to two plasma systems.
First, we revisit the linear Vlasov-Poisson system as a time-reversal-symmetric Schr\"{o}dinger system to which the fluctuation theorem and detailed fluctuation theorem apply.
Second, we show that a linear gyrokinetic system~\cite{Antonsen,Catto} in a uniform magnetic field belongs to the same class.
The gyrokinetic equation in a uniform magnetic field is widely used to analyze kinetic Alfv\'{e}n waves and related phenomena in space plasmas.~\cite{Schekochihin,Zocco2011}
The present results provide a nonequilibrium statistical-mechanical perspective on collisionless plasma dynamics and support future applications of quantum-computing algorithms~\cite{Ameri,Dalzell2025} to plasma simulations.

The remainder of this paper is organized as follows.
Section~II formulates the fluctuation theorem and the detailed fluctuation theorem for general classical systems whose governing equations have Schr\"{o}dinger form and time-reversal symmetry.
Section~III treats the linear Vlasov-Poisson system as the first example.
There, the governing equations are transformed into Schr\"{o}dinger form, the unitary and antiunitary time-reversal operators are identified, and Case-Van Kampen (CVK) modes~\cite{Case,VK,VK&F,Nicholson} are represented as Hamiltonian eigenvectors.
These eigenvectors are used to construct the exact solution and to evaluate the stochastic relative entropy.
We also derive and numerically verify a new expression for the probability density function of the stochastic relative entropy for a specified initial probability distribution.
Section~IV considers a linear gyrokinetic system in a uniform background magnetic field as the second example.
For this system, we derive the Schr\"{o}dinger equation and the corresponding time-reversal operators, using an invariant analogous to that of the linear Vlasov-Poisson system to define the squared norm of the state vector.
The state-vector space is written as the tensor product of species, perpendicular-velocity, and parallel-velocity spaces.
We show that the state vector decomposes into two mutually orthogonal components: one associated with perpendicular-velocity-space structures generated by the zeroth- and first-order Bessel functions, and the other corresponding to rapidly decaying ballistic modes independent of electromagnetic fluctuations. 
Section~V summarizes the conclusions.

\section{Fluctuation and Detailed Fluctuation Theorems for Classical Systems Described by Schr\"{o}dinger Equations}

\subsection{Schr\"{o}dinger equation and time-reversal symmetry}

We consider a classical system whose state at time $\tau$ is represented by the ket vector $| \psi (\tau) \rangle$ and satisfies a Schr\"{o}dinger equation 
 (with $\hbar = 1$), 
\begin{equation}
\label{Schreq}
i \frac{d}{d \tau}
| \psi (\tau) \rangle
= 
\widehat{H} 
| \psi (\tau) \rangle
, 
\end{equation}
where $\widehat{H}$ is a Hermitian Hamiltonian operator. 
We use  $\widehat{\cdot}$  to represent operators acting on vectors 
throughout this work. 
Examples of such classical systems are given in Secs.~III and IV.
The state vector $| \psi (\tau) \rangle$  at time $\tau$ is related to the initial state vector 
$| \psi (0) \rangle$  by %
\begin{equation}
| \psi (\tau) \rangle
= 
\widehat{U}(\tau)
| \psi (0) \rangle
, 
\end{equation}
where 
\begin{equation}
\widehat{U}(\tau)
\equiv 
\exp 
( - i \tau \widehat{H} )
, 
\end{equation}
is the time evolution operator that satisfies
\begin{eqnarray}
& & 
\widehat{U}( \tau ) \widehat{U}( \tau' )
= \widehat{U}( \tau + \tau' )
,
\hspace{3mm}
\widehat{U}(0)
= 
\widehat{1}
,
\nonumber \\ & & 
[\widehat{U}(\tau)]^\dagger 
=
[\widehat{U}(\tau)]^{-1}
=
\widehat{U}(- \tau )
,
\end{eqnarray}
and 
\begin{equation}
i \frac{d}{d \tau} 
\widehat{U}(\tau)
=
\widehat{H} 
\widehat{U}(\tau)
.
\end{equation}
We consider the vector space spanned by 
the $N_c$ ket vectors 
$\{ | n \rangle \}_{n = 0, 1, \cdots, N_c -1}$ and 
represent the state vector $|\psi (\tau) \rangle$ in this space as the $N_c$-dimensional complex vector, 
\begin{equation}
\label{boldpsi}
\boldsymbol{\psi} (\tau) \equiv
\mbox{}^t 
[ \psi_0 (\tau), \psi_1 (\tau), \cdots, \psi_{N_c-1} (\tau)]
,
\end{equation}
where 
$\psi_n (\tau) = \langle n | \psi (\tau) \rangle$  for 
$ n = 0, 1, \cdots, N_c -1 $, and 
$\mbox{}^t [\cdots]$ denotes the transpose of a row vector to express it as a column vector.  
Here, we impose the orthonormality and closure conditions,  $\langle n | n' \rangle = \delta_{n n'}$ 
and 
$\sum_{n=0}^{N_c -1} | n \rangle \langle n | = \widehat{1}$. 
The Schr\"{o}dinger equation is expressed for $\boldsymbol{\psi} (\tau)$ as 
\begin{equation}
\label{SchNc}
i \frac{d}{d \tau} \boldsymbol{\psi} (\tau)
=
{\bf H} \boldsymbol{\psi} (\tau)
,
\end{equation}
where ${\bf H}$ is a Hermitian $N_c \times N_c$ Hamiltonian matrix 
defined from the Hamiltonian operator $\widehat{H}$
by 
${\bf H} \equiv [  H_{n n'}  ]_{n,n'= 0, 1, \cdots, N_c -1}$ 
and 
$ H_{n n'} \equiv \langle n | \widehat{H} |n' \rangle$. 
The solution of Eq.~(\ref{SchNc}) is given by 
\begin{equation}
\boldsymbol{\psi} (\tau)
=
{\bf U} (\tau) 
\boldsymbol{\psi} (0)
,
\end{equation}
where 
\begin{equation}
{\bf U} (\tau) = \exp (-i \tau \mathbf{H})
\end{equation}
is the unitary $N_c \times N_c$ time evolution matrix. 
Thus, the squared norm, 
$
\langle \psi(\tau) | \psi(\tau) \rangle 
\equiv
|| \boldsymbol{\psi} (\tau) ||^2
\equiv
\sum_{n=0}^{N_c  - 1} | \psi_n (\tau) |^2
$, 
remains constant in time $\tau$. 
The system considered here is classical, not quantum mechanical. 
Here, $\psi_n(\tau)$ $(n=0,1,\cdots,N_c-1)$ are 
classical variables specifying the system state, and $|\psi_n(\tau)|^2$ 
is not the probability that the system occupies the $n$th state.

We now consider the unitary operator $\widehat{T}$ representing the time-reversal transformation 
described in Appendix~A. 
Then, the $N_c \times N_c$  unitary time-reversal matrix is defined by 
${\bf T} \equiv [  T_{n n'}  ]_{n,n'= 0, 1, \cdots, N_c -1}$ 
with  
$T_{n n'} \equiv \langle n | \widehat{T} |n' \rangle$. 
It is shown from Eqs.~(\ref{TK}) and (\ref{THK}) that 
the unitary time-reversal matrix ${\bf T}$ satisfies 
\begin{equation}
{\bf T}^\dagger = {\bf T}^{-1}
,
\hspace*{5mm}
{\bf T}  {\bf H} = - {\bf H}  {\bf T}
.
\end{equation}
It follows that 
\begin{equation}
\label{TUUT}
{\bf T}  {\bf U} (\tau) =  {\bf U}(- \tau) {\bf T}
\end{equation}
holds so that, 
when $\boldsymbol{\psi} (\tau) = {\bf U}(\tau) \boldsymbol{\psi} (0)$ is a solution to 
the Schr\"{o}dinger equation, 
${\bf T}  \boldsymbol{\psi} (-\tau)  =  {\bf U} (\tau) {\bf T}  \boldsymbol{\psi} (0)$ 
becomes a solution as well. 

We next consider the antiunitary operator $\widehat{K}$ representing another time-reversal transformation 
described in Appendix~A. 
From 
the antiunitary time-reversal operator $\widehat{K}$, 
we define a $N_c \times N_c$ matrix, 
${\bf K} \equiv [  K_{n n'}  ]_{n,n'= 0, 1, \cdots, N_c -1}$ 
with  
$K_{n n'} \equiv \langle n | (\widehat{K} |n' \rangle) 
\equiv (( \langle n | \widehat{K}) |n' \rangle )^*$
where $\mbox{}^*$ represents the complex conjugate. 
As seen in the definition of $K_{nn'}$ above, 
parentheses are generally required when defining matrix elements of antilinear operators.~\cite{QM}
This defines another unitary matrix associated with time reversal, ${\bf K}$. 
From Eq.~(\ref{TK}), we have
\begin{equation}
{\bf K}^\dagger = {\bf K}^{-1}
. 
\end{equation}
The antiunitary operator $\widehat{K}$ transforms the state vector  
$| \psi (\tau) \rangle$ to  
\[
\widehat{K} | \psi (\tau) \rangle 
=
\sum_{n=0}^{N_c -1} | n \rangle
 \langle n | (\widehat{K} | \psi (\tau) \rangle )
 ,
 \]
 which is represented by the $N_c$-dimensional vector 
\begin{eqnarray}
\widehat{K} [\boldsymbol{\psi} (\tau)]
& \equiv & 
\mbox{}^t
[  \langle 0 | (\widehat{K} | \psi (\tau) \rangle ), 
\cdots, 
\langle N_c - 1 | (\widehat{K} | \psi (\tau) \rangle ) ]
\nonumber \\
& = & 
{\bf K} \boldsymbol{\psi}^* (\tau) 
.
\end{eqnarray}
Here, $\mbox{}^*$ also denotes complex conjugation of vectors and matrices; 
for 
$\boldsymbol{\psi} = \mbox{}^t [\psi_0, \psi_1, \cdots, \psi_{N_c -1}]$ 
and 
${\bf K} = [K_{mn}]_{m,n = 0, 1, \cdots, N_c -1}$, 
we have
$\boldsymbol{\psi}^* = \mbox{}^t [\psi_0^*, \psi_1^*, \cdots, \psi_{N_c -1}^*]$ 
and 
${\bf K}^* = [K_{mn}^*]_{m,n = 0, 1, \cdots, N_c -1}$, 
respectively. 
Also, the condition 
$
\widehat{K} \widehat{H} = 
\widehat{H} \widehat{K} 
$
in Eq.~(\ref{THK}) yields 
\begin{equation}
{\bf K} {\bf H}^* = {\bf H}  {\bf K}
,  
\end{equation}
from which we have
\begin{equation}
{\bf K} ( i {\bf H} )^* =  - i {\bf H}  {\bf K}
,  
\end{equation}
and 
\begin{equation}
\label{KUUK}
{\bf K} {\bf U}^* (\tau)  = 
 {\bf U} (- \tau) {\bf K}
. 
\end{equation}
Therefore, 
when $\boldsymbol{\psi} (\tau) = {\bf U}(\tau) \boldsymbol{\psi} (0)$ is a solution to 
the Schr\"{o}dinger equation, 
${\bf K}   \boldsymbol{\psi}^* (-\tau)   =  {\bf U} ( \tau) {\bf K} \boldsymbol{\psi}^* (0)$ 
becomes a solution as well.

\subsection{Fluctuation theorem}

We now assume the initial vector $\boldsymbol{\psi} (0)$
to be given randomly.  
Then, the vector $\boldsymbol{\psi} (\tau)$ at time $\tau$, 
which is uniquely determined by $\boldsymbol{\psi}(0)$, is also a random variable.
Hereafter, 
\begin{equation}
\boldsymbol{\Psi} (\tau) \equiv   
\mbox{}^t
[
\Psi_0 (\tau), \Psi_1 (\tau), \cdots, 
\Psi_{N_c-1} (\tau)
]
\end{equation}
denotes the state vector as a random (or stochastic) variable while 
$
\boldsymbol{\psi} (\tau)
$ 
represents a specific realization of $\boldsymbol{\Psi} (\tau)$. 
%
%===revision===
%\textcolor{red}{
More specifically, according to probability theory, the random variable 
$\boldsymbol{\Psi}(\tau)$ can also be regarded as a function of a hidden variable $\omega$ 
that represents the outcome of a random trial, and can be written as $\boldsymbol{\Psi}(\tau, \omega)$.
When $\omega$ takes a specific value as a result of the trial,
the realization of the random variable is expressed as
$\boldsymbol{\psi}(\tau) = \boldsymbol{\Psi}(\tau, \omega)$,
which represents the relation 
between the random variable $\boldsymbol{\Psi}$ and its realization $\boldsymbol{\psi}$.
%
%}
%===
%
The probability that the real and imaginary parts of the random variables
$\Psi_n (\tau)\equiv 
\Psi_{r,n} (\tau) + i \Psi_{i,n} (\tau)$ $(n=0, 1, 2, \cdots, N_c - 1)$ lie
within 
the infinitesimal intervals 
%
%===revision===
%\textcolor{red}{
$[\psi_{r,n}, \psi_{r,n} + d\psi_{r,n} )$ and 
$[\psi_{i,n}, \psi_{i,n} + d\psi_{i,n} )$, respectively, 
is given by 
\begin{equation}
P ( \boldsymbol{\psi} ; \tau )
\,
d \Gamma
,
\end{equation}
where 
the volume element is defined as  
\begin{equation}
\label{velment}
d \Gamma 
\equiv
\prod_{n = 0}^{N_c -1}
d\psi_{r,n} d\psi_{i,n}
.
\end{equation}
%}
%===
%
%
To treat the probability distribution of 
$\boldsymbol{\Psi} (\tau) = {\bf U} (\tau) \boldsymbol{\Psi} (0)$ at time $\tau$, 
we use the probability density functional 
$P  [ \boldsymbol{\psi} (\tau) ; \tau]$, which  
denotes the value of the above-mentioned probability density $P ( \boldsymbol{\psi} ; \tau )$
evaluated at 
 $\boldsymbol{\psi} = \boldsymbol{\psi}(\tau)$ 
and time $\tau$. 
The probability that the real and imaginary parts of the random state variables
$\Psi_n (\tau)\equiv 
\Psi_{r,n} (\tau) + i \Psi_{i,n} (\tau)$ $(n=0, 1, 2, \cdots, N_c - 1)$ lie
in the infinitesimal intervals $[\psi_{r,n} (\tau), \psi_{r,n}(\tau) + d\psi_{r,n}(\tau))$ and 
$[\psi_{i,n}(\tau), \psi_{i,n}(\tau) + d\psi_{i,n}(\tau))$, respectively, 
is given by 
\begin{equation}
P [ \boldsymbol{\psi} (\tau) ; \tau]
d \Gamma (\tau) 
,
\end{equation}
where 
$d \Gamma (\tau) \equiv
\prod_{n = 0}^{N_c -1}
d\psi_{r,n} (\tau) d\psi_{i,n} (\tau)$. 
The above probability is conserved along the trajectory 
drawn by the vector $\boldsymbol{\psi} (\tau)$ so that 
we have 
\begin{equation}
P [\boldsymbol{\psi}(\tau); \tau ]  d \Gamma (\tau)  
=  
P [\boldsymbol{\psi}(0); 0 ]  d \Gamma (0) 
.
\end{equation}

Since ${\bf U}(\tau)$ is the unitary matrix, 
$d \Gamma (\tau) \equiv
\prod_{n = 0}^{N_c -1}
d\psi_{r,n} (\tau) d\psi_{i,n} (\tau)$
remains constant along the 
trajectory of the vector $\boldsymbol{\psi}(\tau)$ 
\begin{equation}
\label{dGdG}
d \Gamma (\tau)
= 
d \Gamma (0)
\end{equation}
and therefore the probability density 
$
P [ \boldsymbol{\psi}(\tau) ; \tau] 
$
%
%===revision===
%\textcolor{red}{
is independent of $\tau$. 
%}
%===
%
\begin{equation}
P[ \boldsymbol{\psi}(\tau) ; \tau ]
= 
P[ \boldsymbol{\psi}(0) ; 0 ]
\end{equation}
This is the analog of Liouville's theorem in Hamiltonian mechanics.

An example of a steady-state probability density functional of 
$\boldsymbol{\psi}(\tau)$ is given by 
\begin{equation}
 P^{(ss)} [ \boldsymbol{\psi}(\tau) ]
= 
\frac{ 
\exp 
\left[
- \beta
|| \boldsymbol{\psi}(\tau) ||^2
\right]
}{
\int 
d \Gamma (\tau) 
 \;
\exp 
\left[
- \beta
|| \boldsymbol{\psi}(\tau) ||^2
\right]
}
, 
\end{equation}
where $\beta$ is a positive constant. 
In the above steady-state distribution, 
the average of 
$|| \boldsymbol{\psi}(\tau) ||^2$
 is given by 
\begin{equation}
\int 
d\Gamma (\tau) \;
P^{(ss)} [  \boldsymbol{\psi}(\tau)  ]
\; 
 || \boldsymbol{\psi}(\tau) ||^2 
= 
\frac{N_c}{\beta}
.
\end{equation}

The stochastic relative entropy 
of the distribution 
$
P[ \boldsymbol{\Psi}(\tau) ; \tau ]
= 
P[ \boldsymbol{\Psi}(0) ; 0 ]
$ 
with respect to $P[ \boldsymbol{\Psi}(\tau) ; 0 ]$,  
%
%===revision===
%\textcolor{red}{
which represents the initial probability density at the point $\boldsymbol{\Psi}(\tau)$ in the space of state vectors, 
%}
%===
% 
is defined by 
\begin{equation}
\label{srentropy}
\Delta S [ \boldsymbol{\Psi}(0) ; \tau ]
\equiv 
\log \left[
 \frac{  P [\boldsymbol{\Psi}(\tau); \tau ] }{
   P [ \boldsymbol{\Psi}(\tau) ; 0 ]
 }
\right]
\equiv
\log \left[
 \frac{  P [\boldsymbol{\Psi}(0); 0 ] }{
   P [ \boldsymbol{\Psi}(\tau) ; 0 ]
 }
\right]
,
\end{equation}
where $\boldsymbol{\Psi}(\tau)$ are related to 
$\boldsymbol{\Psi}(0)$
by 
$\boldsymbol{\Psi}(\tau) =  {\bf U}(\tau) \boldsymbol{\Psi}(0)$. 
%
%===revision===
%\textcolor{red}{
Note that 
the difference between $P[\boldsymbol{\Psi}(\tau); \tau] = P[\boldsymbol{\Psi}(0);0]$ and 
$P[\boldsymbol{\Psi}(\tau); 0]$ causes $\Delta S [\boldsymbol{\Psi}(0) ; \tau]$ to become nonzero. 
%}
%===
%
We then define $P(\Delta S)$ as the probability density such that 
$P(\Delta S) d(\Delta S)$ gives 
the probability for the stochastic relative entropy 
$\Delta S [ \boldsymbol{\Psi}(0) ; \tau ] $  to 
take a value in the infinitesimal interval $[\Delta S, \Delta S + d(\Delta S) )$. 
The probability density $P(\Delta S)$ is given by 
\begin{eqnarray}
\label{PDSdef}
P(\Delta S)
 & = & 
\int 
d \Gamma (0) \, 
 P[ \boldsymbol{\psi}(0) ; 0 ] 
   \delta [ \Delta S [ \boldsymbol{\psi}(0) ; \tau ] -  \Delta S  ]
\nonumber \\ 
& =  & 
\langle 
   \delta [ \Delta S [ \boldsymbol{\psi}(0) ; \tau ] -  \Delta S  ]
\rangle_{\rm ens}
.
\end{eqnarray}
where 
$\langle \cdots \rangle_{\rm ens}
\equiv \int d\Gamma(0) P [ \boldsymbol{\psi}(0) ; 0 ]  \cdots$ represents the ensemble average. 
Now, assume that there exists a unitary time-reversal matrix ${\bf T}$ as described earlier  
and that the initial probability density  
$P [ \boldsymbol{\psi}(0) ; 0 ]$ satisfies a symmetry condition, 
\begin{equation}
\label{PT0}
P [ {\bf T} \boldsymbol{\psi}(0); 0 ] 
= 
P [ \boldsymbol{\psi}(0) ; 0 ] 
.
\end{equation}
Then, as shown in Appendix~B,  
we can prove the fluctuation theorem, 
\begin{equation}
\label{FT}
\frac{
P(\Delta S  )}{
P(- \Delta S  )
}
= 
\exp \Delta S 
. 
\end{equation}
It is also shown that 
the fluctuation theorem, Eq.~(\ref{FT}), holds 
when there exists a time-reversal antiunitary transformation 
$\widehat{K}$ 
as described earlier  
and the initial probability density  
$P [ \boldsymbol{\psi}(0) ; 0 ]$ satisfies 
\begin{equation}
\label{PK0}
P [ \widehat{K} [\boldsymbol{\psi}(0)] ; 0 ] 
= 
P [ \boldsymbol{\psi}(0) ; 0 ] 
, 
\end{equation}
where $\widehat{K} [\boldsymbol{\psi}(0)] = {\bf K} \boldsymbol{\psi}^* (0)$. 
Equation~(\ref{FT}) also leads to 
the integral fluctuation theorem,~\cite{Shiraishi} 
\begin{equation}
\langle
\exp ( - \Delta S [ \boldsymbol{\Psi}(0) ; \tau ]  )
\rangle_{\rm ens}
= 
1
. 
\end{equation}

The ensemble average 
$\langle \Delta S [ \boldsymbol{\Psi}(0) ; \tau ]  \rangle_{\rm ens}$ of the 
stochastic relative entropy 
is never negative, 
which corresponds to the second law of thermodynamics. 
It is given by 
\begin{eqnarray}
& & 
\hspace*{-5mm}
\langle \Delta S [ \boldsymbol{\Psi}(0) ; \tau ]  \rangle_{\rm ens}
=
\int_{-\infty}^{+ \infty} d (\Delta S) \, 
P(\Delta S) \Delta S
\nonumber 
\\
& & 
=
\int 
d \Gamma (\tau) \;
 P[  \boldsymbol{\psi}(\tau)  ; \tau ] 
 \log \left[
 \frac{  P [  \boldsymbol{\psi}(\tau)  ; \tau ] }{
   P  [  \boldsymbol{\psi}(\tau) ; 0 ]
 }
 \right]
 \geq 0
, 
\end{eqnarray}
indicating that $\langle \Delta S [ \boldsymbol{\Psi}(0) ; \tau ]  \rangle_{\rm ens}$ 
is the relative entropy 
(Kullback-Leibler divergence)~\cite{Shiraishi} of the probability distribution 
$P [ \boldsymbol{\psi}(\tau)  ; \tau ]$ at time $\tau$
with respect to 
$P [ \boldsymbol{\psi}(\tau) ; 0 ]$. 
Thus,  $\langle \Delta S [ \boldsymbol{\Psi}(0) ; \tau ]  \rangle_{\rm ens}$ represents 
the information loss incurred when using the initial probability density distribution 
as a surrogate for the true distribution at time $\tau$.

\subsection{Detailed fluctuation theorem}

The detailed fluctuation theorem~\cite{Jarzynski} 
can be shown to be valid in systems which are governed by 
the Schr\"{o}dinger equation, Eq.~(\ref{SchNc}). 
We now express the $N_c$-dimensional complex vector $\boldsymbol{\psi}(\tau)$ 
as 
\begin{eqnarray}
\boldsymbol{\psi}(\tau)
& = & 
\mbox{}^t 
[ \psi_0 (\tau), \psi_1 (\tau), \cdots, \psi_{N_c - 1} (\tau)]
\nonumber \\
& = & 
\left[
\begin{array}{c}
{\bf z} (\tau) 
\\
{\bf y} (\tau) 
\end{array}
\right]
=
{\bf U} (\tau)
\left[
\begin{array}{c}
{\bf z} (0) 
\\
{\bf y} (0) 
\end{array}
\right]
,
\end{eqnarray}
where ${\bf z} (\tau)$ 
and ${\bf y} (\tau)$  are given by 
\begin{eqnarray}
{\bf z} (\tau)
& = & 
\mbox{}^t 
[  \psi_0 (\tau),  \cdots, \psi_{N_z - 1} (\tau)]
\nonumber \\
& = & 
\mbox{}^t 
[ z_0 (\tau), \cdots, z_{N_z - 1} (\tau)]
,
\end{eqnarray}
and 
\begin{eqnarray}
{\bf y} (\tau)
& = & 
\mbox{}^t 
[  \psi_{N_z} (\tau),  \cdots, \psi_{N_c - 1} (\tau)]
\nonumber \\
& = & 
\mbox{}^t 
[ y_0 (\tau), \cdots, y_{N_y - 1} (\tau)]
,
\end{eqnarray}
respectively. 
Here, $N_z < N_c$ and $N_y \equiv N_c - N_z > 0$. 
We regard 
the $N_z$-dimensional complex vector ${\bf z} (\tau)$ 
and the $N_y$-dimensional 
complex vector ${\bf y} (\tau)$ as 
representing the state of 
the system of interest and that of the thermal reservoir, 
respectively, 
as treated in Ref.~\cite{Jarzynski}  
to derive the detailed fluctuation theorem. 

We also employ the following expressions, 
\begin{eqnarray}
\label{ztau}
{\bf z} (\tau)
& = & 
\boldsymbol{\zeta}  [  \tau ; \boldsymbol{\psi} (0) ]
\nonumber \\
& = & 
\mbox{}^t
[
\;
\zeta_0  [  \tau ; \boldsymbol{\psi} (0) ], 
\cdots,
\zeta_{N_z - 1}  [  \tau ; \boldsymbol{\psi} (0) ]
\;
]
,
\end{eqnarray}
and 
\begin{eqnarray}
\label{ytau}
{\bf y} (\tau)
& = & 
\boldsymbol{\eta} [  \tau ; \boldsymbol{\psi} (0) ]
\nonumber \\
& = & 
\mbox{}^t 
[ 
\;
\eta_0 [  \tau ; \boldsymbol{\psi} (0) ], 
\cdots,
\eta_{N_y - 1} [  \tau ; \boldsymbol{\psi} (0) ]
\;
]
,
\end{eqnarray}
to explicitly represent 
the ${\bf z}$ and ${\bf y}$ components of 
the state vector $\boldsymbol{\psi} (\tau)$ at time $\tau$, 
which starts from $\boldsymbol{\psi} (0)$ at the initial time $0$ and 
evolves in time according to the Schr\"{o}dinger equation. 
Since 
\begin{equation}
{\bf U}(\tau) = {\bf U} (\tau - \tau') {\bf U}(\tau')
, 
\end{equation}
we obtain 
\begin{equation}
\boldsymbol{\psi}(\tau) = {\bf U} (\tau - \tau')\boldsymbol{\psi}(\tau') 
, 
\end{equation}
from which we find 
\begin{eqnarray}
\label{zzetayeta}
{\bf z} (\tau)
& = & 
\boldsymbol{\zeta}   [  \tau ; \boldsymbol{\psi} (0) ]
=
\boldsymbol{\zeta}  [   \tau - \tau' ; \boldsymbol{\psi} (\tau') ] 
=
\boldsymbol{\zeta}   [  0 ; \boldsymbol{\psi} (\tau) ] 
,
\nonumber \\ 
{\bf y} (\tau)
& = & 
\boldsymbol{\eta}   [  \tau ; \boldsymbol{\psi} (0) ]
=
\boldsymbol{\eta}  [   \tau - \tau' ; \boldsymbol{\psi} (\tau') ] 
=
\boldsymbol{\eta}  [  0 ; \boldsymbol{\psi} (\tau) ] 
. 
\hspace*{5mm}
\end{eqnarray}
Equation~(\ref{zzetayeta}) implies that, when
selecting  different initial state vectors
$\boldsymbol{\psi} (0)$, $\boldsymbol{\psi} (\tau')$, and 
$\boldsymbol{\psi} (\tau)$ with taking 
elapsed time $\tau$, $\tau - \tau'$, and $0$, respectively,  
we have the same state vector $\boldsymbol{\psi} (\tau)$ at time $\tau$, which has  
the same ${\bf z}$ and ${\bf y}$ components.  

We here use the capital letters, 
$\boldsymbol{\Psi}(\tau) \equiv \mbox{}^t [{\bf Z}(\tau), {\bf Y}(\tau)]$,  
and the small letters, 
$\boldsymbol{\psi}(\tau) \equiv \mbox{}^t [{\bf z}(\tau), {\bf y}(\tau)]$,  
 to represent the random variables and their realizations, respectively. 
We now consider the conditional probability density function 
$P({\bf z}_B | {\bf z}_A)$ that the random variable ${\bf Z} (\tau)$ 
takes the specified value 
${\bf z}(\tau) = {\bf z}_B = \mbox{}^t [ (z_B)_0, \cdots,  (z_B)_{N_y -1} ]$ 
at time $\tau$, 
given that the initial random variable ${\bf Z} (0)$ 
takes the value 
${\bf z}(0) = {\bf z}_A  = \mbox{}^t [ (z_A)_0, \cdots,  (z_A)_{N_z -1} ]$. 
We also assume that at the initial time $t=0$, 
the system of interest and the thermal reservoir, which are randomly 
given as 
${\bf Z} (0)$  
and 
${\bf Y}  (0)$, 
respectively, 
are statistically independent of each other. 
Therefore, the initial probability density of 
$\boldsymbol{\Psi}(0) = \mbox{}^t [{\bf Z} (0), {\bf Y}  (0)]$  is written as 
\begin{equation}
 P[\boldsymbol{\psi}(0) ; 0 ]
 = 
  P_{\rm Z} [ {\bf z} (0) ; 0 ]
  P_{\rm Y} [ {\bf y} (0) ; 0 ]
.
\end{equation}
 Then, the conditional probability density 
$P({\bf z}_B | {\bf z}_A)$ 
is given by 
\begin{eqnarray}
\label{PBA}
P( {\bf z}_B | {\bf z}_A )
& = & 
\int 
d \Gamma (0) 
\;    P_Y [ {\bf y} (0) ; 0 ]
\nonumber \\ 
& & 
\mbox{}
\times
\delta^{2N_z}
[ {\bf z} (0) - {\bf z}_A ] 
\;
\delta^{2N_z}
[ {\bf z} (\tau) - {\bf z}_B ]
,
\hspace*{5mm}
\end{eqnarray}
where
\begin{eqnarray}
\delta^{2N_z}
[ {\bf z} (0) - {\bf z}_A ]
&  \equiv  & 
\prod_{n = 0}^{N_z-1}
\delta [z_{r,n} (0) - (z_A)_{r,n} ]
\nonumber \\ & & 
\hspace*{-3mm}
\mbox{} \times
\delta [z_{i,n} (0)  - (z_A)_{i,n} ]
,
\end{eqnarray}
and
\begin{eqnarray}
\delta^{2N_z}
[ {\bf z} (\tau) - {\bf z}_B ]
& \equiv & 
\prod_{n = 0}^{N_z-1}
\delta [z_{r,n} (\tau) - (z_B)_{r,n} ]
\nonumber \\ & & 
\hspace*{-3mm}
\mbox{} \times
\delta [z_{i,n} (\tau) - (z_B)_{i,n} ]
. 
\end{eqnarray}
Here,  
$(z_A)_{r,n} \equiv {\rm Re} \,(z_A)_n$, 
$(z_A)_{i,n} \equiv {\rm Im} \,(z_A)_n$, 
$(z_B)_{r,n} \equiv {\rm Re} \,(z_B)_n$, 
$(z_B)_{i,n} \equiv {\rm Im} \,(z_B)_n$, 
$
z_{r,n} (\tau) \equiv 
{\rm Re} \; z_n (\tau) 
$
and 
$
z_{i,n} (\tau) \equiv 
{\rm Im} \; z_n (\tau)
$. 
To explicitly show that the integrand on the right-hand side of 
Eq.~(\ref{PBA})
depends on $\boldsymbol{\psi}(0)$, 
the real and imaginary components of which are used as integration 
variables, 
we use Eq.~(\ref{ztau}) to rewrite Eq.~(\ref{PBA}) as 
\begin{eqnarray}
\label{PBA2}
P( {\bf z}_B | {\bf z}_A )
& = & 
\int 
d \Gamma [\boldsymbol{\psi}(0)] 
\;    P_Y ( \boldsymbol{\eta}   [  0 ; \boldsymbol{\psi} (0) ]  ; 0 )
\nonumber \\ 
& & 
\mbox{}
\times
\delta^{2N_z}
( \boldsymbol{\zeta}   [  0 ; \boldsymbol{\psi} (0) ] - {\bf z}_A) 
\nonumber \\ 
& & 
\mbox{}
\times
\delta^{2N_z}
( \boldsymbol{\zeta}   [  \tau ; \boldsymbol{\psi} (0) ] - {\bf z}_B) 
,
\end{eqnarray}
where 
$
d \Gamma [\boldsymbol{\psi}(0)]
\equiv
d \Gamma (0)
\equiv
\prod_{n = 0}^{N_c -1}
d\psi_{r,n}  (0) d\psi_{i,n} (0) 
$.

Under the condition that the initial random variable ${\bf Z} (0)$ takes 
the specified value ${\bf z} (0) = {\bf z}_A$, 
we denote 
the difference between the stochastic entropies of the probability 
distributions 
$P_Y [ {\bf y} (\tau)   ; 0 ]$ and 
$P_Y [ {\bf y} (0)  ; 0 ]$
by 
\begin{eqnarray}
\label{DSYdef}
\Delta S_Y [ {\bf z}_A, {\bf y} (0)  ; \tau ]
& \equiv & 
\log \left[
 \frac{  P_Y [ {\bf y} (0)   ; 0 ] }{
   P_Y [ {\bf y} (\tau)  ; 0 ]
 }
\right]
\nonumber \\ 
& = &  
\log \left[
 \frac{  P_Y [ \boldsymbol{\eta}   [  0 ; \boldsymbol{\psi} (0) ]   ; 0 ] }{
   P_Y [ \boldsymbol{\eta}   [  \tau ; \boldsymbol{\psi} (0) ]  ; 0 ]
 }
\right]
\nonumber \\ 
& = &  
\log \left[
 \frac{  P_Y [ \boldsymbol{\eta}   [  0 ; \boldsymbol{\psi} (0) ]   ; 0 ] }{
   P_Y [ \boldsymbol{\eta}   [  0 ; \boldsymbol{\psi} (\tau) ]  ; 0 ]
 }
\right]
, 
\end{eqnarray}
where 
${\bf y}  (\tau) = \boldsymbol{\eta}   [  \tau ; \boldsymbol{\psi} (0) ] 
= \boldsymbol{\eta}   [  0 ; \boldsymbol{\psi} (\tau) ]$
depends on ${\bf z}_A = {\bf z} (0)$ and ${\bf y} (0)$
through  $\boldsymbol{\psi}(0)$. 
The above quantity 
$\Delta S_Y [ {\bf z}_A, {\bf y} (0)  ; \tau ]$
corresponds to the entropy generated by the heat transfer from 
the system of interest to the thermal reservoirs 
in Ref.~\cite{Jarzynski}. 
Then, we use $P({\bf z}_B, \Delta S' | {\bf z}_A)$ to denote the conditional 
probability density function that 
${\bf Z} (\tau)$ and $\Delta S_Y [ {\bf Z} (0), {\bf Y} (0)  ; \tau ]$   
take the specified values ${\bf z} (\tau) = {\bf z}_B$ and $\Delta S'$, respectively, 
given that
${\bf Z}  (0)$ takes the value ${\bf z} (0) = {\bf z}_A$. 
The conditional probability density $P({\bf z}_B, \Delta S' | {\bf z}_A)$  is given by 
\begin{eqnarray}
\label{PBDSA}
& & 
\hspace*{-8mm}
P({\bf z}_B, \Delta S' | {\bf z}_A)
\nonumber \\ 
& = & 
\int 
d \Gamma (0)
\;    P_Y [ {\bf y} (0) ; 0 ]
\;
\delta^{2N_z}
( {\bf z} (0) - {\bf z}_A) 
\nonumber \\ 
& & 
\mbox{}
\times
\delta^{2N_z}
( {\bf z} (\tau) - {\bf z}_B) 
\; 
   \delta \bigl(
   \Delta S_Y [ {\bf z}_A, {\bf y} (0)  ; \tau ]
    -  \Delta S' \bigr)
\nonumber \\ 
& = & 
\int 
d \Gamma [\boldsymbol{\psi}(0)] 
\;    P_Y [ \boldsymbol{\eta}   [  0 ; \boldsymbol{\psi} (0) ]  ; 0 ]
\nonumber \\ 
& & 
\mbox{}
\times
\delta^{2N_z}
( \boldsymbol{\zeta}   [  0 ; \boldsymbol{\psi} (0) ]  - {\bf z}_A) 
\;
\delta^{2N_z}
( \boldsymbol{\zeta}   [  \tau ; \boldsymbol{\psi} (0) ] - {\bf z}_B) 
\nonumber \\ & & 
\mbox{} \times 
   \delta \bigl(
   \Delta S_Y [ \boldsymbol{\zeta}   [  0 ; \boldsymbol{\psi} (0) ] , 
   \boldsymbol{\eta}   [  0 ; \boldsymbol{\psi} (0) ]  ; \tau ]
    -  \Delta S' \bigr)
.
\end{eqnarray}
Again, 
the dependence of the integrand on $\boldsymbol{\psi}(0)$, 
is explicitly shown in the last expression of the integral in Eq.~(\ref{PBDSA}).

We here assume that there exists the time-reversal 
$N_c \times N_c$ unitary matrix ${\bf T}$ as mentioned earlier and 
that ${\bf T}$ are a block-diagonal matrix 
which acts on the state vector $\boldsymbol{\psi} (\tau)$ as 
\begin{equation}
\label{TzTy}
{\bf T} \boldsymbol{\psi} (\tau)
= 
\left[
\begin{array}{cc}
{\bf T}_z & 0
\\
0  & {\bf T}_y
\end{array}
\right]
\left[
\begin{array}{c}
{\bf z} (\tau) 
\\
{\bf y} (\tau) 
\end{array}
\right]
=
\left[
\begin{array}{c}
{\bf T}_z {\bf z} (\tau) 
\\
{\bf T}_y {\bf y} (\tau) 
\end{array}
\right]
, 
\end{equation}
where ${\bf T}_z$ and ${\bf T}_y$ are $N_z \times N_z$ and $N_y \times N_y$ 
unitary matrices, respectively. 

The detailed fluctuation theorem holds 
under the assumption that the initial probability density distribution 
$P_Y [  {\bf y} (0) ; 0 ]$ is symmetric with respect to 
the time-reversal transformation, 
\begin{equation}
\label{PYTY0}
P_Y [ {\bf T}_y {\bf y} (0) (0) ; 0 ] 
= 
P_Y [ {\bf y} (0) ; 0 ] 
. 
\end{equation}
Then, 
as shown in Appendix~C, 
we can follow a procedure similar to that 
in Ref.~\cite{Jarzynski} to 
prove that the present system obeys the detailed fluctuation theorem, 
\begin{equation}
\label{DFT}
\frac{
P({\bf z}_B, \Delta S' | {\bf z}_A)}{
P(  {\bf T}_z {\bf z}_A, - \Delta S' | {\bf T}_z {\bf z}_B )
}
= 
\exp \Delta S'
.
\end{equation}
Using Eqs.~(\ref{DSYdef}) and (\ref{PBDSA}), we obtain 
\begin{eqnarray}
\langle \Delta S_Y \rangle_{{\bf z}_A}
& = & 
\int d \Gamma_{zB}
\int_{-\infty}^{+ \infty} d (\Delta S') \, \Delta S'
\nonumber \\
&  &
\mbox{}
\times
P({\bf z}_B, \Delta S' | {\bf z}_A) 
\nonumber \\
& = &
\int d \Gamma_y(0)
 P_Y [ {\bf y} (0) ; 0 ] 
 \log \left[
 \frac{  P_Y [ {\bf y} (0)  ; 0 ] }{
   P_Y [ {\bf y} (\tau) ; 0 ]
 }
 \right]
 \nonumber \\
 &  \geq  & 
 0
,
\end{eqnarray}
where 
$\langle \cdots \rangle_{{\bf z}_A}$ represents the ensemble average 
under the condition that 
${\bf Z} (0)$ takes the specified value ${\bf z} (0) = {\bf z}_A$. 
The volume elements $d \Gamma_{zB}$ and $d \Gamma_y(0)$ are defined by 
$
 d \Gamma_{zB}
\equiv
\prod_{n=0}^{N_z -1} d (z_B)_{r, n} d (z_B)_{i, n}
$ 
and 
$
 d \Gamma_y (0)
\equiv
\prod_{n=0}^{N_y -1} d y_{r, n}(0)  d y_{i, n} (0)
$ 
respectively, 
where $y_{r, n} (0) \equiv {\rm Re} \; y_n (0)$ 
and $y_{i, n} (0) \equiv {\rm Im} \; y_n (0)$. 
We see that $\langle \Delta S_Y \rangle_{{\bf z}_A}$ is the relative entropy 
(Kullback-Leibler divergence) of the initial probability distribution 
$P_Y [ {\bf y} (0) ; 0 ]$ 
with respect to that with 
${\bf y} (0)$ replaced by
${\bf y} (\tau)$. 

We can extend the above argument to the case in which the system of interest passes through multiple specified states between times $0$ and $\tau$.
We now use $P({\bf z}_1, {\bf z}_2, \cdots, {\bf z}_M, \Delta S' | {\bf z}_0)$ 
to denote the conditional probability density function that 
${\bf Z} (\tau)$ takes the specified values ${\bf z}_1$, ${\bf z}_2$, $\cdots$, ${\bf z}_M$ at 
times $\tau_1$, $\tau_2$, $\cdots$, $\tau_M (= \tau)$, respectively, and that 
$\Delta S_Y [ {\bf Z} (0), {\bf Y} (0)  ; \tau ]$   
takes the value $\Delta S'$, given that 
${\bf Z}  (0)$ takes the specified value ${\bf z}_0$. 
Then, $P({\bf z}_1, {\bf z}_2, \cdots, {\bf z}_M, \Delta S' | {\bf z}_0)$ is given by
\begin{eqnarray}
& & 
\hspace*{-5mm}
P({\bf z}_1, {\bf z}_2, \cdots, {\bf z}_M, \Delta S' | {\bf z}_0) 
\nonumber \\ 
& = & 
\int 
d \Gamma (0)
\;    P_Y [ {\bf y} (0) ; 0 ]
\left(
\prod_{m=0}^M
\delta^{2N_z}
( {\bf z} ( \tau_m) - {\bf z}_m) 
\right)
\nonumber \\  
& & 
\mbox{}
\times
   \delta \bigl(
   \Delta S_Y [ {\bf z} (0), {\bf y} (0)  ; \tau ]
    -  \Delta S' \bigr)
\nonumber \\ 
& = & 
\int 
d \Gamma [\boldsymbol{\psi}(0)] 
\;    P_Y [ \boldsymbol{\eta}   [  0 ; \boldsymbol{\psi} (0) ]  ; 0 ]
\nonumber \\ 
& & 
\mbox{}
\times
\left(
\prod_{m=0}^M
\delta^{2N_z}
( \boldsymbol{\zeta}   [  \tau_m ; \boldsymbol{\psi} (0) ]  - {\bf z}_m) 
\right)
\nonumber \\ & & 
\mbox{} \times 
   \delta \bigl(
   \Delta S_Y [ \boldsymbol{\zeta}   [  0 ; \boldsymbol{\psi} (0) ] , 
   \boldsymbol{\eta}   [  0 ; \boldsymbol{\psi} (0) ]  ; \tau ]
    -  \Delta S' \bigr)
. 
\end{eqnarray}
The procedure used to derive Eq.~(\ref{DFT})
can be extended to show 
\begin{equation}
\label{DFTM}
\frac{
P({\bf z}_1, {\bf z}_2, \cdots, {\bf z}_M, \Delta S' | {\bf z}_0) 
}{
P({\bf T}_z {\bf z}_{M-1}, {\bf T}_z {\bf z}_{M-2}, 
\cdots, {\bf T}_z {\bf z}_0, - \Delta S' | {\bf T}_z {\bf z}_M) 
}
=
\exp \Delta S'
,
\end{equation}
where 
$P({\bf T}_z {\bf z}_{M-1}, {\bf T}_z {\bf z}_{M-2}, 
\cdots, {\bf T}_z {\bf z}_0, - \Delta S' | {\bf T}_z {\bf z}_M)$ 
denotes the conditional probability density function that 
${\bf Z}(\tau-\tau_{M-i})={\bf T}_z{\bf z}_{M-i}$
for $i=1,2,\cdots,M-1$,
${\bf Z}(\tau)={\bf T}_z{\bf z}_0$,
and 
$\Delta S_Y [ {\bf Z} (0), {\bf Y} (0)  ; \tau ]= 
- \Delta S'$, given that 
${\bf Z}  (0) = {\bf T}_z {\bf z}_M$. 
Furthermore, taking the limit $M\rightarrow \infty$ in Eq.~(\ref{DFTM}), 
we obtain 
\begin{equation}
\label{DFTcp}
\frac{
P[  \{  {\bf z}_+ (\tau') \}_{0 \leq \tau' \leq \tau}, \Delta S' | {\bf z}_+ (0) ]
}{
P[ \{ {\bf z}_- (\tau') \}_{0 \leq \tau' \leq \tau}, - \Delta S' | {\bf z}_- (0) ]
}
=
\exp \Delta S'
\end{equation}
where 
$\{ {\bf z}_+ (\tau') \}_{0 \leq \tau' \leq \tau}$ denotes a specified 
continuous path of ${\bf Z}  (\tau')$ over the interval
$0 \leq \tau' \leq \tau$, 
and $\{ {\bf z}_- (\tau') \}_{0 \leq \tau' \leq \tau}$ 
denotes the corresponding time-reversed path defined by
\begin{equation}
\label{zmp}
{\bf z}_- (\tau')
\equiv
{\bf T}_z  {\bf z}_+ (\tau - \tau')
. 
\end{equation}
It should be noted that the detailed fluctuation theorem remains valid
when ${\bf T}(\cdot)$ is replaced by ${\bf K}(\cdot)^*$
throughout the above arguments.

\section{Linear Vlasov-Poisson System}

Following Ref.~\cite{Sugama2025}, we reformulate the linearized Vlasov-Poisson system as a Schr\"{o}dinger equation.
Details about Hamiltonian eigenvectors associated with the Case-Van Kampen (CVK) modes, time-reversal operators, 
and a newly derived approximate expression for the probability density function of the stochastic relative entropy 
(which are not included in Ref.~\cite{Sugama2025})
are provided in this section and Appendixes~D and E. 

\subsection{Linearized Vlasov-Poisson equations and invariant}

The electron distribution function in the two-dimensional phase space $(x,v)$ at time $t$ 
is denoted by
 $f(x, v, t)$. 
It is obtained from the six-dimensional distribution function $F(x, y, z, v_x, v_y, v_z, t)$ 
as $f(x, v_x, t) = \int_{-\infty}^{+\infty} dv_y \int_{-\infty}^{+\infty} dv_z 
F(x, y, z, v_x, v_y, v_z, t)$,  
and $F$ is assumed to be independent of $y$ and $z$. 
We decompose $f(x,v,t)$ into a Maxwellian equilibrium,
$
f_0 (v)
= 
\pi^{-1/2} (n_0/v_T)
\exp
( - v^2/v_T^2 )
$ 
and a perturbation $f_1(x,v,t)$. 
Here, 
$
v_T \equiv \sqrt{2} v_t \equiv 
\sqrt{2T/m}
$, 
where $n_0$ and $T$ are the equilibrium electron density and temperature, respectively, 
and $m$ is the electron mass. 
Assuming a collisionless plasma, 
we employ 
the linearized Vlasov equation~\cite{Nicholson}, 
\begin{equation}
\frac{\partial f_1(x, v, t)}{\partial t}
+
v
\frac{\partial f_1(x, v, t)}{\partial x}
- 
\frac{e}{m} E(x, t) 
\frac{\partial f_0(v)}{\partial v}
=
0
, 
\end{equation}
where $-e$ is the electron charge and 
the nonlinear term $-  (e / m) E(x, t) \partial f_1(x, v, t) / \partial v$ is neglected. 
Ions are assumed to have a uniform density $n_0$, and the ion motion 
is ignored because the ion mass is much larger than the electron mass.
The electric field $E(x, t)$ in the $x$-direction is determined 
from Poisson's equation, 
\begin{equation}
\label{PoissonEq}
\frac{\partial E (x, t) }{ \partial x }
= - 4 \pi e
 \int_{-\infty}^{+\infty} d v
\,  f _1 (x, v, t)
.
\end{equation}
We assume the system to be periodic with period $L$ in the $x$-direction,
and impose the constraint condition,  $\int_{-L/2}^{L/2} dx \, E(x,t) = 0$. 
Here, we do not consider an equilibrium electric field that could 
give rise to an inhomogeneous equilibrium distribution of electrons.

It can be shown that
\begin{equation}
\label{Df1}
D[f_1]
\equiv 
\int_{-L/2}^{+L/2}\frac{dx}{L}
\biggl[
 \frac{ [E(x, t)]^2 }{8 \pi n_0 T}
+ 
\frac{1}{n_0}
\int_{-\infty}^{+\infty} dv \; \frac{ [f_1(x,v,t)]^2 }{2 f_0(v)}
\biggr]
, 
\end{equation}
is conserved for any solution of the linearized Vlasov-Poisson equations.
From Eqs.~(\ref{PoissonEq}) and (\ref{Df1}), one finds that 
the invariant functional $D[f_1]$ takes a quadratic form with respect to $f_1$. 
It is well known that 
the energy and the Gibbs entropy per single electron, 
defined by 
$
{\cal E}
\equiv
(n_0 L)^{-1}
\int_{-L/2}^{L/2} dx \,
[
\int_{-\infty}^{+\infty} dv 
f m v^2/ 2
+
E^2/ 8\pi 
]
$
and 
$
S_f
\equiv
-
(n_0 L)^{-1}
\int_{-L/2}^{L/2} dx \int_{-\infty}^{+\infty} dv 
f \log f
$,
respectively, 
are conserved in the nonlinear Vlasov-Poisson system 
although they are not in the linear system. 
As noted in Ref.~\cite{Sugama2025}, 
we obtain the relation,  
$
D[f_1] =   {\cal E}^{(2)} / T - S_f^{(2)} 
$, 
where ${\cal E}^{(2)}$ and $S_f^{(2)}$ denote the second-order terms in the expansions of 
the ${\cal E}$ and  $S_f$, 
respectively, 
with respect to the ordering parameter $\alpha \sim f_1/f_0$, 
characterizing the perturbation amplitude~\cite{Maekaku}.

\subsection{Representing functions in velocity space as ket vectors}

We consider the solution 
$ f_1 (x, v, t)$ of the linearized Vlasov-Poisson equations written as 
$
f_1 (x, v, t) = \mbox{Re}[ f_1(k, v, t) \exp ( i k  x ) ]
$, 
and define the normalized time and velocity by 
$\tau \equiv k v_T t$
and $\xi \equiv  v / v_T$, respectively.
Using the Hermite polynomials, 
$
H_n(\xi) \equiv (-1)^n e^{\xi^2} d^n (e^{-\xi^2}) /d\xi^n 
\;
( n = 0, 1,  2, \cdots )
$, 
we introduce 
\begin{equation}
h_n (\xi)
\equiv
\frac{e^{-\xi^2/2}}{\pi^{1/4} }
 \frac{H_n(\xi)}{
 (2^n n!)^{1/2}
 }
,
\end{equation}
which satisfy 
$
\int_{-\infty}^{+\infty} d\xi \;  
h_n (\xi) h_{n'} (\xi) 
= \delta_{n n'}
$.
We also define the dimensionless function 
$\widetilde{f} (\kappa, \xi, \tau)$ by 
\begin{equation}
f_1(k, v, t) 
= (n_0/v_T) h_0 (\xi ) \widetilde{f} (\kappa, \xi, \tau) 
,
\end{equation}
where the normalized wavenumber is defined by $\kappa \equiv k \lambda_D$ 
with the Debye length $\lambda_D \equiv \omega_p/ v_t$ 
and the plasma frequency $\omega_p = (4\pi n_0 e^2/m)^{1/2}$. 
The linearized Vlasov--Poisson equations then reduce to 
\begin{eqnarray}
\label{tildefeq}
i \frac{\partial}{\partial \tau}
\widetilde{f} (\kappa, \xi, \tau) 
& = & 
\xi \widetilde{f} (\kappa, \xi, \tau) 
+
\kappa^{-2} \xi h_0(\xi) 
\nonumber \\ & & \mbox{}
\times
\int_{-\infty}^{+\infty} d\xi' \;
h_0(\xi' ) \widetilde{f} (\kappa, \xi', \tau) 
.
\end{eqnarray}

Following the notation of quantum mechanics,~\cite{QM}
we associate complex-valued functions of the normalized velocity $\xi$
with ket vectors denoted by $|\;\rangle$.
Thus, the ket vectors represent electron distributions in velocity space rather than in position space. 
The basis vectors $|n\rangle$ and $|\xi'\rangle$ correspond to
the functions $h_n(\xi)$ and $\delta(\xi-\xi')$, respectively.
The bra vector conjugate to $|u\rangle$ is denoted by $\langle u|$.
 Then,  $\delta (\xi - \xi')$ and $h_n (\xi)$ are expressed as 
\begin{equation}
\langle \xi | \xi' \rangle 
= 
\delta (\xi - \xi')
\end{equation}
and 
\begin{equation}
\langle \xi | n \rangle 
= 
h_n (\xi) 
,
\end{equation}
respectively, and the orthonormality condition satisfied by $h_n (\xi)$ is rewritten as 
\begin{equation}
\langle n | n' \rangle 
= 
\int_{-\infty}^{+\infty} d\xi \;  
\langle n | \xi \rangle \langle \xi | n' \rangle 
= \delta_{n n'}
.
\end{equation}
The sets $\{ | \xi  \rangle \}_{-\infty < \xi < + \infty}$ and 
$\{ | n  \rangle \}_{n = 0, 1, 2, \cdots}$ 
form two orthonormal bases satisfying the closure relation, 
\begin{equation}
\int_{-\infty}^{+\infty} 
| \xi \rangle d\xi \langle \xi | 
= 
\sum_{n=0}^\infty
| n \rangle  \langle n | 
= 
\widehat{1}
,
\end{equation}
where $\widehat{1}$ denotes the identity operator.
We define the operators 
\begin{equation}
\label{hatXi0}
\widehat{\Xi}
\equiv 
\int_{-\infty}^{+\infty} d\xi \; | \xi \rangle 
\xi d \xi \langle \xi |
\end{equation}
and 
\begin{equation}
\widehat{N}
\equiv 
\sum_{n=0}^\infty  | n \rangle 
n \langle n |
,
\end{equation}
which satisfy 
$
\widehat{\Xi} | \xi \rangle 
= 
\xi | \xi \rangle 
$
and 
$
\widehat{N}  | n \rangle 
= 
n  | n \rangle 
$.
A representation of state vectors and operators refers to expressing them as column vectors and matrices of complex numbers with respect to a chosen set of basis vectors, and it depends on that choice.~\cite{QM}
The representations associated with
$\{| \xi\rangle\}_{-\infty<\xi<+\infty}$
and
$\{|n\rangle\}_{n=0,1,2,\cdots}$
are referred to as the
$\{\Xi\}$ and $\{N\}$ representations, respectively.

We define the ket vector
$|\widetilde{f}(\tau)\rangle$
by
\begin{equation}
\langle \xi  | \widetilde{f} (\tau) \rangle
= 
\widetilde{f} (\kappa, \xi, \tau) 
,
\end{equation}
where the $\kappa$-dependence is omitted in the notation $| \widetilde{f} (\tau) \rangle$ for simplicity.  
Equation~(\ref{tildefeq}) can then be written as
\begin{equation}
\label{fketeq}
i \frac{d}{d \tau}
| \widetilde{f} (\tau) \rangle
= 
\widehat{\Xi} 
\bigl( \widehat{1} + \kappa^{-2} | 0 \rangle \langle 0 | \bigr) 
| \widetilde{f} (\tau) \rangle
.
\end{equation}
Since
$
\widehat{\Xi} 
( \widehat{1} + \kappa^{-2} | 0 \rangle \langle 0 | ) 
$ 
is not Hermitian, 
 Eq.~(\ref{fketeq}) does not 
take a form of the Schr\"{o}dinger equation, 
and $\langle \widetilde{f} (\tau) | \widetilde{f} (\tau) \rangle$ 
is not conserved.
The invariant $D[f_1]$ of the linearized Vlasov-Poisson system is expressed as
\begin{eqnarray}
\label{Df1a}
D[f_1] 
& = & 
\frac{1}{4}
\biggl[
\int_{-\infty}^{+\infty} d\xi \; [ \widetilde{f}(\kappa, \xi, \tau) ]^2
\nonumber
\\ & & 
\mbox{} 
+ \kappa^{-2} 
\Big| \int_{-\infty}^{+\infty} d\xi \; 
h_0 (\xi) \widetilde{f}(\kappa, \xi, \tau) \Bigr|^2
\biggr]
\nonumber
\\ 
& = & 
 \frac{1}{4}
\left[  
\langle  \widetilde{f} (\tau) |  \widetilde{f} (\tau) \rangle
+ \kappa^{-2} 
\langle  \widetilde{f} (\tau) | 0 \rangle
\langle 0 | \widetilde{f} (\tau) \rangle
\right]
\nonumber
\\
& = & 
 \frac{1}{4}
\langle \widetilde{f} (\tau) | 
\bigl( \widehat{1} + \kappa^{-2}  | 0 \rangle \langle 0 | \bigr) 
 | \widetilde{f} (\tau) \rangle
. 
\end{eqnarray}

We define the Hermitian operator $\widehat{A}$ by
\begin{eqnarray}
\label{hatA0}
\widehat{A}
& = & 
\widehat{A}^\dagger
=
\sum_{n=0}^\infty
| n \rangle
\left( 1 + \kappa^{-2} \delta_{n0} \right)^{1/2} 
\langle n | 
\nonumber
\\ 
& = & 
\widehat{1}
+ 
| 0 \rangle
\bigl[
\left( 1 + \kappa^{-2} \right)^{1/2} - 1
\bigr]
\langle 0 | 
,
\end{eqnarray}
where $\dagger$ represents the Hermitian-conjugate operator. 
Its inverse and square are
\begin{eqnarray}
\widehat{A}^{-1}
& = & 
\sum_{n=0}^\infty
| n \rangle
\left( 1 + \kappa^{-2} \delta_{n0} \right)^{-1/2} 
\langle n | 
\nonumber
\\ 
& = & 
\widehat{1}
+ 
| 0 \rangle
\bigl[
\left( 1 + \kappa^{-2} \right)^{-1/2} - 1
\bigr]
\langle 0 | 
\end{eqnarray}
and 
\begin{equation}
\label{Asquare}
\widehat{A}^2
= 
\sum_{n=0}^\infty
| n \rangle
\left( 1 + \kappa^{-2} \delta_{n0} \right)
\langle n | 
= 
\widehat{1}
+ 
| 0 \rangle
\kappa^{-2}
\langle 0 | 
,
\end{equation}
respectively. 

Defining the state vector 
$
| \psi (\tau) \rangle \equiv 
\widehat{A} |\widetilde{f}(\tau)\rangle
$,
and using Eqs.~(\ref{Df1a}) and (\ref{Asquare}), 
the invariant becomes 
\begin{equation}
D[f_1]
= 
 \frac{1}{4}
\langle \widetilde{f} (\tau) | 
\widehat{A}^2
 | \widetilde{f} (\tau) \rangle
 = 
  \frac{1}{4}
\langle \psi (\tau) 
 | \psi (\tau) \rangle
.
\end{equation}
Therefore, 
$\langle \psi (\tau) | \psi (\tau) \rangle$ 
is conserved, and 
the time evolution operator $\widehat{U}(\tau)$ defined by 
$
 | \psi (\tau) \rangle
= 
\widehat{U}(\tau)
 | \psi (0) \rangle
$
is unitary. 
We define the Hamiltonian operator by 
\begin{equation}
\label{hatH0}
\widehat{H} 
=
\widehat{A} \, \widehat{\Xi} \, \widehat{A}
, 
\end{equation}
which is manifestly Hermitian. 
Applying $\widehat{A}$ to both sides of Eq.~(\ref{fketeq}) 
and using Eq.~(\ref{hatH0}), we find that 
$| \psi (\tau) \rangle$ satisfies 
the Schr\"{o}dinger equation shown in Eq.~(\ref{Schreq}) and that 
the time evolution operator is expressed as 
$
\widehat{U} (\tau)
= 
\exp ( - i \tau \widehat{H} )
$. 

Using 
$
2 \xi H_n (\xi) = H_{n+1} (\xi) + 2 n H_{n-1} (\xi)
$, 
we can express 
the Hamiltonian $\widehat{H}$ in Eq.~(\ref{hatH0}) and the Schr\"{o}dinger equation 
in the $\{ N \}$ representation as 
\begin{equation}
\label{HN}
\widehat{H}
=
\frac{1}{\sqrt{2}}
\sum_{n=0}^\infty 
\sqrt{n + 1 + \kappa^{-2} \delta_{n0} }
\Bigl(
| n + 1 \rangle \langle n | + | n \rangle \langle n + 1 |
\Bigr)
, 
\end{equation}
and
$
i  \, 
d \psi_n (\tau)  / d\tau
=
\sum_{n'=0}^\infty H_{n n'} \psi_{n'} (\tau) 
$,
respectively, 
where 
$ \psi_n (\tau)  \equiv \langle n | \psi (\tau) \rangle$ 
and
\begin{eqnarray}
\label{Hnn}
& & 
\hspace*{-3mm}
 H_{n n'} 
\equiv 
\langle n | \widehat{H} | n' \rangle 
\nonumber \\ 
&  & 
=
\frac{1}{\sqrt{2}}
\big[
\delta_{n, n'+1}
\sqrt{n + \kappa^{-2} \delta_{n' 0}} 
+
\delta_{n+1, n'}
\sqrt{n' + \kappa^{-2} \delta_{n0} }
\big]
.
\nonumber \\ & & 
\end{eqnarray} 
In Eq.~(\ref{HN}),  
$| n+1 \rangle \langle n |$ and  $| n \rangle \langle n+1 |$ 
act as creation and annihilation operators, respectively, on the basis states
basis vectors $|n\rangle$ $(n=0, 1, 2, \cdots)$. 
They describe the energy transfer in Landau damping 
from macroscopic to increasingly fine velocity-space structures 
through transitions among the discrete states 
$| n \rangle$ $(n=0, 1, 2, \cdots)$.

\subsection{Time-reversal operators in the linear Vlasov-Poisson system}

We here define the time-reversal unitary and antiunitary operators, 
$\widehat{T}$ and $\widehat{K}$, for the linear Vlasov-Poisson system.
They need to satisfy the conditions given by Eqs.~(\ref{TK}) 
and (\ref{THK}) in Appendix~A. 

The time-reversal unitary operator $\widehat{T}$ is 
the linear operator [see Eq.~(\ref{A1})] defined by 
\begin{equation}
\widehat{T} | \xi \rangle 
=  
| - \xi \rangle 
\hspace*{3mm}
( - \infty < \xi < + \infty )
.
\end{equation}
Equivalently, it is defined by 
\begin{equation}
\label{Tn}
\widehat{T} | n \rangle 
=  
(- 1)^n
| n \rangle 
\hspace*{3mm}
(  n = 0, 1, 2, \cdots )
.
\end{equation}
It follows that 
$\widehat{T} = \widehat{T}^{-1} = \widehat{T}^\dagger$ 
as required from Eq.~(\ref{TK}). 
Furthermore, 
\begin{equation}
\label{TXiA}
\widehat{T} \, \widehat{\Xi} = - \widehat{\Xi} \, \widehat{T}
,
\hspace*{3mm}
\widehat{T} \, \widehat{A} =  \widehat{A} \, \widehat{T}
, 
\end{equation}
where $\widehat{\Xi}$ and $\widehat{A}$ are defined by 
Eqs.~(\ref{hatXi0}) and (\ref{hatA0}), respectively. 
Using Eqs.~(\ref{hatH0}) and (\ref{TXiA}) then yields 
$\widehat{T} \, \widehat{H} = - \widehat{H} \, \widehat{T}$,  
which is required from Eq.~(\ref{THK}). 
Note that the state vector and the perturbed distribution function 
are written as 
$| \psi (\tau) \rangle = \sum_{n=0}^\infty | n \rangle \psi_n (\tau)$ 
and 
$f_1(k, v, t) 
= 
(n_0/v_T) \pi^{-1/2} e^{-\xi^2}
\sum_{n=0}^\infty (1+ \kappa^{-2} \delta_{n0} )^{-1/2} \psi_n (\tau) H_n (\xi) / (2^n n!)
$, respectively. 
Therefore, 
the transformation 
$| \psi (\tau) \rangle = \sum_{n=0}^\infty | n \rangle \psi_n (\tau)$ $\rightarrow$ 
$\widehat{T} | \psi (\tau) \rangle 
= \sum_{n=0}^\infty (-1)^n | n \rangle \psi_n (\tau)$ 
corresponds to $f_1(k, v, t)$ $\rightarrow$ 
$f_1(k, - v, t)$. 

The time-reversal antiunitary operator $\widehat{K}$ is defined as the antilinear operator [see Eq.~(\ref{A2})] satisfying 
\begin{equation}
\widehat{K} | \xi \rangle 
=  
| \xi \rangle 
\hspace*{3mm}
( - \infty < \xi < + \infty )
.
\end{equation}
Equivalently, 
\begin{equation}
\label{Kn}
\widehat{K} | n \rangle 
=  
| n \rangle 
\hspace*{3mm}
(  n = 0, 1, 2, \cdots )
, 
\end{equation}
because 
$| n \rangle  = \int_{-\infty}^{+\infty} | \xi \rangle 
d \xi  \langle \xi | n \rangle$ where 
 $\langle \xi | n \rangle$ $( - \infty < \xi < + \infty )$ are real.   
Since $\widehat{K}$ is antilinear, 
\begin{equation}
\widehat{K} | \psi \rangle 
=  
\int_{-\infty}^{+\infty} | \xi \rangle 
( \langle \xi | \psi \rangle )^*
=
\sum_{n=0}^\infty 
| n \rangle 
( \langle n | \psi \rangle )^*
,  
\end{equation}
for an arbitrary state vector $| \psi \rangle$. 
Thus, in both $\{ \Xi \}$ and $\{ N \}$ representations, 
 $\widehat{K}$ acts as 
complex conjugation of the components of the state vector.
It follows that 
$\widehat{K} = \widehat{K}^{-1} = \widehat{K}^\dagger$ 
as required from Eq.~(\ref{TK}).
Furthermore, 
\begin{equation}
\label{KXiA}
\widehat{K} \, \widehat{\Xi} =  \widehat{\Xi} \, \widehat{K}
,
\hspace*{3mm}
\widehat{K} \, \widehat{A} =  \widehat{A} \, \widehat{K}
.
\end{equation}
Using Eqs.~(\ref{hatH0}) and (\ref{KXiA}) then yields 
$\widehat{K} \, \widehat{H} =  \widehat{H} \, \widehat{K}$, 
which is required from Eq.~(\ref{THK}).

\subsection{Case-Van Kampen modes} 

The perturbed distribution function for 
the Case-Van Kampen (CVK) mode~\cite{Case,VK,VK&F,Nicholson}  
is defined by
$f_{{\rm CVK}, \zeta} (k, v, t) \equiv (n_0/v_T) h_0(\xi) 
\widetilde{f}_{{\rm CVK}, \zeta}(\kappa, \xi, \tau)$ 
$(-\infty < \zeta < + \infty)$, 
where $\widetilde{f}_{{\rm CVK}, \zeta}(\kappa, \xi, \tau)$ 
is represented as 
\begin{eqnarray}
 & & 
 \hspace*{-7mm}
  \widetilde{f}_{{\rm CVK}, \zeta}(\kappa, \xi, \tau)
   \equiv  
 \langle \xi | \widetilde{f}_{{\rm CVK}, \zeta} \rangle
\nonumber 
\\ & & 
\hspace*{-7mm}
= \frac{1}{h_0 (\xi)} 
\left[ 
\delta ( \xi  - \zeta )  \, {\rm Re} [ \epsilon (\xi)]
-  \frac{1}{\pi} 
P \left( \frac{1}{\xi - \zeta} \right) 
\, {\rm Im} [ \epsilon (\xi)] 
\right]
. 
\hspace*{4mm}
\end{eqnarray}
Here, 
$
\epsilon(\zeta)
 \equiv 
1 + \kappa^{-2}
[ 1 + \zeta Z(\zeta) ]
$, 
where $Z(\zeta)$ is the plasma dispersion function,
$
Z(\zeta)
=  \pi^{-1/2} P
\int_{-\infty}^{+\infty} dz \,
e^{-z^2}/ (z - \zeta)
+ i \pi^{1/2} e^{-\zeta^2}
$
defined for real $\zeta$.
The CVK state vector is defined by
\begin{equation}
| {\rm CVK}, \zeta \rangle
\equiv 
\frac{h_0(\zeta)} {|\epsilon(\zeta)|}
\widehat{A} 
| \widetilde{f}_{{\rm CVK}, \zeta} \rangle
,
\end{equation}
and satisfies the eigenvalue equation, 
\begin{equation}
\label{Heveq}
\widehat{H} 
| {\rm CVK}, \zeta \rangle
=
\zeta 
| {\rm CVK}, \zeta \rangle
.
\end{equation}
The set 
$| {\rm CVK}, \zeta \rangle$ $(-\infty < \zeta < + \infty)$
constitutes a complete orthonormal basis satisfying 
\begin{equation}
\langle {\rm CVK}, \zeta
| {\rm CVK}, \zeta' \rangle
= 
 \delta (\zeta - \zeta') 
\end{equation}
and 
\begin{equation}
\int_{-\infty}^{+\infty}
| {\rm CVK}, \zeta \rangle
d\zeta
\langle {\rm CVK}, \zeta |
= 
\widehat{1}
.
\end{equation}
In the $\{{\rm CVK}\}$ representation, 
the Hamiltonian and time-evolution operator are diagonalized as
\begin{equation}
\widehat{H}
= 
\int_{-\infty}^{+\infty}
| {\rm CVK}, \zeta \rangle
\zeta  d\zeta
\langle {\rm CVK}, \zeta |
\end{equation}
and 
\begin{equation}
 \widehat{U} (\tau)
= 
\int_{-\infty}^{+\infty}
| {\rm CVK}, \zeta \rangle
e^{-i \zeta \tau} d\zeta
\langle {\rm CVK}, \zeta |
,
\end{equation}
respectively. 
Accordingly, 
$|\psi (\tau)\rangle = \int_{-\infty}^{+\infty}
| {\rm CVK}, \zeta \rangle e^{-i \zeta \tau}  d\zeta
\langle {\rm CVK}, \zeta | \psi (0) \rangle$
and
$\langle {\rm CVK}, \zeta |\psi (\tau)\rangle$ $=$
$e^{-i \zeta \tau}  \langle {\rm CVK}, \zeta | \psi (0) \rangle$. 
Appendix~D details the relation between the
$\{{\rm CVK}\}$ and $\{N\}$ representations, 
which is useful for expressing CVK-mode solutions in terms of the Hermite expansion.

\subsection{State-vector subspace generated by a finite number of Case-Van Kampen modes} 

Hereafter, we consider a finite set of CVK state vectors,
$\{ | {\rm CVK}, \zeta_j \rangle \}_{j =0, 1, \cdots, N_{\rm cvk}-1 }$ 
for a given positive integer $N_{\rm cvk}$. 
The quantities
$\{\zeta_j\}_{j=0,1,\cdots,N_{\rm cvk}-1}$
are the $N_{\rm cvk}$ real solutions of the $N_{\rm cvk}$th-order algebraic equation
given from the condition 
$
\langle N_{\rm cvk} | {\rm CVK}, \zeta \rangle = 0
$
[see Eqs.~(\ref{D2}) and (\ref{npsiH}) in Appedix~D], 
where $\langle N_{\rm cvk}|$ denotes the $N_{\rm cvk}$th basis bra vector in the $\{N\}$ representation.
Rather than treating the full state space, we restrict ourselves to the subspace spanned by these CVK state vectors. 
Since they are eigenvectors of the Hamiltonian $\widehat{H}$, this subspace is invariant under both $\widehat{H}$ and the time-evolution operator $\widehat{U}(\tau)$.
Any state vector in this subspace can be written as
\begin{equation}
| \psi (\tau) \rangle
= \sum_{j=0}^{ N_{\rm cvk} - 1 }
c_j (\tau)  | {\rm CVK}, \zeta_j \rangle
, 
\end{equation}
where 
$
c_j (\tau) = c_j (0) \exp ( - i \zeta_j \tau )
$.
The components of $|\psi(\tau)\rangle$ in the $\{N\}$ representation are therefore
$
\psi_n (\tau) 
  \equiv  
\langle n 
| \psi (\tau) \rangle
=
 \sum_{j=0}^{ N_{\rm cvk} - 1 }
c_j (\tau)  \langle n | {\rm CVK}, \zeta_j \rangle
$. 
Note that 
$
\psi_{N_{\rm cvk}} (\tau) = 0
$
holds for any $\tau$. 
Extracting the first $N_{\rm cvk}$ components,
$\{\psi_n(\tau)\}_{n=0,1,\cdots,N_{\rm cvk}-1}$,
we define the $N_{\rm cvk}$-dimensional complex column vector, 
\begin{equation}
\boldsymbol{\psi} (\tau)
\equiv   
\mbox{}^t
[
\psi_0 (\tau), \psi_1 (\tau), \cdots, 
\psi_{N_{\rm cvk}-1} (\tau)
]
.
\end{equation}
This construction establishes a one-to-one correspondence between
$\boldsymbol{\psi}(\tau)$
and the vectors in the subspace spanned by the $N_{\rm cvk}$ CVK state vectors.

The vector 
$\boldsymbol{\psi} (\tau)$ 
satisfies 
the Schr\"{o}dinger equation, Eq.~(\ref{SchNc}), 
whose Hermitian Hamiltonian matrix
${\bf H} = [  H_{n n'}  ]_{n,n'= 0, 1, \cdots, N_{\rm cvk}-1}$ 
is the $N_{\rm cvk}\times N_{\rm cvk}$ submatrix of the
infinite-dimensional matrix
$[H_{nn'}]_{n,n'=0,1,2,\cdots}$
defined by Eq.~(\ref{Hnn}). 
Accordingly, $\boldsymbol{\psi}(\tau)$ may be viewed 
as the solution obtained by truncating 
the infinite-dimensional Schr\"odinger equation 
in the $\{N\}$ representation to dimension $N_{\rm cvk}$.
At the same time, through its one-to-one correspondence 
with the state vectors in the subspace spanned by the $N_{\rm cvk}$ CVK states, 
$\boldsymbol{\psi}(\tau)$ also represents the exact solution within that subspace.
The solution of Eq.~(\ref{SchNc}) is given by 
$
\boldsymbol{\psi} (\tau)
=
{\bf U} (\tau) 
\boldsymbol{\psi} (0)
$
with the unitary matrix ${\bf U} (\tau) = \exp (-i \tau \mathbf{H})$.  
Hence, the squared norm, 
$
|| \boldsymbol{\psi} (\tau) ||^2
\equiv
\sum_{n=0}^{N_{\rm cvk} - 1} | \psi_n (\tau) |^2
$, 
is conserved. 
Note, however, that unlike ${\bf H}$, the matrix ${\bf U}(\tau)$ is not the
$N_{\rm cvk}\times N_{\rm cvk}$ submatrix of the infinite-dimensional evolution 
matrix,
$[  \langle n | \widehat{U}(\tau) | n' \rangle  ]_{n,n'= 0, 1, 2, \cdots}
= [  \langle n | \exp ( - i \tau \widehat{H})| n' \rangle  ]_{n,n'= 0, 1, 2, \cdots}$.

The actions of the time-reversal unitary and antiunitary operators,
$\widehat{T}$ and $\widehat{K}$,
on the $N_{\rm cvk}$-dimensional complex vector
$\boldsymbol{\psi}$
are represented by the
$N_{\rm cvk}\times N_{\rm cvk}$ matrices
${\bf T}=[T_{nn'}]$
and
${\bf K}=[K_{nn'}]$
through
$\widehat{T} [\boldsymbol{\psi}] = {\bf T} \boldsymbol{\psi}$ and 
$\widehat{K} [\boldsymbol{\psi}] = {\bf K} \boldsymbol{\psi}^*$, 
respectively. 
From Eqs.~(\ref{Tn}) and (\ref{Kn}), we obtain
${\bf T} = [ (-1)^n \delta_{n n'}]_{n, n' = 0, 1, 2, \cdots, N_{\rm cvk} - 1}$ 
and 
${\bf K} = {\bf I} =  [\delta_{n n'}]_{n n' = 0, 1, 2, \cdots, N_{\rm cvk} - 1}$,  
where ${\bf I}$ denotes the identity matrix.

\subsection{Evaluation of the probability density function $P(\Delta S)$}

We consider the same example of the distribution of the initial state vector 
as in Ref.~\cite{Sugama2025}, which is given by 
\begin{equation}
\label{initialP}
 P [ \boldsymbol{\psi}(0) ; 0 ]
= 
\frac{1}{Z}
\exp 
\biggl[
- 
\sum_{n=0}^{N_{\rm cvk}  - 1}
\beta_n
| \psi_n (0) |^2
\biggr]
, 
\end{equation}
where $\beta_n > 0$ and 
\begin{equation}
Z 
\equiv 
\int
d \Gamma (0) 
\exp 
\biggl[
- 
\sum_{n=0}^{N_{\rm cvk}  - 1}
\beta_n
| \psi_n (0) |^2
\biggr]
.
\end{equation}
Note that 
the initial probability density given by Eq.~(\ref{initialP}) is symmetric under 
the time-reversal unitary and antiunitary transformations, 
\begin{equation}
P [ {\bf T} \boldsymbol{\psi}(0); 0 ] 
=
P [  {\bf K} \boldsymbol{\psi}^* (0) ; 0 ] 
= 
P [ \boldsymbol{\psi}(0) ; 0 ] 
,
\end{equation} 
where the definitions of ${\bf T}$ and ${\bf K}$ are given in 
the last paragraph in Sec.~III.E. 
Therefore, when using the initial probability density in Eq.~(\ref{initialP}), 
the fluctuation theorem and the detailed fluctuation theorem 
described in Secs.~II.B and II.C hold. 
The fluctuation theorem is numerically confirmed in 
Ref.~\cite{Sugama2025}
where a total of $10^6$ initial vectors 
$
\boldsymbol{\psi} (0) 
$ 
are randomly generated according to 
$P [ \boldsymbol{\psi}(0) ; 0 ]$
in Eq.~(\ref{initialP}) 
to evaluate $P(\Delta S)$ in Eq.~(\ref{PDSdef}). 

Equation~(\ref{initialP}) yields a stationary distribution,
$
P [ \boldsymbol{\psi}(\tau); 0 ] 
= 
P [ \boldsymbol{\psi}(0) ; 0 ] 
$,
if all $\beta_n$ are equal.
Here, we assume 
$
\beta_n  = \beta_0  / \rho
$
for 
$
n = 1, 2, \cdots, N_{\rm cvk} - 1
$, 
where $\beta_0 > 0$ and $0 < \rho  < 1$. 
Under this assumption,
$
\langle || \boldsymbol{\Psi}(\tau) ||^2 \rangle_{\rm ens}
=
\beta_0^{-1}
[ 1 + \rho ( N_{\rm cvk} - 1 ) ]
$ 
and the stochastic relative entropy becomes
\begin{equation}
\Delta S [ \boldsymbol{\psi}(0) ; \tau ]
=
\log
\left[
\frac{
P [ \boldsymbol{\psi}(0) ; 0 ] 
}{
P [ \boldsymbol{\psi}(\tau) ; 0 ] 
}
\right]
=
Q
\left( 
\frac{1}{T_{\rm res}} 
-
\frac{1}{T_0}  
\right)
, 
\end{equation}
where 
\begin{equation}
Q \equiv (8\pi n_0 L)^{-1}
 \int_{-L/2}^{+L/2} dx  \left(  | E(x, 0) |^2  -  | E(x, t) |^2 \right) 
\end{equation}
is the decrease in electric field energy per electron. 
The effective inverse temperatures of the $n=0$ state and the reservoir states ($n\ge1$) are defined by
$
1/T_0  \equiv
4 \beta_0  ( 1 + \kappa^2 ) / T
$
and 
$
1/ T_{\rm res} \equiv 1 / (T_0 \, \rho)
$, 
respectively, where 
$
|\psi_0 (\tau) |^2 =  (2 \pi n_0 T L )^{-1} 
( 1 + \kappa^2 )  \int_{-L/2}^{+L/2} dx | E(x, t) |^2 
$. 
Thus, 
$
\Delta S [ \boldsymbol{\psi}(0) ; \tau ]
$
can be interpreted as the entropy generated per electron during the interval $[0,\tau]$ by Landau damping, which transfers electric field energy from the $n=0$ state at temperature $T_0$ to the reservoir composed of the $n\ge1$ states at the lower temperature
$T_{\rm res} = T_0 \, \rho < T_0$. 
The fluctuation theorem implies that both damping and growth of the electric field energy are possible, with their relative probabilities constrained by Eq.~(\ref{FT}).
In the nonlinear Vlasov-Poisson system, 
conservation of total energy implies that $Q$ is equal to the increase in kinetic energy per electron.

The probability density function $P(\Delta S)$ can be approximately 
expressed by an analytical formula when some conditions are 
satisfied as shown below. 
We consider the case in which  
\begin{equation}
\label{rhocond}
\frac{\rho}{1 - | U_{00} (\tau) |^2}  
\ll 
 1 
\end{equation}
 holds.
 Then,  when $\Delta S$ is positive and satisfies
\begin{equation}
\label{DScond}
 1
 \ll 
\frac{ \Delta S }{ 1 - \rho }
 \ll 
 \bigg(
 \frac{ 1 - | U_{00} (\tau) |^2 }{
 \rho | U_{00} (\tau) | }
 \bigg)^2
 , 
\end{equation}
$P (\Delta S)$ is well approximated by 
\begin{equation}
\label{PDSaprox}
P (\Delta S) 
=
C (\tau) 
 \exp 
[
-
C (\tau)
\Delta S 
]
,
\end{equation}
where $C (\tau)$ is defined by 
\begin{equation}
C (\tau) 
\equiv 
\frac{\rho}{ 
(1 - \rho) ( 1 - | U_{00} (\tau ) |^2 )
}
.
\end{equation}
Derivation of Eq.~(\ref{PDSaprox}) is given in Appendix~E. 
Combining the fluctuation theorem 
with Eqs.~(\ref{DScond}) and (\ref{PDSaprox}), 
we obtain 
\begin{equation}
\label{PDSnegative}
P (\Delta S) 
=
C (\tau)  
\exp 
[
-
\{ 1 + C (\tau) \}
|\Delta S |
]
\end{equation}
when $\Delta S < 0$ and 
\begin{equation}
 1 
 \ll 
\frac{ | \Delta S | }{ 1 - \rho }
 \ll 
 \bigg(
 \frac{ 1 - | U_{00} (\tau) |^2 }{
 \rho | U_{00} (\tau) | }
 \bigg)^2
.  
\end{equation}
We see that $\rho \ll 1$ is required for Eq.~(\ref{rhocond})  to be valid 
and that 
the validity range is wider as $|U_{00}(\tau)|$ approaches to zero. 
Especially, in the limit 
$
\tau \rightarrow + \infty
$, 
$|U_{00}(\tau)| \rightarrow + 0$,  
and accordingly, 
Eqs.~(\ref{PDSaprox}) and (\ref{PDSnegative}) give 
\begin{equation}
\label{PDSinfty}
P(\Delta S)
=
  \frac{\rho}{1 - \rho}
\exp
\bigg(
-
  \frac{\rho \Delta S}{1 - \rho}
\bigg)
\end{equation}
for $\Delta S / (1-\rho) \gg 1$, 
and 
\begin{equation}
\label{PDSinfty2}
P(\Delta S)
=
  \frac{\rho}{1 - \rho}
\exp
\bigg(
-
  \frac{|\Delta S|}{1 - \rho}
\bigg)
, 
\end{equation}
for $- \Delta S / (1-\rho) \gg 1$, respectively.

\begin{figure}
\vspace*{-25mm}
\centering
\mbox{}\hspace*{15mm}
\includegraphics[width=0.8\linewidth]{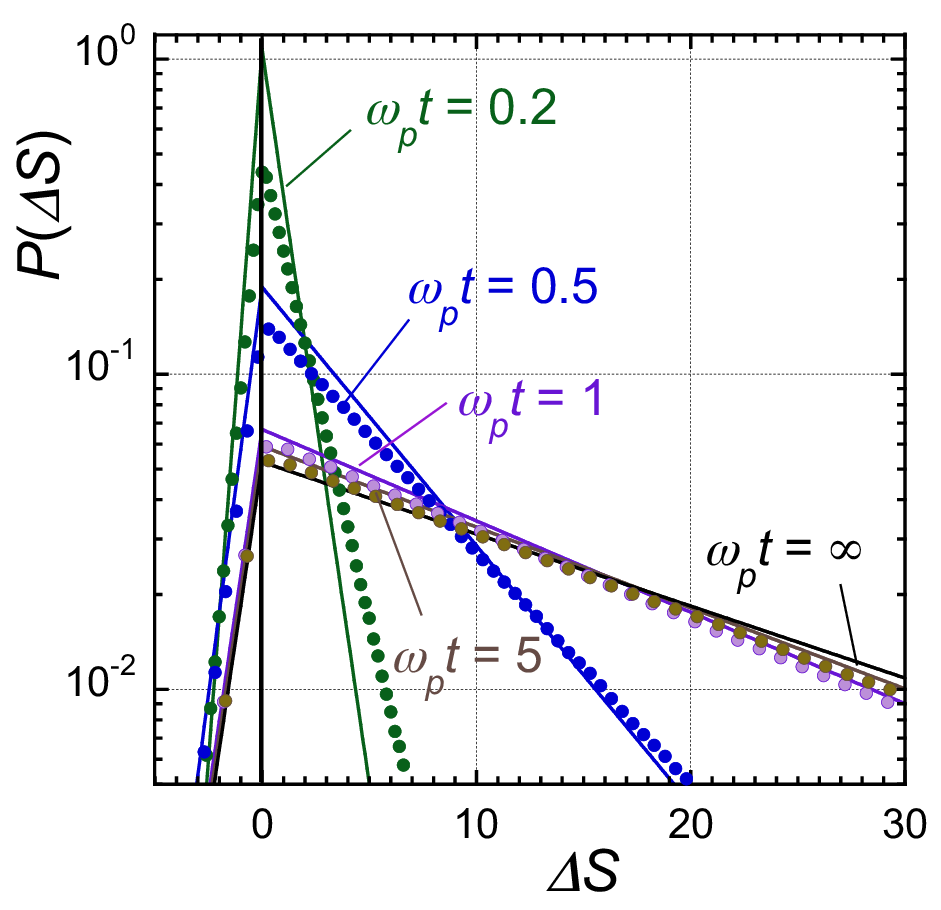}
\\[20mm] \mbox{}
 \caption{
 \label{Fig2}
Probability density function $P(\Delta S)$ of the stochastic relative entropy.
Solid circles represent $P(\Delta S)$
at times
$\omega_p t \equiv \tau /(\sqrt{2} \kappa) =  0.2$, 0.5, 1, and 5, 
for which the integral in Eq.~(\ref{PDSdef}) 
is evaluated by randomly generating 
a total of $10^6$ initial vectors 
$
\boldsymbol{\psi} (0) 
$ 
according to 
$P [ \boldsymbol{\psi}(0) ; 0 ]$
in Eq.~(\ref{initialP}). 
Solid lines are obtained from Eqs.~(\ref{PDSaprox}) and 
(\ref{PDSnegative}) at the corresponding times, 
while the line at $t = + \infty$ is given using 
Eqs.~(\ref{PDSinfty}) and (\ref{PDSinfty2}). 
}
\end{figure}

In Fig~1, 
the probability density function $P(\Delta S)$ evaluated at 
$\omega_p t \equiv \tau /(\sqrt{2} \kappa) =  0.2$, 0.5, 1, and 5 
in the same way as in Ref.~\cite{Sugama2025} and that given using 
Eqs.~(\ref{PDSaprox}) and (\ref{PDSnegative})
are shown by solid circles and lines, respectively. 
The line obtained from Eqs.~(\ref{PDSinfty}) and (\ref{PDSinfty2}) 
at $t = + \infty$ is also shown. 
Here, 
$\kappa = k \lambda_D = 1/2$, $\rho = 1/20$, and $N_{\rm cvk} = 20$ 
are used. 
It is confirmed in Ref.~\cite{Sugama2025} that 
solid circles in Fig.~1 obtained at each time for $N_{\rm cvk} = 20$
can be regarded as equal to
the limit that $P(\Delta S)$ converge as $N_{\rm cvk} \rightarrow \infty$. 
In the present case, we have 
$U_{00}(\tau) = 0.975145$,  0.849354, 0.460399, and  0.330348
at 
$\omega_p t \equiv \tau /(\sqrt{2} \kappa) =  0.2$, 0.5, 1, and 5, respectively. 
Then, using Eq.~(\ref{DScond}), the lower bound of $\Delta S$ in the validity range for 
the approximate expression in Eq.~(\ref{PDSaprox}) is 
given by $1- \rho = 0.95$, while the upper bound 
is obtained as 
$(1 - \rho) ( 1- |U_{00}(\tau) |^2 )^2 / ( \rho |U_{00} (\tau) | )^2 = 0.963098$, 
40.8848, 113.28, and 2763.56
at 
$\omega_p t \equiv \tau /(\sqrt{2} \kappa) =  0.2$, 0.5, 1, and 5, respectively. 
We can confirm from Fig.~1 that, 
as time $t$ increases, 
the validity range is wider and the approximate expression 
are in better agreement with the results obtained by numerically 
evaluating $P(\Delta S)$ according to the definition in Eq.~(\ref{PDSdef}).

\section{Linear Collisionless Gyrokinetic System in a Uniform Background Magnetic Field} 

\subsection{Governing equations for the linear gyrokinetic system}

We consider a linear collisionless gyrokinetic system 
in a uniform background magnetic field ${\bf B}$.  
The perturbed particle distribution function 
for species $a$ with the wavenumber vector ${\bf k}$
is given by the sum of the adiabatic and nonadiabatic parts as
\begin{equation}
\label{ad-nad}
f_{a{\bf k}}
=
- f_{aM} \frac{e_a}{T_a}
\phi_{{\bf k}}
+ h_{a{\bf k}}
e^{-i {\bf k} \cdot \boldsymbol{\rho}_a}
.
\end{equation}
Here, $\phi_{\bf k}$ is the electrostatic potential, 
and $f_{aM}$ represents the Maxwellian equilibrium distribution function, 
$
f_{aM} 
\equiv 
\pi^{-3/2} 
( n_a / v_{Ta}^3 )
\exp 
(
- v^2 / v_{Ta}^2
)
$, 
where $v_{Ta}\equiv \sqrt{2T_a/m_a}$ denotes the thermal velocity. 
The mass, electric charge, gyrofrequency, and 
gyroradius vector 
of the particle species $a$ are denoted by 
$m_a$, $e_a$, $\Omega_a \equiv e_a B / m_a c$, and 
$\boldsymbol{\rho}_a 
\equiv {\bf b}\times {\bf v} / \Omega_a$, respectively, 
where ${\bf v}$ is the particle velocity and ${\bf b}$ is the unit vector parallel to ${\bf B}$. 
The background density and temperature, $n_a$ and $T_a$, are assumed to be constant. 
The nonadiabatic part $h_{a{\bf k}}$ of the distribution function is independent of 
the gyrophase. 
Here, the gyrophase is defined by the angle of 
the direction of the perpendicular velocity ${\bf v}_\perp$ 
(or the gyroradius vector $\boldsymbol{\rho}_a$) 
around the magnetic field line. 
The linearized collisionless gyrokinetic equation for $h_{a{\bf k}_\perp}$ 
is given by~\cite{Antonsen,Catto}
\begin{equation}
\label{GKE0}
\left(
\frac{\partial }{\partial t} 
+ i  \omega_E
+ i k_\parallel v_\parallel 
\right)
h_{a{\bf k}}
=
\frac{e_a}{T_a} f_{aM}
\left(
\frac{\partial }{\partial t} +
 i   \omega_E 
\right)
\psi_{a{\bf k}}
,
\end{equation}
where 
$\omega_E \equiv {\bf k} \cdot ( c {\bf E} \times {\bf b} / B)$. 
Here, ${\bf E}$ is a uniform background electric field, 
and the effect of $\omega_E$ appears as a Doppler-shift frequency. 
Hereafter, we use a reference frame moving with the ${\bf E} \times {\bf B}$ drift due to the background electric field ${\bf E}$, so that we set $\omega_E=0$. 
In Eq.~(\ref{GKE0}), 
$h_{a{\bf k}}$ is regarded as a function of time $t$ and 
$(v_\parallel, v_\perp)$, where 
$v_\parallel$ and $v_\perp$ are the components of the velocity ${\bf v}$ parallel and perpendicular to the background magnetic field ${\bf B}$, respectively. 
The gyrophase-averaged potential $\psi_{a {\bf k}}$ 
associated with the turbulent electromagnetic 
fields is defined in terms of 
the electrostatic potential  $\phi_{\bf k}$ and the
vector potential ${\bf A}_{\bf k}$ as 
\begin{eqnarray}
\label{psi}
& & 
\psi_{a {\bf k}}
 \equiv 
\oint \frac{d \varphi}{2\pi}
e^{i {\bf k} \cdot
\boldsymbol{\rho}_a}
\left(
\phi_{\bf k}- \frac{\bf v}{c}
\cdot {\bf A}_{\bf k}
\right)
\nonumber 
\\ &  & 
=
J_0 \left(
\frac{k_\perp v_\perp}{\Omega_a}
\right)
\left(
\phi_{\bf k} - \frac{v_\parallel}{c}
A_{\parallel {\bf k}}
\right)
+ 
J_1 \left(
\frac{k_\perp v_\perp}{\Omega_a}
\right)
\frac{v_\perp}{c}
\frac{B_{\parallel {\bf k}}}{k_\perp}
,
\nonumber 
\\ &  & 
\end{eqnarray}
where 
$A_{\parallel {\bf k}} \equiv {\bf b}\cdot {\bf A}_{\bf k}$
, 
$B_{\parallel {\bf k}} \equiv i {\bf b}\cdot( {\bf k}
\times {\bf A}_{\bf k} )$, and 
$J_0$ and $J_1$ denote 
the zeroth- and first-order Bessel functions, respectively. 
The wavenumber vector ${\bf k}$ is given by the sum of 
parallel and perpendicular components as 
${\bf k} = k_\parallel {\bf b} + {\bf k}_\perp$. 
In gyrokinetic theory, $k_\perp \equiv | {\bf k}_\perp |$ 
is assumed to be of order of the thermal gyroradius, 
the parallel wavenumber 
$k_\parallel \equiv {\bf k} \cdot {\bf b}$ is considered much 
smaller than $k_\perp$. 

The electrostatic potential $\phi_{\bf k}$ 
is determined by Poisson's equation, 
\begin{equation}
\label{GKP}
\left( k_\perp^2
+ \lambda_D^{-2}
\right)  \phi_{\bf k}
=
4\pi \sum_a e_a 
\int d^3 v \; 
h_{a{\bf k}} 
  J_0 (k_\perp v_\perp / \Omega_a) 
. 
\end{equation}
where $\lambda_D \equiv (\sum_a 4\pi n_a e_a^2/T_a)^{-1/2}$ 
and 
$\int d^3 v = 2 \pi 
\int_{-\infty}^{+\infty} d v_\parallel 
\int_0^{+\infty} v_\perp d v_\perp$. 
In addition, $A_{\parallel{\bf k}}$ and $B_{\parallel{\bf k}}$ 
are determined by the parallel and perpendicular components 
of Amp\'{e}re's law, 
\begin{equation}
\label{GKA1}
 k_\perp^2 A_{\parallel{\bf k}}
=
\frac{4\pi}{c} \sum_a e_a 
\int d^3 v \; 
v_\parallel 
h_{a {\bf k}_\perp} 
J_0 (k_\perp v_\perp / \Omega_a) 
,
\end{equation}
and 
\begin{equation}
\label{GKA2}
 - k_\perp B_{\parallel{\bf k}}
=
\frac{4\pi}{c} \sum_a e_a 
\int d^3 v \; 
h_{a{\bf k}} 
v_\perp J_1 (k_\perp v_\perp / \Omega_a)
,
\end{equation}
respectively. 
We now 
find that~\cite{Sugama1996,Sugama2009} 
\begin{equation}
\label{GKC}
\sum_a 
T_a 
\int d^3 v 
\frac{| f_{a{\bf k}}|^2}{2 f_{aM}}
+ \frac{1}{8\pi}
 ( |{\bf E}_{\bf k}|^2 + |{\bf B}_{\bf k}|^2 )
\end{equation}
is conserved by any solutions of 
Eqs.~(\ref{GKE0}), (\ref{GKP}), (\ref{GKA1}), and (\ref{GKA2}). 
Here, 
${\bf E}_{\bf k} = - i {\bf k}_\perp \phi_{\bf k}$ 
and 
${\bf B}_{\bf k} =  i ({\bf k} \times {\bf b}) A_{\parallel {\bf k}}
+ B_{\parallel {\bf k}} {\bf b}$. 
Equation~(\ref{GKC}) presents the gyrokinetic version 
of the conserved quantity given in Eq.~(\ref{Df1}) for 
the linear Vlasov-Poisson system. 

We now define
the perturbed {\it gyrocenter} distribution 
function $f_{a{\bf k}}^{(g)}$ by 
\begin{equation}
\label{deltafg} 
 f_{a{\bf k}}^{(g)} 
 =
  - 
  \frac{e_a \psi_{a {\bf k}}}{T_a} f_{aM} + 
h_{a{\bf k}},
\end{equation}
which is 
independent of the gyrophase. 
Then, using Eqs.~(\ref{ad-nad}) and (\ref{deltafg}), 
the perturbed particle density $f_{a{\bf k}_\perp}$ is related to 
$f_{a{\bf k}}^{(g)}$ by  
\begin{equation} 
\label{deltafp}
 f_{a{\bf k}} 
=
e^{-i {\bf k} \cdot \boldsymbol{\rho}_a}
 f_{a{\bf k}}^{(g)}  -  f_{aM} \frac{e_a}{T_a}
(
\phi_{\bf k}
 -  
e^{-i {\bf k} \cdot \boldsymbol{\rho}_a}
\psi_{a{\bf k}}
)
.
\end{equation}
On the right-hand side of Eq.~(\ref{deltafp}), 
the factor $e^{-i {\bf k} \cdot \boldsymbol{\rho}_a}$ 
in the first term results from the difference between the particle 
and gyrocenter positions while the second group of terms 
represents the polarization, that is the variation of the particle 
distribution due to the potential perturbation. 

Using $f_{a{\bf k}}^{(g)}$ instead of $f_{a{\bf k}}$, 
Eqs.~(\ref{GKE0}), (\ref{GKP}), (\ref{GKA1}), and (\ref{GKA2})
 are rewritten as 
\begin{equation}
\label{GKEg}
\left(
\frac{\partial}{\partial t} +
i k_\parallel v_\parallel  
\right)
 f_{a{\bf k}}^{(g)}
=
- \frac{e_a}{T_a} f_{aM}
i  k_\parallel v_\parallel 
\psi_{a{\bf k}}
,
\end{equation}
\begin{equation}
\label{GKPg}
K_{00} 
\,
 \phi_{\bf k}
- 
K_{02}
\frac{B_{\parallel {\bf k}}}{k_\perp}
=
4\pi \sum_a e_a 
\int d^3 v \; 
  J_0 (k_\perp \rho_a) f^{(g)}_{a{\bf k}} 
,
\end{equation}
\begin{equation}
\label{GKA1g}
K_{11} \, 
 A_{\parallel{\bf k}}
=
\frac{4\pi}{c} \sum_a e_a 
\int d^3 v \; 
v_\parallel 
J_0 (k_\perp v_\perp / \Omega_a) 
 f^{(g)}_{a{\bf k}} 
,
\end{equation}
and 
\begin{eqnarray}
\label{GKA2g}
& & 
\hspace*{-5mm}
- 
K_{02} \, 
\phi_{{\bf k}}
 - 
K_{22}
 \frac{B_{\parallel{\bf k}}}{k_\perp}
\nonumber \\
& & 
\hspace*{-3mm}
=
\frac{4\pi}{c} \sum_a e_a 
\int d^3 v \; 
v_\perp J_1 (k_\perp v_\perp / \Omega_a)
f^{(g)}_{a{\bf k}} 
,
\end{eqnarray}
respectively. 
Here, the coefficients $K_{00}$, $K_{02}$, $K_{11}$, and $K_{22}$ 
are positive and defined by 
\begin{eqnarray}
\label{K00}
K_{00} 
& \equiv &
k_\perp^2
+ \sum_a \lambda_{Da}^{-2}
\{ 1 - \Gamma_0(b_a) \}
\nonumber \\
K_{02} 
& \equiv & 
\sum_a \lambda_{Da}^{-2}
b_a^{1/2}
\{ \Gamma_0(b_a) - \Gamma_1(b_a) \}
\frac{v_{Ta}}{\sqrt{2}c} 
\nonumber \\
K_{11} 
& \equiv & 
 k_\perp^2 
 +
 \sum_a
\frac{\omega_{pa}^2}{c^2}
\Gamma_0(b_a)
\nonumber \\
K_{22} 
& \equiv & 
 k_\perp^2
 + \sum_a \lambda_{Da}^{-2} 
 b_a \{ \Gamma_0(b_a) - \Gamma_1(b_a) \}
 \frac{v_{Ta}^2}{c^2} 
,
\end{eqnarray}
where 
$
b_a \equiv k_\perp^2 T_a /(m_a  \Omega_a^2)
$ 
$
\omega_p^2 \equiv \sum_a \omega_{pa}^2
$,
$
\omega_{pa}^2 \equiv 4\pi n_a e_a^2 / m_a
$,
$
\lambda_{Da}^2
\equiv
T_a / (4\pi n_a e_a^2 )
$, 
$
\Gamma_0(b_a) \equiv I_0 (b_a) \exp (-b_a)
$, 
$
\Gamma_1(b_a) \equiv I_1 (b_a) \exp (-b_a)
$, 
and 
$I_0$ and $I_1$ denote 
the modified Bessel functions.

\subsection{Derivation of the Schr\"{o}dinger equation for the linear gyrokinetic system}

We define dimensionless variables 
$\xi$ and $X$ by 
$\xi \equiv v_\parallel / v_{Ta}$ and 
$X \equiv v_\perp^2 / v_{Ta}^2$ 
to represent 
parallel and perpendicular coordinates in the velocity-space. 
Then, the Maxwellian equilibrium distribution function $f_{aM}$ is 
expressed as 
\begin{equation}
f_{aM} ({\bf v})= \frac{n_a}{\pi^{3/2} v_{Ta}^3} e^{- X - \xi^2 }
= \frac{n_a}{\pi v_{Ta}^3}
 [ l_0 (X) h_0  ( \xi ) ]^2
, 
\end{equation}
where 
$l_0 (X) \equiv e^{ - X / 2 }$ and 
$h_0 (\xi) \equiv \pi^{-1/4} e^{- \xi^2 / 2 }$. 
Now,
we express the perturbed gyrocenter distribution function 
$f_{a{\bf k}}^{(g)}(v_\parallel, v_\perp, t)$ by 
a dimensionless function 
$\widetilde{f}_a
( X, \xi, \tau)$ 
as 
\begin{equation}
f_{a{\bf k}}^{(g)} 
(v_\parallel, v_\perp, t)
\equiv
\frac{1}{\pi}
\frac{n_a}{v_{Ta}^3}
l_0 (X ) 
h_0 (\xi) 
\widetilde{f}_a
( X, \xi, \tau)
.
\end{equation}
Here, we employ the normalized time $\tau \equiv k_\parallel v_c t$
where $v_c$ is an arbitrarily chosen characteristic time
(for example, we may use the thermal velocity of some 
specific particle species 
for $v_c$). 
The gyrokinetic equation in Eq.~(\ref{GKEg}) is 
rewritten for $\widetilde{f}_a ( X, \xi, \tau)$ 
as
\begin{eqnarray}
\label{LGKf1}
 &  & 
i
\frac{\partial}{\partial \tau}
\widetilde{f}_a
( X, \xi, \tau)
\nonumber \\ 
& & 
= 
\frac{v_{Ta}}{v_c}
\xi 
\bigg[
\widetilde{f}_a ( X, \xi, \tau)
+ 
l_0 (X) 
h_0 (\xi) 
\frac{e_a}{T_a}
\bigg\{
J_0 (\sqrt{2 b_a X} )
\nonumber \\
& & 
\hspace*{12mm}
\mbox{}
\times 
\bigg( 
\phi_{\bf k} (\tau)
- \frac{v_{Ta}}{c}
\xi 
A_{\parallel {\bf k}}  (\tau)
\bigg)
+
J_1 (\sqrt{2 b_a X} )
\nonumber \\ 
& & 
\mbox{} 
\hspace*{12mm} 
\times 
\frac{v_{Ta}}{c} \sqrt{X}
\frac{B_{\parallel {\bf k}} (\tau)}{k_\perp}
\bigg\}
\bigg]
\end{eqnarray}

We define the ket vector 
$| \widetilde{f} (\tau) \rangle$
associated with the distribution function 
$\widetilde{f}_a ( X, \xi, \tau)$
by 
\begin{equation}
\widetilde{f}_a
( X, \xi, \tau)
\equiv
\langle a  \, X \, \xi |
\widetilde{f} (\tau) \rangle 
. 
\end{equation}
The ket vector $|\widetilde{f} (\tau) \rangle$ 
belongs to the vector space 
\begin{equation}
{\cal E} \equiv 
{\cal E}_{ps} \otimes {\cal E}_{v_\perp} \otimes {\cal E}_{v_\parallel} 
,
\end{equation}
which represents  the tensor product of 
the vector spaces 
${\cal E}_{ps}$, ${\cal E}_{v_\perp}$, and ${\cal E}_{v_\parallel}$. 
Here, ${\cal E}_{ps}$ is a vector space which 
is generated by basis vectors $\{ | a \rangle \}$, 
where the ket vector $| a \rangle$ is defined 
analogously with an isospin state vector of a nucleon~\cite{QM} 
to represent the state vector associated with the particle species $a$. 
Let $N_{ps}$ denote the number of particle species. 
Then, ${\cal E}_{ps}$ is an $N_{ps}$-dimensional complex vector space. 
The basis vectors $\{ | a \rangle \}$ satisfy the orthonormality condition 
and the closure relation given by 
\begin{equation}
\langle a | a' \rangle = \delta_{a a'}
,
\hspace*{5mm}
\widehat{1}_{ps}
=
\sum_a 
| a \rangle \langle a |
,
\end{equation}
where $\widehat{1}_{ps}$ is the identity operator in ${\cal E}_{ps}$. 
For example, for a plasma consisting of electrons and a single-species 
ions, $a = e, i$, $N_{ps} = 2$, and the basis vectors of ${\cal E}_{ps}$ 
are given by $\{ | e \rangle ,  | i \rangle \}$. 

Vectors in the vector spaces 
${\cal E}_{v_\perp}$ and ${\cal E}_{v_\parallel}$ are associated with 
perpendicular and parallel velocity-space distributions, respectively. 
Using $X \equiv v_\perp^2 / v_{Ta}^2$ and 
$\xi \equiv v_\parallel / v_{Ta}$, 
${\cal E}_{v_\perp}$ and ${\cal E}_{v_\parallel}$ are 
 generated by the basis vectors 
$\{ | X \rangle \}_{0\leq X < + \infty}$ and 
$\{ | \xi \rangle \}_{- \infty < \xi < + \infty}$, respectively. 
These basis vectors satisfy 
\begin{equation}
\langle X | X' \rangle 
=
\delta (X - X')
,
\hspace*{5mm}
\int_0^{+\infty} 
| X \rangle dX  \langle X |
=
\widehat{1}_{v_\perp}
\end{equation}
and 
\begin{equation}
\langle \xi | \xi' \rangle 
=
\delta (\xi - \xi')
,
\hspace*{5mm}
\int_{-\infty}^{+\infty} 
| \xi \rangle d\xi  \langle \xi |
=
\widehat{1}_{v_\parallel}
, 
\end{equation}
where $\widehat{1}_{v_\perp}$ and $\widehat{1}_{v_\parallel}$ 
denote the identity operators in ${\cal E}_{v_\perp}$ and 
${\cal E}_{v_\parallel}$, respectively. 
The vector space ${\cal E}$ is generated by 
the basis vectors $\{ | a \, X \, \xi \rangle \}$, where 
$| a \, X \, \xi \rangle$ is given  by  the tensor product 
of $| a \rangle$, $| X \rangle$, and $| \xi \rangle$ as 
\begin{equation}
| a \, X \, \xi \rangle
=
| a \rangle \otimes | X \rangle \otimes | \xi \rangle
. 
\end{equation}
The basis vectors 
$\{ | a \, X \, \xi \rangle \}$
satisfy
\begin{equation}
\langle a \, X \, \xi | a' \, X' \, \xi' \rangle
=
\delta_{a a'} 
\delta (X - X')
\delta (\xi - \xi')
,
\end{equation}
and 
\begin{equation}
\sum_a \int_0^{+\infty} dX \int_{-\infty}^{+\infty} d\xi
| a \, X \, \xi \rangle
\langle a \, X \, \xi |
=
\widehat{1}_{ps}
\otimes 
\widehat{1}_{v_\perp}
\otimes 
\widehat{1}_{v_\parallel}
\equiv
\widehat{1}
.
\end{equation}
where $\widehat{1}$ is the identity operator in ${\cal E}$. 

We consider three vectors denoted by
\begin{eqnarray}
| \chi_\nu  \rangle 
& \equiv & 
\sum_a 
\int_0^{+ \infty} dX  
\int_{- \infty}^{+ \infty} d\xi
\; 
| a \, X  \, \xi \rangle
\langle  a \, X  \, \xi | \chi_\nu  \rangle 
\nonumber \\
& & 
( \nu = 0, 1, 2 )
,
\end{eqnarray}
which are used to express 
$\phi_{{\bf k}_\perp} (\tau)$, 
$A_{\parallel {\bf k}_\perp} (\tau)$, and 
$B_{\parallel {\bf k}_\perp} (\tau)$ 
by 
\begin{eqnarray}
& & 
\hspace*{-3mm}
\phi_{{\bf k}_\perp} (\tau)
=  
\langle \chi_0
|
\widetilde{f} (\tau) \rangle 
\nonumber \\
&  & 
=
\sum_a 
\int_0^{+ \infty} dX
\int_{- \infty}^{+ \infty} d\xi
\,
\langle \chi_0 | a \, X \, \xi \rangle
\langle  a \, X \, \xi | \widetilde{f} (\tau) \rangle 
,
\nonumber \\
&  & 
\end{eqnarray}
\begin{eqnarray}
& & 
\hspace*{-3mm}
A_{\parallel {\bf k}_\perp} (\tau)
=  
\langle \chi_1
|
\widetilde{f} (\tau) \rangle 
\nonumber \\
&  & 
=
\sum_a 
\int_0^{+ \infty} dX 
\int_{- \infty}^{+ \infty} d\xi
\,
\langle \chi_1 | a \, X \, \xi \rangle
\langle  a \, X \, \xi | \widetilde{f} (\tau) \rangle 
,
\nonumber \\
&  & 
\end{eqnarray}
and 
\begin{eqnarray}
& & 
\hspace*{-3mm}
\frac{1}{k_\perp} B_{\parallel {\bf k}_\perp} (\tau)
=  
\langle \chi_2
|
\widetilde{f} (\tau) \rangle 
\nonumber \\
&  & 
=
\sum_a 
\int_0^{+ \infty} dX
\int_{- \infty}^{+ \infty} d\xi
\,
\langle \chi_2 | a \, X \, \xi \rangle
\langle  a \, X \, \xi | \widetilde{f} (\tau) \rangle 
, 
\nonumber \\
&  & 
\end{eqnarray}
respectively. 
From 
Eqs.~(\ref{GKPg}), (\ref{GKA1g}), and (\ref{GKA2g}), we find that 
$\{ | \chi_\nu  \rangle \}_{j=1,2,3}$ should satisfy
\begin{equation}
\label{Kchi02}
\left[
\begin{array}{cc}
K_{00} & - K_{02} 
\\
- K_{02} &  - K_{22}
\end{array}
\right]
\left[
\begin{array}{c}
 | \chi_0 \rangle
 \\
  | \chi_2 \rangle
\end{array}
\right]
=
4\pi 
\left[
\begin{array}{c}
 | \sigma_0 \rangle 
 \\
 | \sigma_2  \rangle 
\end{array}
\right]
,
\end{equation}
and 
\begin{equation}
\label{Kchi1}
K_{11}
 | \chi_1 \rangle
=  
4\pi 
 | \sigma_1 \rangle 
 ,
\end{equation}
where $\{ | \sigma_\nu  \rangle \}_{\nu = 1,2,3}$ are defined by 
\begin{eqnarray}
| \sigma_\nu  \rangle 
& \equiv & 
\sum_a 
\int_0^{+ \infty} dX  
\int_{- \infty}^{+ \infty} d\xi
\; 
| a \, X  \, \xi \rangle
\langle  a \, X  \, \xi | \sigma_\nu  \rangle 
\nonumber \\
& & 
(\nu = 0, 1, 2 )
,
\end{eqnarray}
with 
\begin{equation}
\langle a \, X  \, \xi | \sigma_0 \rangle 
\equiv 
n_a e_a 
l_0 (X ) 
J_0 ( \sqrt{ 2 b_a X  } )
h_0 (\xi)
\end{equation}
\begin{equation}
\langle a \, X  \, \xi | \sigma_1 \rangle 
\equiv 
\frac{n_a e_a v_{Ta} }{\sqrt{2}c}
l_0 (X ) 
J_0 ( \sqrt{ 2 b_a X  } )
h_1 (\xi)
\end{equation}
and 
\begin{equation}
\langle a \, X  \, \xi | \sigma_2 \rangle 
\equiv 
\frac{n_a e_a v_{Ta} }{c}
l_0 (X ) 
 \sqrt{X }
J_1 ( \sqrt{ 2 b_a X  } )
h_0 (\xi)
.
\end{equation}

Here, we can consider the basis vectors $\{ | n \rangle \}_{n = 0, 1, 2, \cdots}$ 
in ${\cal E}_{v_\parallel}$, where 
$\langle \xi | n \rangle \equiv h_n (\xi)$. 
Then, 
we can use 
$
| 0 \rangle 
\equiv 
\int_{-\infty}^{+\infty} 
| \xi \rangle d\xi \langle \xi | 0 \rangle
\equiv 
\int_{-\infty}^{+\infty} 
h_0 (\xi)
| \xi \rangle d\xi 
$ 
in ${\cal E}_{v_\parallel}$ 
to express 
$ | \sigma_0 \rangle$ and $ | \sigma_2 \rangle $ 
 in the form of the tensor product as
\begin{equation}
\label{sigma02timesh02}
 | \sigma_0 \rangle 
 = 
 | \sigma^\times_0\rangle 
 \otimes
 | 0 \rangle
 , 
 \hspace{5mm}
 | \sigma_2 \rangle 
 = 
 | \sigma^\times_2\rangle 
 \otimes
 | 0 \rangle
\end{equation}
In this section, 
we use $ | \, \cdot^\times \rangle$ to represent
a vector in ${\cal E}^\times \equiv {\cal E}_{ps} \otimes {\cal E}_{v_\perp}$. 
Similarly, using 
$
| 1 \rangle 
\equiv 
\int_{-\infty}^{+\infty} 
| \xi \rangle d\xi \langle \xi | 1 \rangle
\equiv 
\int_{-\infty}^{+\infty} 
h_1 (\xi)
| \xi \rangle d\xi 
$, 
$ | \sigma_1 \rangle$  is expressed as 
\begin{equation}
\label{sigma1timesh1}
 | \sigma_1 \rangle 
 = 
 | \sigma^\times_1\rangle 
 \otimes
 | 1 \rangle
. 
\end{equation}
Here, $\{ | \sigma^\times_\nu \rangle \}_{\nu = 0, 1, 2}$ are the 
vectors in ${\cal E}^\times \equiv {\cal E}_{ps} \otimes {\cal E}_{v_\perp}$, 
which are given by 
\begin{equation}
 | \sigma^\times_\nu\rangle 
  \equiv  
  \sum_a 
  \int_0^{+\infty} 
d X 
  | a \, X  \rangle 
  \langle a \, X  | \sigma^\times_\nu\rangle 
 \hspace*{5mm}
 (j= 0,1,2)
 ,
\end{equation}
with 
\begin{equation}
\langle a \, X  | \sigma^\times_0\rangle 
 \equiv  
 n_a e_a 
l_0 (X ) 
J_0 (\sqrt{2 b_a X  } )
,
\end{equation}
\begin{equation}
\langle a \, X  | \sigma^\times_1\rangle 
 \equiv  
 \frac{ v_{Ta}}{\sqrt{2} c}
\langle a \, X  | \sigma^\times_0\rangle 
\end{equation}
and 
\begin{equation}
\langle a \, X  | \sigma^\times_2 \rangle 
 \equiv  
 n_a e_a  \frac{v_{Ta}}{c}
l_0 (X ) \sqrt{X }
J_1 (\sqrt{2 b_a X  } )
. 
\end{equation}
Correspondingly to Eqs.~(\ref{sigma02timesh02}) and (\ref{sigma1timesh1}), 
we have
\begin{equation}
\label{chi02chitimes}
 | \chi_0 \rangle 
 = 
 | \chi^\times_0 \rangle 
 \otimes
 | 0 \rangle
 , 
 \hspace{5mm}
 | \chi_2 \rangle 
 = 
 | \chi^\times_2\rangle 
 \otimes
 | 0 \rangle
\end{equation}
and 
\begin{equation}
\label{chi1chitimes}
 | \chi_1 \rangle 
 = 
 | \chi^\times_1\rangle 
 \otimes
 | 1 \rangle
. 
\end{equation}
Here, $\{ | \chi^\times_\nu \rangle \}_{\nu =0,1,2}$ 
are the vectors in 
${\cal E}^\times \equiv {\cal E}_{ps} \otimes {\cal E}_{v_\perp}$, which 
are determined from 
$\{ | \sigma^\times_\nu \rangle \}_{\nu =0,1,2}$
by 
\begin{equation}
\label{Kchi02times}
\left[
\begin{array}{cc}
K_{00} & - K_{02} 
\\
- K_{02} &  - K_{22}
\end{array}
\right]
\left[
\begin{array}{c}
 | \chi^\times_0 \rangle
 \\
  | \chi^\times_2 \rangle
\end{array}
\right]
=
4\pi 
\left[
\begin{array}{c}
 | \sigma^\times_0 \rangle 
 \\
 | \sigma^\times_2  \rangle 
\end{array}
\right]
,
\end{equation}
and 
\begin{equation}
\label{Kchi1times}
K_{11}
 | \chi^\times_1 \rangle
=  
4\pi 
 | \sigma^\times_1 \rangle 
 .
\end{equation}

We denote the equilibrium pressure of particle species $a$ by 
$p_a \equiv n_a  T_a$ and define the operator $\widehat{p}$ in 
${\cal E}$ by 
\begin{equation}
\widehat{p} 
\equiv 
\widehat{p}^\times 
\otimes
\widehat{1}_{v_\parallel}
\equiv
\bigg(
\sum_a 
| a \rangle p_a \langle a |
\bigg)
\otimes
\widehat{1}_{v_\perp}
\otimes 
\widehat{1}_{v_\parallel}
\end{equation}
where $\widehat{p}^\times$ is the operator in 
${\cal E}^\times \equiv {\cal E}_{ps} \otimes {\cal E}_{v_\perp}$ defined by 
\begin{equation}
\widehat{p}^\times 
\equiv
\bigg(
\sum_a 
| a \rangle p_a \langle a |
\bigg)
\otimes
\widehat{1}_{v_\perp}
. 
\end{equation}
We see that 
the operators $\widehat{p}$ and $\widehat{p}^\times$ are Hermitian and positive-definite.  
The inverse operators of $\widehat{p}$  and $\widehat{p}^\times$ are 
given by 
\begin{equation}
\label{pinv}
\widehat{p}^{-1}
\equiv 
( \widehat{p}^\times) ^{-1}
\otimes 
\widehat{1}_{v_\parallel}
\equiv 
\bigg(
\sum_a 
| a \rangle p_a^{-1} \langle a |
\bigg)
\otimes
\widehat{1}_{v_\perp}
\otimes 
\widehat{1}_{v_\parallel}
\end{equation}
and 
\begin{equation}
( \widehat{p}^\times )^{-1}
\equiv
\bigg(
\sum_a 
| a \rangle p_a^{-1} \langle a |
\bigg)
\otimes
\widehat{1}_{v_\perp}
,
\end{equation}
respectively. 

For the vectors $\{ | \sigma_\nu \rangle \}_{\nu =0,1,2}$ 
and the operator $\widehat{p}^{-1}$ defined above, 
we can derive the following formulas, 
\begin{equation}
\langle \sigma_0 | 
\widehat{p}^{-1}
 | \sigma_0 \rangle 
 = 
\langle \sigma^\times_0 | 
(\widehat{p}^\times)^{-1}
 | \sigma^\times_0 \rangle 
=
\frac{1}{4\pi}
( k_\perp^2 + \lambda_D^{-2} - K_{00} )
,
\end{equation}
\begin{eqnarray}
& & 
 \hspace*{-3mm}
\langle \sigma_0 | 
\widehat{p}^{-1}
 | \sigma_2 \rangle 
=
 \langle \sigma_2 | 
\widehat{p}^{-1}
 | \sigma_0 \rangle 
 \nonumber \\
 &  & =
\langle \sigma^\times_0 | 
(\widehat{p}^\times)^{-1}
 | \sigma^\times_2 \rangle 
=
\langle \sigma^\times_2 | 
(\widehat{p}^\times)^{-1}
 | \sigma^\times_0 \rangle 
 =
\frac{1}{4\pi}
 K_{02} 
 ,
 \hspace*{5mm}
\end{eqnarray}
\begin{equation}
\langle \sigma_2 | 
\widehat{p}^{-1}
 | \sigma_2 \rangle 
 = 
\langle \sigma^\times_2 | 
(\widehat{p}^\times)^{-1}
 | \sigma^\times_2 \rangle 
=
\frac{1}{4\pi}
( K_{22} - k_\perp^2 )
,
\end{equation}
and 
\begin{equation}
\langle \sigma_1 | 
\widehat{p}^{-1}
 | \sigma_1 \rangle 
 = 
\langle \sigma^\times_1 | 
(\widehat{p}^\times)^{-1}
 | \sigma^\times_1 \rangle 
=
\frac{1}{4\pi}
( K_{11} - k_\perp^2 )
, 
\end{equation}
where $K_{00}$, $K_{02}$, $K_{22}$, 
and $K_{11}$ are given in Eq.~(\ref{K00}). 

We can now rewrite Eq.~(\ref{LGKf1}) as 
\begin{equation}
\label{LGKf2}
i \frac{d}{d\tau}
 | \widetilde{f} (\tau) \rangle 
= 
\widehat{L}
 | \widetilde{f} (\tau) \rangle 
 , 
\end{equation}
where $\widehat{L}$ is defined by
\begin{equation}
\label{hatL}
\widehat{L}
 \equiv 
\frac{\widehat{v}_T}{v_c} \widehat{\Xi}
\,
\big(
\widehat{1}
+
\widehat{p}^{-1}
\widehat{\Pi}
\big)
=
\frac{\widehat{v}_T}{v_c}  \widehat{p}^{-1} 
\widehat{\Xi}
\,
\big(
\widehat{p}
+
\widehat{\Pi}
\big)
. 
\end{equation}
The operators $\widehat{v}_T$, $\widehat{\Xi}$, and 
$\widehat{\Pi}$ are defined by 
\begin{equation}
\label{vT}
\widehat{v}_T
\equiv 
\bigg(
\sum_a 
| a \rangle 
v_{Ta} 
\langle a |
\bigg)
\otimes
\widehat{1}_{v_\perp}
\otimes 
\widehat{1}_{v_\parallel}
,
\end{equation}
\begin{equation}
\label{hatXi}
\widehat{\Xi}
\equiv 
\widehat{1}_{ps}
\otimes
\widehat{1}_{v_\perp}
\otimes 
\bigg(
\int 
|\xi \rangle 
\xi d\xi
\langle \xi |
\bigg)
,
\end{equation}
and 
\begin{eqnarray}
\label{Pi}
\widehat{\Pi} 
& \equiv & 
\widehat{\Pi}_{0} - \widehat{\Pi}_{1} + \widehat{\Pi}_{2} 
\nonumber \\
& = & 
( \widehat{\Pi}^\times_0 + \widehat{\Pi}^\times_2 )
\otimes
| 0 \rangle \langle 0 |
- 
\widehat{\Pi}^\times_1  
\otimes
| 1 \rangle \langle 1 |
,
%\nonumber \\ &  &  
\end{eqnarray}
respectively, where 
\begin{equation}
\label{PiPitimes}
\widehat{\Pi}_\nu
\equiv 
| \sigma_\nu \rangle \langle \chi_\nu |
,
\hspace*{2mm}
\widehat{\Pi}^\times_\nu 
\equiv 
| \sigma^\times_\nu\rangle \langle \chi^\times_\nu |
\hspace*{2mm}
( j = 0, 1, 2)
. 
\end{equation}
The operators $\widehat{v}_T$ and $\widehat{\Xi}$ which are defined in 
Eqs.~(\ref{vT}) and (\ref{hatXi}), respectively, 
are Hermitian, and furthermore, $\widehat{v}_T$ is positive-definite. 
Here, $\{ \widehat{\Pi}_j \}_{j=0,1,2}$ are related to 
$\{ \widehat{\Pi}^\times_j \}_{j=0,1,2}$ by 
\begin{equation}
\label{Pi02Pitimes}
\widehat{\Pi}_0
=
\widehat{\Pi}^\times_0  
\otimes
| 0  \rangle \langle 0 |
, 
\hspace*{2mm}
\widehat{\Pi}_2
=
\widehat{\Pi}^\times_2  
\otimes
| 0  \rangle \langle 0 |
,
\end{equation}
and 
\begin{equation}
\label{Pi1Pitimes}
\widehat{\Pi}_1
=
\widehat{\Pi}^\times_1  
\otimes
| 1 \rangle \langle 1 |
.
\end{equation}

Using Eqs.~(\ref{Kchi02}), (\ref{Kchi02times}), 
(\ref{PiPitimes}), and (\ref{Pi02Pitimes}),
we can express 
$\widehat{\Pi}_{0}  + \widehat{\Pi}_{2}$ 
as 
\begin{eqnarray} 
\label{Pi0plusPi2}
\widehat{\Pi}_{0}  + \widehat{\Pi}_{2} 
& = & 
| \sigma_0 \rangle \langle \chi_0 |
+
| \sigma_2 \rangle \langle \chi_2 |
=
| \chi_0 \rangle \langle \sigma_0 |
+
| \chi_2 \rangle \langle \sigma_2 |
\nonumber \\
& = & 
\frac{1}{4 \pi}
\Big[
| \chi_0 \rangle 
\hspace*{2mm}
| \chi_2 \rangle 
\Big]
\left[
\begin{array}{cc}
K_{00} 
&
- K_{02} 
\\
- K_{02} 
&
- K_{22} 
\end{array}
\right]
\left[
\begin{array}{c}
\langle \chi_0 |
\\
\langle \chi_2 |
\end{array}
\right]
\nonumber \\
& = & 
4 \pi 
\Big[
| \sigma_0 \rangle 
\hspace*{2mm}
| \sigma_2 \rangle 
\Big]
\left[
\begin{array}{cc}
\Lambda_{00} 
&
- \Lambda_{02} 
\\
- \Lambda_{02} 
&
- \Lambda_{22} 
\end{array}
\right]
\left[
\begin{array}{c}
\langle \sigma_0 |
\\
\langle \sigma_2 |
\end{array}
\right]
\nonumber \\
& = & 
( 
\widehat{\Pi}^\times_0 
+
\widehat{\Pi}^\times_2 
)
\otimes
| 0 \rangle \langle 0 |
,
\end{eqnarray}
where 
\begin{eqnarray} 
\label{Pi0timesplusPi2times}
\widehat{\Pi}^\times_{0}  + \widehat{\Pi}^\times_{2} 
& = & 
| \sigma^\times_0 \rangle \langle \chi^\times_0 |
+
| \sigma^\times_2 \rangle \langle \chi^\times_2 |
=
| \chi^\times_0 \rangle \langle \sigma^\times_0 |
+
| \chi^\times_2 \rangle \langle \sigma^\times_2 |
\nonumber \\
& = & 
\frac{1}{4 \pi}
\Big[
| \chi^\times_0 \rangle 
\hspace*{2mm}
| \chi^\times_2 \rangle 
\Big]
\left[
\begin{array}{cc}
K_{00} 
&
- K_{02} 
\\
- K_{02} 
&
- K_{22} 
\end{array}
\right]
\left[
\begin{array}{c}
\langle \chi^\times_0 |
\\
\langle \chi^\times_2 |
\end{array}
\right]
\nonumber \\
& = & 
4 \pi 
\Big[
| \sigma^\times_0 \rangle 
\hspace*{2mm}
| \sigma^\times_2 \rangle 
\Big]
\left[
\begin{array}{cc}
\Lambda_{00} 
&
- \Lambda_{02} 
\\
- \Lambda_{02} 
&
- \Lambda_{22} 
\end{array}
\right]
\left[
\begin{array}{c}
\langle \sigma^\times_0 |
\\
\langle \sigma^\times_2 |
\end{array}
\right]
, 
\nonumber \\ & & 
\end{eqnarray}
and 
\begin{equation}
\label{matrixLambda}
\left[
\begin{array}{cc}
\Lambda_{00} & - \Lambda_{02} 
\\
- \Lambda_{02} &  - \Lambda_{22}
\end{array}
\right]
\equiv
\left[
\begin{array}{cc}
K_{00} & - K_{02} 
\\
- K_{02} &  - K_{22}
\end{array}
\right]^{-1}
. 
\end{equation}
We see from Eqs.~(\ref{Pi0plusPi2}) and 
(\ref{Pi0timesplusPi2times}) that 
the operators $\widehat{\Pi}_{0}  + \widehat{\Pi}_{2}$ 
and $\widehat{\Pi}^\times_{0}  + \widehat{\Pi}^\times_{2}$
are Hermitian 
although $\widehat{\Pi}_{0}$, $\widehat{\Pi}_{2}$, 
 $\widehat{\Pi}^\times_{0}$, and $\widehat{\Pi}^\times_{2}$ are
 not Hermitian independently. 
We can also use Eqs.~(\ref{Kchi1}), (\ref{Kchi1times}), 
(\ref{PiPitimes}), and (\ref{Pi1Pitimes}),
to express 
$\widehat{\Pi}_{1}$ and $\widehat{\Pi}^\times_{1}$ as 
\begin{eqnarray}
\label{Pi1Kcc}
\widehat{\Pi}_{1}  
& = & 
| \sigma_1 \rangle \langle \chi_1 |
=
| \chi_1 \rangle \langle \sigma_1 |
\nonumber \\ 
& = & 
\frac{1}{4 \pi}
K_{11} 
| \chi_1 \rangle \langle \chi_1 |
=
4 \pi 
\Lambda_{11} 
| \sigma_1 \rangle \langle \sigma_1 |
,
\end{eqnarray}
and 
\begin{eqnarray}
\label{Pi1Kcctimes}
\widehat{\Pi}^\times_{1}  
& = & 
| \sigma^\times_1 \rangle \langle \chi^\times_1 |
=
| \chi^\times_1 \rangle \langle \sigma^\times_1 |
\nonumber \\ 
& = & 
\frac{1}{4 \pi}
K_{11} 
| \chi^\times_1 \rangle \langle \chi^\times_1 |
=
4 \pi 
\Lambda_{11} 
| \sigma^\times_1 \rangle \langle \sigma^\times_1 |
, 
\end{eqnarray}
respectively, where 
\begin{equation}
\label{Lambda11}
\Lambda_{11} = 1 / K_{11}
.
\end{equation}
We see from Eqs.~(\ref{Pi1Kcc}) and (\ref{Pi1Kcctimes}) that 
$\widehat{\Pi}_{1}$ and $\widehat{\Pi}^\times_{1}$ 
are Hermitian positive-definite operators. 

The conserved quantity in Eq.~(\ref{GKC}) is expressed as 
\begin{eqnarray}
& & 
\hspace*{-3mm}
\sum_a 
T_a 
\int d^3 v 
\frac{| f_{a{\bf k}(\tau) }|^2}{2 f_{aM}}
+ \frac{1}{8\pi}
 ( |{\bf E}_{\bf k}(\tau) |^2 + |{\bf B}_{\bf k}(\tau) |^2 )
 \nonumber \\ & &
=
\frac{1}{2}
p_{\rm tot} 
\langle 
\widetilde{f} (\tau) 
| \widehat{S} |
\widetilde{f} (\tau) \rangle 
, 
\end{eqnarray}
where 
\begin{equation}
\label{ptot}
p_{\rm tot} 
\equiv 
\sum_a p_a 
\end{equation}
is the total equilibrium pressure and 
the operator $\widehat{S}$ is defined by 
\begin{eqnarray}
\label{ptotS}
p_{\rm tot} 
\widehat{S} 
&  \equiv  & 
\big( \widehat{p} -  \widehat{\Pi}_{1}
+ \widehat{\Pi}_{2}   \big)^\dagger 
 \widehat{p}^{-1}
\big( \widehat{p} 
-  \widehat{\Pi}_{1} + \widehat{\Pi}_{2}
 \big) 
\nonumber \\ 
& & \mbox{}
+ \frac{1}{4\pi} 
\Big[
K_{00} \widehat{\Phi}_0
+ k_\perp^2 
\big( \widehat{\Phi}_1
+ \widehat{\Phi}_2 \big)
\Big]
, 
\end{eqnarray}
with 
\begin{equation}
\label{Phinu}
\widehat{\Phi}_\nu
\equiv 
| \chi_\nu \rangle \langle \chi_\nu |
\hspace*{2mm}
( \nu = 0, 1, 2)
. 
\end{equation}
We see from Eqs.~(\ref{ptotS}) and (\ref{Phinu}) that 
$\widehat{S}$ and $\widehat{\Phi}_j$ $(j=0,1,2)$ 
are Hermitian positive-definite operators. 
Therefore, there exists a unique Hermitian positive-definite operator 
$\widehat{A} (= \widehat{A}^\dagger)$ which satisfies 
\begin{equation}
\label{SA2}
\widehat{S}
= 
\widehat{A}^2
.
\end{equation}
It is shown later how to obtain a specific expression of $\widehat{A}$. 

In the same way as in Sec.~III.B, 
we define the state vector 
\begin{equation}
\label{psiAf}
| \psi (\tau) \rangle \equiv 
\widehat{A} |\widetilde{f}(\tau)\rangle
. 
\end{equation}
Then, we see from Eqs.~(\ref{ptotS}), (\ref{SA2}), and (\ref{psiAf}) 
that 
\begin{equation}
\langle \psi (\tau) 
 | \psi (\tau) \rangle
\equiv 
\langle 
\widetilde{f} (\tau) 
| \widehat{S} |
\widetilde{f} (\tau) \rangle 
\end{equation}
is independent of $\tau$ 
and 
that 
the time evolution operator $\widehat{U}(\tau)$ defined by 
$
 | \psi (\tau) \rangle
= 
\widehat{U}(\tau)
 | \psi (0) \rangle
$
is unitary. 
Then, $\widehat{U}(\tau)$ can be written 
in terms of a Hermitian operator $\widehat{H}$, 
which is the Hamiltonian, 
as 
$
\widehat{U} (\tau)
= 
\exp ( - i \tau \widehat{H} )
$, 
and 
$| \psi (\tau) \rangle$ satisfies 
the Schr\"{o}dinger equation shown in Eq.~(\ref{Schreq}). 
From Eqs.~(\ref{LGKf2}) and (\ref{psiAf}), we find that 
$\widehat{H}$ is expressed as 
\begin{equation}
\label{hatH}
\widehat{H}
\equiv 
\widehat{A}
\,
\widehat{L}
\,
\widehat{A}^{-1}
. 
\end{equation}

Using Eqs.~(\ref{Pi}), (\ref{Pi0plusPi2}) and (\ref{Pi1Kcc}), 
we can reduce Eq.~(\ref{ptotS}) to 
a more compact form as 
\begin{equation}
p_{\rm tot} 
\widehat{S} 
=
\widehat{p} + \widehat{\Pi}
. 
\end{equation}
Then,  Eq.~(\ref{hatL}) is rewritten as 
\begin{equation}
\label{hatL2}
\widehat{L}
 \equiv 
\frac{p_{\rm tot}}{v_c}
\,
\widehat{v}_T
\,
 \widehat{p}^{-1}
 \,
 \widehat{\Xi}
 \,
\widehat{S}
, 
\end{equation}
from which 
we obtain 
\begin{equation}
\label{SL}
\widehat{S} 
\,
\widehat{L} 
=
\frac{p_{\rm tot}}{v_c}
\,
\widehat{S} 
\,
\widehat{v}_T
\,
 \widehat{p}^{-1}
 \,
 \widehat{\Xi}
 \,
\widehat{S}
.
\end{equation}
On the right-hand side of Eq.~(\ref{SL}), 
the three Hermitian operators
$\widehat{v}_T$,
$ \widehat{p}^{-1}$, and 
$ \widehat{\Xi}$ commute with one another, 
and their product  is 
expressed as 
\begin{eqnarray}
\label{vpXi}
& & 
\hspace*{-5mm}
\widehat{v}_T
\,
 \widehat{p}^{-1}
 \,
 \widehat{\Xi}
\nonumber \\
& & 
\hspace*{-3mm}
=
\bigg(
\sum_a 
| a \rangle 
\frac{v_{Ta}}{p_a}
\langle a |
\bigg)
\otimes 
\widehat{1}_{v_\perp}
\otimes 
\int_{-\infty}^{+\infty}
| \xi \rangle 
\xi d\xi
\langle \xi |
\nonumber \\
& & 
\hspace*{-3mm}
=
\bigg(
\sum_a 
| a \rangle 
\frac{v_{Ta}}{p_a}
\langle a |
\bigg)
\otimes 
\widehat{1}_{v_\perp}
\nonumber \\
& & \mbox{}
\otimes 
\sum_{n=0}^\infty
\bigg(
\frac{n+1}{2}
\bigg)^{1/2}
\big(
| n + 1 \rangle  \langle n |
+
| n \rangle  \langle n + 1 |
\big)
,
\hspace*{3mm}
\end{eqnarray}
which is apparently 
a Hermitian operator as well. 
Since $\widehat{S}$ and $\widehat{v}_T  \widehat{p}^{-1} \widehat{\Xi}$ are 
Hermitian, we find from Eqs.~(\ref{SL}) and (\ref{vpXi}) that 
$\widehat{S} \widehat{L}$ is Hermitian, too, 
\begin{equation} 
\label{SLLS}
\widehat{S} \widehat{L} 
= 
\widehat{L}^\dagger \widehat{S} 
.
\end{equation}
Substituting Eq.~(\ref{hatL2})  into Eq.~(\ref{hatH}) and using Eq.~(\ref{SA2})
we can express the Hamiltonian operator $\widehat{H}$ as 
\begin{equation}
\label{hatHGK}
\widehat{H}
=
\frac{p_{\rm tot}}{v_c}
\,
\widehat{A}
\,
\widehat{v}_T
\,
 \widehat{p}^{-1}
 \,
 \widehat{\Xi}
 \,
\widehat{A}
, 
\end{equation}
which is easily found to be Hermitian because $\widehat{A}$ and 
$\widehat{v}_T \widehat{p}^{-1} \widehat{\Xi}$ are Hermitian.

\subsection{Matrix representation of state vectors and operators
for the gyrokinetic system}

Hereafter, to represent the operators  by matrices, 
we use the orthonormal basis vector systems 
$\{ |a \, j \, \xi \rangle \}$ 
or 
$\{ |a \, j \, n \rangle \}$ in ${\cal E}$, where 
$
|a \, j \, \xi \rangle \equiv | a \, j  \rangle \otimes | \xi \rangle
$
and 
$
|a \, j \, n \rangle \equiv | a  \, j  \rangle \otimes | n \rangle
$. 
Here, $| a \, j  \rangle$ $(a = a_1, \cdots , a_{N_{ps}} ; j = 0, 1, 2, \cdots)$ are the basis vectors in 
${\cal E}^\times \equiv {\cal E}_{ps} \otimes {\cal E}_{v_\perp}$ 
as defined in Appendix~F. 
The basis vectors
$| n  \rangle$ $(n =0, 1,  2, \cdots )$ in ${\cal E}_{v_\parallel}$ 
are defined by 
$\langle \xi | n \rangle = 
h_n (\xi)
\equiv
\pi^{-1/4} e^{-\xi^2/2}
 H_n(\xi) /  
 (2^n n!)^{1/2}
$. 
We see that, 
\begin{equation}
\langle a \, j \, \xi | 
\Pi_0
| a \, j' \, \xi \rangle 
=
\langle a \, j \, \xi | 
\Pi_2
| a' \, j' \, \xi' \rangle 
= 0 
\; 
 \mbox{for} \; ( j , j' ) \neq ( 0 , 0 )
, 
\end{equation}
\begin{equation}
\langle a \, j \, \xi | 
\Pi_1
| a' \, j' \, \xi' \rangle 
= 0 
\; 
 \mbox{for} \; ( j , j' ) \neq ( 1 , 1 )
, 
\end{equation}
and 
\begin{equation}
\langle a \, j \, \xi | 
\Pi
| a' \, j' \, \xi' \rangle 
= 0 
\; 
\mbox{for} \; ( j , j' ) \neq ( 0 , 0 ), ( 1 , 1 )
. 
\end{equation}
We consider 
the subspaces ${\cal E}^\times_{\rm I}$ and 
${\cal E}^\times_{\rm I\!I}$ of 
${\cal E}^\times \equiv {\cal E}_{ps} \otimes {\cal E}_{v_\perp}$, 
which are spanned by 
$\{ | a \, j \rangle \}_{a = a_1, \cdots, a_{N_{ps}} ; j=0,1}$ and 
$\{ | a \, j \rangle \}_{a = a_1, \cdots, a_{N_{ps}} ; j\geq 2}$, respectively. 
Therefore, these subspaces are orthogonal to each other.
Then, ${\cal E}^\times = {\cal E}^\times_{\rm I} \oplus {\cal E}^\times_{\rm I\!I}$ 
where  $\oplus$ represents the direct sum.  
The projectors $\widehat{P}^\times_{\rm I}$ and 
$\widehat{P}^\times_{\rm I\!I}$ associated 
with ${\cal E}^\times_{\rm I}$ and ${\cal E}^\times_{\rm I\!I}$ 
are given by 
\begin{equation}
\widehat{P}^\times_{\rm I} \equiv 
\sum_a 
\sum_{j=0}^1  | a \, j \rangle \langle a \, j |
\hspace*{3mm}
\mbox{and}
\hspace*{3mm}
\widehat{P}^\times_{\rm I\!I} \equiv 
\sum_a
\sum_{j=2}^\infty  | a \, j \rangle \langle a \, j |
,
\end{equation}
respectively, which satisfy 
$\widehat{P}^\times_{\rm I} \oplus \widehat{P}^\times_{\rm I\!I} = 
\widehat{1}^\times \equiv \widehat{1}_{ps} \otimes\widehat{1}_{v_\perp}$
and
$\widehat{P}^\times_{\rm I} \, \widehat{P}^\times_{\rm I\!I} = 
\widehat{P}^\times_{\rm I\!I} \, \widehat{P}^\times_{\rm I} =
0$. 
Then, the subspaces ${\cal E}_{\rm I}$ and ${\cal E}_{\rm I\!I}$ 
in ${\cal E} \equiv {\cal E}^\times 
\otimes {\cal E}_{v_\parallel} \equiv 
{\cal E}_{\rm I} \oplus {\cal E}_{\rm I\!I}$ are defined by 
\begin{equation}
{\cal E}_{\rm I} \equiv 
{\cal E}^\times_{\rm I} 
\otimes {\cal E}_{v_\parallel}
\hspace*{3mm}
\mbox{and}
\hspace*{3mm}
{\cal E}_{\rm I\!I} \equiv 
{\cal E}^\times_{\rm I\!I} 
\otimes {\cal E}_{v_\parallel}
,
\end{equation}
which are orthogonal to each other. 
The corresponding projectors 
$\widehat{P}_{\rm I}$ and 
$\widehat{P}_{\rm I\!I}$ are given by 
\begin{equation}
 \widehat{P}_{\rm I} \equiv 
 \widehat{P}^\times_{\rm I}
\otimes \widehat{1}_{v_\parallel}
\hspace*{3mm}
\mbox{and}
\hspace*{3mm}
\widehat{P}_{\rm I\!I} \equiv 
\widehat{P}^\times_{\rm I\!I} 
\otimes \widehat{1}_{v_\parallel}
,
\end{equation}
which satisfy
$\widehat{P}_{\rm I} \oplus \widehat{P}_{\rm I\!I} = 
\widehat{1} \equiv 
\widehat{1}^\times 
\otimes \widehat{1}_{v_\parallel}
$
and 
$\widehat{P}_{\rm I} \, \widehat{P}_{\rm I\!I} 
= \widehat{P}_{\rm I\!I} \, \widehat{P}_{\rm I} = 0$.

We now express the operator $\widehat{p}$ associated with the equilibrium pressure 
as the direct sum of the operators $\widehat{p}_{\rm I}$ and $\widehat{p}_{\rm I\!I}$, 
\begin{equation}
\label{hatp}
\widehat{p}
= 
\widehat{p}_{\rm I} \oplus \widehat{p}_{\rm I\!I}
,  
\end{equation}
where 
\begin{equation}
\label{hatpI}
\widehat{p}_{\rm I} 
\equiv 
\widehat{p}^\times_{\rm I} \otimes 
\widehat{1}_{v_\parallel}
\equiv
\sum_a \sum_{j=0}^1 
| a \, j \rangle p_a \langle a \, j |
\otimes
\widehat{1}_{v_\parallel}
,  
\end{equation}
and 
\begin{equation}
\label{hatpII}
\widehat{p}_{\rm I\!I}
\equiv 
\widehat{p}^\times_{\rm I\!I} \otimes 
\widehat{1}_{v_\parallel}
\equiv
\sum_a \sum_{j=2}^\infty
| a \, j \rangle p_a \langle a \, j |
\otimes
\widehat{1}_{v_\parallel}
.  
\end{equation}
Here and hereafter, 
the subscripts ${\rm I}$ and ${\rm I\!I}$ are used to 
represent the operators of class ${\rm I}$ and class ${\rm I\!I}$, respectively. 
An operator $\widehat{*}_{\rm I}$ of class ${\rm I}$ 
has nonzero matrix components 
$\langle a \, j \, \xi |  \widehat{*}_{\rm I}  | a' \, j' \, \xi' \rangle $ 
only when $j \leq 1$ and $j' \leq 1$. 
Note that $\widehat{\Pi}_0$, $\widehat{\Pi}_1$, $\widehat{\Pi}_2$, and 
$\widehat{\Pi}$ are operators of class ${\rm I}$. 
On the other hand, 
an operator $\widehat{*}_{\rm I\!I}$ of class ${\rm I\!I}$
has nonzero matrix components 
$\langle a \, j \, \xi | \widehat{*}_{\rm I\!I} | a' \, j' \, \xi' \rangle $ 
only when $j \geq 2$ and $j' \geq 2$. 
In other words, 
$\widehat{*}_{\rm I} | {\rm I} \rangle \in  {\cal E}_{\rm I}$, 
$\widehat{*}_{\rm I\!I} | {\rm I\!I} \rangle \in  {\cal E}_{\rm I\!I}$, 
and 
$\widehat{*}_{\rm I} | {\rm I\!I} \rangle
= \widehat{*}_{\rm I\!I} | {\rm I} \rangle = 0$ 
hold for arbitrary vectors 
$ | {\rm I} \rangle \in {\cal E}_{\rm I}$ and 
$  | {\rm I\!I} \rangle \in {\cal E}_{\rm I\!I}$. 
We can also characterize them by 
$ \widehat{P}_{\rm I} \,  \widehat{*}_{\rm I} 
= \widehat{*}_{\rm I} \, \widehat{P}_{\rm I}
= \widehat{*}_{\rm I}$ 
and 
$\widehat{P}_{\rm I\!I} \,  \widehat{*}_{\rm I\!I} 
= \widehat{*}_{\rm I\!I} \, \widehat{P}_{\rm I\!I} 
= \widehat{*}_{\rm I\!I}$. 
We then easily find 
that the product of operators of class ${\rm I}$ and class ${\rm I\!I}$ 
vanish as 
$ \widehat{*}_{\rm I} \, \widehat{*}_{\rm I\!I} =
\widehat{*}_{\rm I\!I} \, \widehat{*}_{\rm I} = 0$. 

Similarly to $\widehat{p}$ in Eq.~(\ref{hatp}), 
$\widehat{S}$  is given by the direct 
sum of two Hermitian operators 
of class ${\rm I}$ and class ${\rm I\!I}$ as
\begin{equation}
\widehat{S}
\equiv 
\widehat{S}_{\rm I}
\oplus
\widehat{S}_{\rm I\!I}
,
\end{equation}
where $\widehat{S}_{\rm I}$ and
$\widehat{S}_{\rm I\!I}$ are given by 
\begin{equation}
\label{ptotSI}
p_{\rm tot}
\widehat{S}_{\rm I}
 \equiv  
\widehat{p}_{\rm I}
+ \widehat{\Pi}
\end{equation}
and
\begin{equation}
\label{ptotSII}
p_{\rm tot}
\widehat{S}_{\rm I\!I}
 \equiv
\widehat{p}_{\rm I\!I}
,
\end{equation}
respectively. 
The operator $\widehat{A}$ is also given by the direct
sum of two Hermitian operators of class ${\rm I}$ 
and  class ${\rm I\!I}$ as 
\begin{equation}
\label{hatA}
\widehat{A}
\equiv 
\widehat{A}_{\rm I}
\oplus
\widehat{A}_{\rm I\!I}
,
\end{equation}
where $\widehat{A}_{\rm I}$ and $\widehat{A}_{\rm I\!I}$ 
are defined as those which satisfy
\begin{equation}
\label{SA2IandII}
\widehat{S}_{\rm I} =  \widehat{A}_{\rm I}^{\; 2}
\hspace{3mm}
\mbox{and}
\hspace{3mm}
\widehat{S}_{\rm I\!I}  = \widehat{A}_{\rm I\!I}^{\; 2}
,
\end{equation}
respectively. 
From Eqs.~(\ref{hatpII}), 
(\ref{ptotSII}) and (\ref{SA2IandII}), we find that  $\widehat{A}_{\rm I\!I}$ is given by 
\begin{eqnarray}
\label{AII}
\widehat{A}_{\rm I\!I}
& \equiv & 
\frac{1}{\sqrt{p_{\rm tot}}} (\widehat{p}_{\rm I\!I})^{1/2}
 \equiv  
\frac{1}{\sqrt{p_{\rm tot}}} (\widehat{p}^\times_{\rm I\!I})^{1/2}
\otimes \widehat{1}_{v_\parallel}
\nonumber \\
& \equiv & 
\frac{1}{\sqrt{p_{\rm tot}}}
\bigg(
\sum_a 
\sum_{j=2}^\infty 
| a \, j \rangle 
\sqrt{p_a}
\langle a \, j |
\bigg)
\otimes
\widehat{1}_{v_\parallel}
\nonumber \\ 
& = & 
\frac{1}{\sqrt{p_{\rm tot}}}
\sum_a 
\sum_{j=2}^\infty 
\int_{-\infty}^{+\infty}
| a \, j \, \xi \rangle 
\sqrt{p_a} d\xi
\langle a \, j \, \xi |
\nonumber \\ 
& = & 
\frac{1}{\sqrt{p_{\rm tot}}}
\sum_a 
\sum_{j=2}^\infty 
\sum_{n=0}^\infty
| a \, j \, n \rangle 
\sqrt{p_a} 
\langle a \, j \, n |
.  
\end{eqnarray}

Before giving a detailed expression of $\widehat{A}_{\rm I}$, 
we use Eq.~(\ref{ptotS}) to express 
the operator $\widehat{S}_{\rm I}$  as
\begin{equation}
\label{SI01n}
\widehat{S}_{\rm I}
=
\widehat{S}^\times_{{\rm I},0}
\otimes
| 0 \rangle   \langle 0 |
+
\widehat{S}^\times_{{\rm I},1}
\otimes
| 1 \rangle   \langle 1 |
+
\widehat{S}^\times_{{\rm I}, \geq 2}
\otimes
\sum_{n=2}^\infty
| n \rangle   \langle n |
, 
\end{equation}
where $\widehat{S}^\times_{{\rm I}, 0}$, $\widehat{S}^\times_{{\rm I}, 1}$, 
and $\widehat{S}^\times_{{\rm I}, \geq 2}$ are defined by 
\begin{eqnarray}
\label{ptotStimesI0}
p_{\rm tot} 
\widehat{S}^\times_{{\rm I}, 0}
&  \equiv  & 
\big(
\widehat{p}^\times_{\rm I}
+ \widehat{\Pi}^\times_{2} 
\big)^\dagger    
\big( \widehat{p}^\times_{\rm I} \big)^{-1}
\big(
\widehat{p}^\times_{\rm I}
+ \widehat{\Pi}^\times_{2} 
\big)
\nonumber \\ 
& & \mbox{}
+ \frac{1}{4\pi} 
\big(
K_{00} \widehat{\Phi}^\times_0
+ k_\perp^2 
 \widehat{\Phi}^\times_2 
\big)
\nonumber \\ 
& = & 
\widehat{p}^\times_{\rm I}
+
\widehat{\Pi}^\times_0
+
\widehat{\Pi}^\times_2
,
\end{eqnarray}
\begin{eqnarray}
\label{ptotStimesI1}
p_{\rm tot} 
\widehat{S}^\times_{{\rm I}, 1}
&  \equiv  & 
\big(
\widehat{p}^\times_{\rm I}
- \widehat{\Pi}^\times_{1} 
\big)^\dagger    
\big( \widehat{p}^\times_{\rm I} \big)^{-1}
\big(
\widehat{p}^\times_{\rm I}
- \widehat{\Pi}^\times_{1} 
\big)
\nonumber \\ 
& & \mbox{}
+ \frac{1}{4\pi} 
 k_\perp^2 
 \widehat{\Phi}^\times_1
 \nonumber \\
& = & 
\widehat{p}^\times_{\rm I}
-
\widehat{\Pi}^\times_1
, 
\end{eqnarray}
and
\begin{equation}
\label{ptotStimesIn}
p_{\rm tot}
\widehat{S}^\times_{{\rm I}, \geq 2}
\equiv
\widehat{p}^\times_{\rm I}
,
\end{equation}
respectively, 
where 
\begin{equation}
\widehat{\Phi}^\times_\nu
\equiv 
| \chi^\times_\nu\rangle \langle \chi^\times_\nu |
\hspace*{2mm}
( \nu  = 0, 1, 2)
. 
\end{equation}
In the same way as $\widehat{S}_{\rm I}$ in 
Eq.~(\ref{SI01n})
the operator $\widehat{A}_{\rm I}$ is
expressed as 
\begin{equation}
\label{AIAAA}
\widehat{A}_{\rm I}
=
\widehat{A}^\times_{{\rm I}, 0}
\otimes
| 0 \rangle   \langle 0 |
+
\widehat{A}^\times_{{\rm I}, 1}
\otimes
| 1 \rangle   \langle 1 |
+
\widehat{A}^\times_{{\rm I}, \geq 2}
\otimes
\sum_{n=2}^\infty
| n \rangle   \langle n |
,
\end{equation}
where
$\widehat{A}^\times_{{\rm I}, 0}$, 
$\widehat{A}^\times_{{\rm I}, 1}$, and 
$\widehat{A}^\times_{{\rm I}, \geq 2}$ 
satisfy
\begin{equation}
\label{SItimesAi2}
\widehat{S}^\times_{{\rm I}, 0}
=
(\widehat{A}^\times_{{\rm I}, 0})^2
,
\hspace*{3mm}
\widehat{S}^\times_{{\rm I}, 1}
=
(\widehat{A}^\times_{{\rm I}, 1})^2
, 
\hspace*{3mm}
\mbox{and}
\hspace*{3mm}
\widehat{S}^\times_{{\rm I}, \geq 2}
=
(\widehat{A}^\times_{{\rm I}, \geq 2})^2
,
\end{equation}
respectively.
Using Eqs.~(\ref{hatpI}), (\ref{ptotStimesIn}) and (\ref{SItimesAi2}), 
we immediately obtain 
\begin{equation}
\sqrt{ p_{\rm tot} }
\,
\widehat{A}^\times_{{\rm I}, \geq 2}
=
( \widehat{p}^\times_{\rm I} )^{1/2}
\equiv
\sum_a 
\sum_{j=0}^1
| a \, j \rangle
\,
\sqrt{p_a}
\,
\langle a  \, j |
.
\end{equation}

The matrix components 
$\langle a \,j | 
\widehat{S}^\times_{{\rm I}, 0}
| a' \, j' \rangle$ 
and 
$\langle a \, j | 
\widehat{S}^\times_{{\rm I}, 1}
| a' \, j' \rangle$ 
are nonzero only if 
$j \leq 1$ and $j' \leq 1$. 
Thus, 
$\widehat{S}^\times_{{\rm I}, 0}$ and 
$\widehat{S}^\times_{{\rm I}, 1}$ defined in 
Eqs.~(\ref{ptotStimesI0}) and (\ref{ptotStimesI1}) are 
expressed as 
\begin{equation}
\widehat{S}^\times_{{\rm I}, 0}
=
\sum_a \sum_{a'}
\sum_{j=0}^1 \sum_{j'=0}^1
| a \, j \rangle
\,
\langle a \, j | 
\widehat{S}^\times_{{\rm I}, 0}
| a' \, j' \rangle
\,
\langle a'  \, j' |
\end{equation}
and 
\begin{equation}
\widehat{S}^\times_{{\rm I}, 1}
=
\sum_a \sum_{a'}
\sum_{j=0}^1 \sum_{j'=0}^1
| a \, j \rangle
\,
\langle a \, j | 
\widehat{S}^\times_{{\rm I}, 1}
| a' \, j' \rangle
\,
\langle a'  \, j' |
,
\end{equation}
respectively, where 
\begin{equation}
\label{StimesI0matrix}
p_{\rm tot} 
\langle a \, j | 
\widehat{S}^\times_{{\rm I}, 0}
| a' \, j' \rangle
=
p_a \delta_{a a'} \delta_{j j'}
+
\langle a \; j | 
( \widehat{\Pi}^\times_0 + \widehat{\Pi}^\times_2 )
| a' \; j' \rangle
\end{equation}
and 
\begin{equation}
\label{StimesI1matrix}
p_{\rm tot} 
\langle a \, j | 
\widehat{S}^\times_{{\rm I}, 1}
| a' \, j' \rangle
=
p_a \delta_{a a'} \delta_{j j'}
-
\langle a \, j | 
\widehat{\Pi}^\times_1 
| a' \, j' \rangle
, 
\end{equation}
for $j \leq 1$ and $j' \leq 1$. 
Thus, $\widehat{S}^\times_{{\rm I}, 0}$ 
and $\widehat{S}^\times_{{\rm I}, 1}$ are 
represented by  $2 N_{ps} \times 2 N_{ps}$ Hermitian positive-definite matrices, 
${\bf S}^\times_{{\rm I}, 0} 
\equiv 
\big[ \langle a \, j | 
\widehat{S}^\times_{{\rm I}, 0}
| a' \, j' \rangle \big ]_{j, j' = 0, 1}$ 
and 
${\bf S}^\times_{{\rm I}, 1} 
\equiv 
\big[ \langle a \, j | 
\widehat{S}^\times_{{\rm I}, 1}
| a' \, j' \rangle \big ]_{j, j' = 0, 1}$, 
the components of which are given by 
Eqs.~(\ref{StimesI0matrix}) and (\ref{StimesI1matrix}). 
The $2 N_{ps}$ positive eigenvalues and the $2 N_{ps}$  corresponding eigenvectors 
can be evaluated relatively easily 
for ${\bf S}^\times_{{\rm I}, 0}$ and ${\bf S}^\times_{{\rm I}, 1}$. 
Using the square roots of the eigenvalues and the eigenvectors 
for ${\bf S}^\times_{{\rm I}, 0}$ and ${\bf S}^\times_{{\rm I}, 1}$, 
we can obtain
these matrices' square roots, 
${\bf A}^\times_{{\rm I}, 0}$ and 
${\bf A}^\times_{{\rm I}, 1}$, 
which are Hermitian positive-definite matrices satisfying 
${\bf S}^\times_{{\rm I}, 0} =( {\bf A}^\times_{{\rm I}, 0} )^2$ 
and 
${\bf S}^\times_{{\rm I}, 1} =( {\bf A}^\times_{{\rm I}, 1} )^2$. 
Using the obtained matrices 
${\bf A}^\times_{{\rm I}, 0}
\equiv 
\big[ \langle a \, j | 
\widehat{A}^\times_{{\rm I}, 0}
| a' \, j' \rangle \big ]_{j, j' = 0, 1}$ 
and 
${\bf A}^\times_{{\rm I}, 1}
\equiv 
\big[ \langle a \, j | 
\widehat{A}^\times_{{\rm I}, 1}
| a' \, j' \rangle \big ]_{j, j' = 0, 1}$, 
the operators 
$\widehat{A}^\times_{{\rm I}, 0}$ 
and  $\widehat{A}^\times_{{\rm I}, 1}$ 
can be expressed as
\begin{equation}
\widehat{A}^\times_{{\rm I}, 0}
=
\sum_a \sum_{a'}
\sum_{j=0}^1 \sum_{j'=0}^1
| a \, j \rangle
\,
\langle a \, j | 
\widehat{A}^\times_{{\rm I}, 0}
| a' \, j' \rangle
\,
\langle a'  \, j' |
\end{equation}
and 
\begin{equation}
\widehat{A}^\times_{{\rm I}, 1}
=
\sum_a \sum_{a'}
\sum_{j=0}^1 \sum_{j'=0}^1
| a \, j \rangle
\,
\langle a \, j | 
\widehat{A}^\times_{{\rm I}, 1}
| a' \, j' \rangle
\,
\langle a'  \, j' |
,
\end{equation}
respectively.

The Hamiltonian operator $\widehat{H}$ is also expressed by
 the sum of operators of class ${\rm I}$ and class ${\rm I\!I}$ as
\begin{equation}
\label{HIpII}
\widehat{H}
\equiv 
\widehat{H}_{\rm I}
\oplus
\widehat{H}_{\rm I\!I}
,
\end{equation}
where, 
\begin{equation}
\label{HI}
\widehat{H}_{\rm I}
\equiv   
\frac{p_{\rm tot}}{v_c}
\,
\widehat{A}_{\rm I} 
\,
\widehat{v}_T
\,
\widehat{p}^{-1}
 \,
 \widehat{\Xi}
 \,
 \widehat{A}_{\rm I} 
\end{equation}
and 
\begin{eqnarray}
\label{HII}
\widehat{H}_{\rm I\!I}
& \equiv & 
\frac{p_{\rm tot}}{v_c}
\,
\widehat{A}_{\rm I\!I}
\,
\widehat{v}_T
\,
 \widehat{p}^{-1}
 \,
 \widehat{\Xi}
 \,
 \widehat{A}_{\rm I\!I}
\nonumber \\ 
& = & 
\bigg(
\sum_a 
\sum_{j=2}^\infty
| a \, j \rangle 
\frac{v_{Ta} }{v_c}
\langle a \, j |
\bigg)
\otimes
\int_{-\infty}^{+\infty}
| \xi \rangle 
\xi \, d\xi
\langle \xi |
\nonumber \\ 
& = & 
\bigg(
\sum_a 
\sum_{j=2}^\infty
| a \, j \rangle 
\frac{v_{Ta} }{v_c}
\langle a \, j |
\bigg)
\nonumber \\ 
& & \mbox{}
\otimes
\sum_{n=0}^\infty
\bigg(
\frac{n+1}{2}
\bigg)^{1/2}
\big(
| n+1 \rangle \langle n |
+ | n \rangle \langle n+1 |
\big)
.
\hspace*{10mm}
\end{eqnarray}

As in Secs.~III.E and III.F, 
to apply the fluctuation theorems presented in Sec.~II 
and validate them numerically, 
we need to truncate the set of basis vectors
$\{ | a\, j\, n \rangle \}$
by restricting the indices to the finite ranges
$j = 0,1,\cdots,J-1$ and
$n = 0,1,\cdots,N-1$.
This yields the
$N_{ps}JN \times N_{ps}JN$ Hamiltonian matrix
$
{\bf H}
\equiv
[
\langle a\, j\, n | \widehat{H} | a'\, j'\, n' \rangle
]
$, 
which is block-diagonal, with blocks separated according to
the conditions $j \leq 1$ and $j \geq 2$. 
We also need to define the
$N_{ps}JN \times N_{ps}JN$ time-reversal matrices
$
{\bf T}
\equiv
[
\langle a\, j\, n | \widehat{T} | a'\, j'\, n' \rangle
]
$
and
$
{\bf K}
\equiv
[
\langle a\, j\, n |
( \widehat{K} | a'\, j'\, n' \rangle )
]
$, 
where the unitary and antiunitary time-reversal operators  
$\widehat{T}$ and $\widehat{K}$ 
are defined in Sec.~IV.D. 
The matrices ${\bf T}$ and ${\bf K}$ thus defined
are both diagonal (see Sec.~IV.D).

\subsection{Time-reversal operators in the linear gyrokinetic system}

We here define time-reversal unitary and antiunitary 
operators, $\widehat{T}$ and $\widehat{K}$, 
in the linear gyrokinetic system
in the same manner as those in the linear Vlasov-Poisson system shown 
in Sec.~II.C.
The operators $\widehat{T}$ and $\widehat{K}$ defined below 
are shown to satisfy the conditions given by Eqs.~(\ref{TK}) and 
(\ref{THK}) in Appendix~A. 

The time reversal unitary operator $\widehat{T}$ is defined as a  
linear operator which satisfies 
\begin{equation}
\label{Tajxi}
\widehat{T} | a \, j \, \xi \rangle 
=  
| a \, j \,  - \xi \rangle 
, 
\end{equation}
which is equivalent to 
\begin{equation}
\label{Tajn}
\widehat{T} | a \, j \, n \rangle 
=  
(- 1)^n
| a \, j \, n \rangle 
.
\end{equation}
In the same way as in the case of the time-reversal unitary operator 
in the linear Vlasov-Poisson system, 
$\widehat{T}$ defined above 
corresponds to the transformation from the perturbed gyrocenter 
distribution function 
$f^{(g)}_{a {\bf k}_\perp} (v_\parallel, v_\perp, t)$ 
to 
$f^{(g)}_{a {\bf k}_\perp} (- v_\parallel, v_\perp, t)$.  
We find from 
Eqs.~(\ref{vpXi}), (\ref{hatA}), (\ref{AII}), (\ref{AIAAA}), 
(\ref{HIpII})--(\ref{HII}), 
and (\ref{Tajn}) that $\widehat{T}$ satisfies 
$\widehat{T} = \widehat{T}^{-1} = \widehat{T}^\dagger$, 
$
\widehat{T} \, \widehat{v}_T \, \widehat{p}^{-1} \,  \widehat{\Xi} 
= 
-
 \widehat{v}_T \,  \widehat{p}^{-1} \,  \widehat{\Xi}  \, \widehat{T}
$,
$
\widehat{T} \, \widehat{A}_{\rm I} =  \widehat{A}_{\rm I}  \, \widehat{T}
$, 
$
\widehat{T} \, \widehat{A}_{\rm I\!I} =  \widehat{A}_{\rm I\!I}  \, \widehat{T}
$, 
$
\widehat{T} \, \widehat{A} =  \widehat{A} \, \widehat{T}
$, 
$
\widehat{T} \, \widehat{H}_{\rm I}  = - \widehat{H}_{\rm I}  \, \widehat{T}
$, 
$
\widehat{T} \, \widehat{H}_{\rm I\!I}  = - \widehat{H}_{\rm I\!I}  \, \widehat{T}
$, 
and 
$\widehat{T} \, \widehat{H} = - \widehat{H} \, \widehat{T}$. 

The time-reversal antiunitary operator $\widehat{K}$ is 
defined as an antilinear operator which satisfies 
\begin{equation}
\widehat{K} | a \, j \, \xi \rangle 
=  
| a \, j \, \xi \rangle 
,  
\end{equation}
which is equivalent to 
\begin{equation}
\label{Kajn}
\widehat{K} | a \, j\, n \rangle 
=  
| a \, j \, n \rangle 
.
\end{equation}
Then, 
we have 
\begin{equation}
\widehat{K} | \psi \rangle 
=  
\int_{-\infty}^{+\infty} | a \, j \, \xi \rangle 
( \langle a \, j \, \xi | \psi \rangle )^*
=
\sum_{n=0}^\infty 
| a \, j \, n \rangle 
( \langle a \, j \, n | \psi \rangle )^*
,  
\end{equation}
for an arbitrary state vector $| \psi \rangle$. 
We also see from 
Eqs.~(\ref{vpXi}), (\ref{hatA}), (\ref{AII}), (\ref{AIAAA}), 
(\ref{HIpII})--(\ref{HII}), 
and (\ref{Kajn}) 
that $\widehat{K}$ satisfies 
$\widehat{K} = \widehat{K}^{-1} = \widehat{K}^\dagger$,  
$
\widehat{K} \, \widehat{v}_T \, \widehat{p}^{-1} \,  \widehat{\Xi}  
=  
\widehat{v}_T \, \widehat{p}^{-1} \,  \widehat{\Xi}  \, \widehat{K}
$,
$
\widehat{K} \, \widehat{A}_{\rm I} =  \widehat{A}_{\rm I}  \, \widehat{K}
$, 
$
\widehat{K} \, \widehat{A}_{\rm I\!I} =  \widehat{A}_{\rm I\!I}  \, \widehat{K}
$, 
$
\widehat{K} \, \widehat{A} =  \widehat{A} \, \widehat{K}
$,
$
\widehat{K} \, \widehat{H}_{\rm I}  = \widehat{H}_{\rm I}  \, \widehat{K}
$, 
$
\widehat{K} \, \widehat{H}_{\rm I\!I}  =  \widehat{H}_{\rm I\!I}  \, \widehat{K}
$, 
and 
$\widehat{K} \, \widehat{H} =  \widehat{H} \, \widehat{K}$.

\subsection{Time evolution of the state vector}
 
Using 
$
\widehat{H}_{\rm I}
\widehat{H}_{\rm I\!I}
=
\widehat{H}_{\rm I\!I}
\widehat{H}_{\rm I}
=
0
$, 
we can express 
the time evolution operator by the 
sum of operators of class ${\rm I}$ and class ${\rm I\!I}$ as 
\begin{eqnarray}
\widehat{U} (\tau) 
& \equiv  &  \exp ( -i \tau \widehat{H} ) 
 \equiv    \exp ( -i \tau  \widehat{H}_{\rm I}
 - i \tau \widehat{H}_{\rm I\!I} ) 
\nonumber \\
&  =  & 
\widehat{1} + 
\sum_{m = 1}^\infty 
\frac{(-i \tau )^m}{m!}
\big[ 
( \widehat{H}_{\rm I} )^m
+ 
( \widehat{H}_{\rm I\!I} )^m
\big]
\nonumber \\
&  \equiv & 
\widehat{U}_{\rm I} (\tau)
\oplus
\widehat{U}_{\rm I\!I} (\tau)
,
\end{eqnarray}
where 
\begin{eqnarray}
\widehat{U}_{\rm I} (\tau)
 & \equiv &
 \exp_{\rm I} ( -i \tau \widehat{H}_{\rm I} ) 
 \equiv  
\widehat{P}_{\rm I} + 
\sum_{m = 1}^\infty 
\frac{(-i \tau \widehat{H}_{\rm I} )^m}{m!}
 \nonumber \\ 
 & \equiv & 
  \sum_a  \sum_{a'}
 \sum_{j=0}^1 \sum_{j'=0}^1 
 \int_{-\infty}^{+\infty}
d\xi 
\int_{-\infty}^{+\infty}
d\xi' 
\, 
| a \, j \, \xi \rangle 
  \nonumber \\ 
 &  & \mbox{}
 \times
\langle a \, j \, \xi |
 \exp ( -i \tau \widehat{H}_{\rm I} ) 
| a' \, j' \, \xi' \rangle 
\langle a' \, j' \, \xi' |
 \nonumber \\ 
 & \equiv & 
  \sum_a  \sum_{a'}
 \sum_{j=0}^1 \sum_{j'=0}^1 
\sum_{n=0}^\infty
\sum_{n'=0}^\infty
| a \, j \, n \rangle 
  \nonumber \\ 
 &  & \mbox{}
 \times
\langle a \, j \, n |
 \exp ( -i \tau \widehat{H}_{\rm I} ) 
| a' \, j' \, n' \rangle 
\langle a' \, j' \, n' |
\end{eqnarray}
and 
\begin{eqnarray}
\label{UII}
&  & 
\hspace*{-6mm}
\widehat{U}_{\rm I\!I} (\tau) \equiv 
 \exp_{\rm I\!I} ( -i \tau \widehat{H}_{\rm I\!I} ) 
  \equiv  
\widehat{P}_{\rm I\!I} + 
\sum_{m = 1}^\infty 
\frac{(-i \tau \widehat{H}_{\rm I\!I} )^m}{m!}
 \nonumber \\
& & 
\hspace*{-5mm}
=
\sum_a 
\sum_{j=2}^\infty
\int_{-\infty}^{+\infty}
d\xi
\,
| a \, j \, \xi \rangle 
 \exp 
 \bigg(
 - i \tau \, \xi 
 \frac{v_{Ta}}{v_c}
\bigg)
\langle a \, j \, \xi |
.
\end{eqnarray}
Then, the state vector 
$| \psi (\tau) \rangle = \widehat{U} (\tau)
| \psi (0) \rangle$ 
is expressed by the direct sum of the vectors in ${\cal E}_{\rm I}$ and 
${\cal E}_{\rm I\!I}$ as 
\begin{eqnarray}
| \psi (\tau) \rangle
& = & 
| \psi_{\rm I} (\tau) \rangle
\oplus
| \psi_{\rm I\!I} (\tau) \rangle
\nonumber \\ 
& = & 
\widehat{U}_{\rm I} (\tau)
| \psi_{\rm I} (0) \rangle
\oplus
\widehat{U}_{\rm I\!I} (\tau)
| \psi_{\rm I\!I} (0) \rangle
,
\end{eqnarray}
where 
$| \psi_{\rm I} (\tau) \rangle = 
\widehat{P}_{\rm I} | \psi (\tau) \rangle
\in {\cal E}_{\rm I}$, 
$| \psi_{\rm I\!I} (\tau) \rangle = 
\widehat{P}_{\rm I\!I} | \psi (\tau) \rangle
\in {\cal E}_{\rm I\!I}$, 
$| \psi_{\rm I} (0) \rangle = 
\widehat{P}_{\rm I} | \psi (0) \rangle
\in {\cal E}_{\rm I}$, 
and
$| \psi_{\rm I\!I} (0) \rangle = 
\widehat{P}_{\rm I\!I} | \psi (0) \rangle
\in {\cal E}_{\rm I\!I}$. 

In the definition of $\widehat{H}_{\rm I}$ in Eq.~(\ref{HI}), 
the Hermitian operators, 
$\widehat{A}_{\rm I}$ and 
$\widehat{v}_T \widehat{p}^{- 1} \widehat{\Xi}$, defined in
Eqs.~(\ref{AIAAA}) and (\ref{vpXi}), respectively,  
can be expressed in matrix form using the basis vectors 
$\{ | a \, j \, n \rangle \}_{a = a_1, \cdots, a_{N_{ps}} ; j = 0, 1 ; n=0, 1, \cdots}$ 
in ${\cal E}_{\rm I}$. 
Note that the matrix components 
$\langle a \, j \, n | \,\widehat{v}_T \widehat{p}^{- 1} \widehat{\Xi} \, | a' \, j' \, n' \rangle$, 
with $j \geq 2$ or $j' \geq 2$ do not contribute to $\widehat{H}_{\rm I}$.
To represent $| \psi_{\rm I} (\tau) \rangle$, 
we truncate the basis
$\{| n\rangle \}$ in ${\cal E}_{v_\parallel}$
to a finite dimension $N_{v_\parallel}$.
The resulting truncated subspace of ${\cal E}_{\rm I}$
is spanned by the
$2N_{ps}N_{v_\parallel}$ basis vectors
$\{ | a \, j \, n \rangle \}_{ a = a_1, \cdots , a_{N_{ps}} ; j = 0, 1 ; n = 0, 1, \cdots, N_{v_\parallel}-1 }$. 
Then, 
$| \psi_{\rm I} (\tau) \rangle$ 
is represented by the column vector with 
$2 N_{ps} N_{v_\parallel}$ complex components, 
while  $\widehat{A}_{\rm I}$ and 
$\widehat{v}_T \widehat{p}^{- 1} \widehat{\Xi}$ 
are represented by 
$2 N_{ps} N_{v_\parallel} \times  2 N_{ps} N_{v_\parallel}$
Hermitian matrices.
Using these matrices to express Eq.~(\ref{HI}) in matrix form, 
we can obtain the Hamiltonian matrix 
${\bf H}_{\rm I}$ and, accordingly, 
the time-evolution matrix 
${\bf U}_{\rm I} (\tau) = \exp ( - i \tau {\bf H}_{\rm I} )$,  
which are $2 N_{ps} N_{v_\parallel} \times  2 N_{ps} N_{v_\parallel}$
Hermitian and unitary matrices, respectively. 

Using 
$| \psi_{\rm I\!I} (\tau) \rangle = 
\widehat{U}_{\rm I\!I} (\tau) | \psi_{\rm I\!I} (0) \rangle$ 
and Eq.~(\ref{UII}), 
we obtain 
\begin{equation}
\langle a \, X \, \xi  | \psi_{\rm I\!I} (\tau) \rangle
=
 \exp 
 \bigg(
 - i \tau \, \xi 
 \frac{v_{Ta}}{ v_c}
\bigg)
\langle  a \, X \, \xi | \psi_{\rm I\!I} (0) \rangle
,
\end{equation}
which corresponds to the so-called ballistic mode. 
Here, we have 
\begin{eqnarray}
\label{psiII}
\langle  a \, X \, \xi | \psi_{\rm I\!I} \rangle
& \equiv & 
\langle  a \, X \, \xi | \widehat{P}_{\rm I\!I} | \psi  \rangle
=
\langle  a \, X \, \xi | 
( \widehat{1} - \widehat{P}_{\rm I} ) 
| \psi  \rangle
\nonumber \\ 
&  =  &
\langle  a \, X \, \xi | \psi  \rangle
-
\sum_{j = 0}^1
\langle  a \, X |  a \, j \rangle
\langle  a \, j \, \xi |  \psi  \rangle
,
\hspace*{8mm}
\end{eqnarray}
which shows that 
$\langle  a \, X \, \xi | \psi_{\rm I\!I} \rangle$ 
can be evaluated from $\langle  a \, X \, \xi | \psi  \rangle$ 
by subtracting its $j = 0, 1$ components so that 
it is independent of the definitions of the basis vectors 
$\{ |  a \, j \rangle \}_{j \geq 2}$ in ${\cal E}^\times_{\rm I\!I}$.
The damping of the ballistic mode due to the phase mixing 
can be more clearly seen by using 
$\{ | n \rangle \}_{n=0, 1, 2, \cdots}$  instead of 
$\{ | \xi \rangle \}_{- \infty < \xi <  + \infty}$
as the basis vectors in ${\cal E}_{v_\parallel}$ 
to represent $| \psi_{\rm I\!I} (\tau) \rangle$ by 
\begin{equation}
\langle a \, X \, n  | \psi_{\rm I\!I} (\tau) \rangle
=
\sum_{n' = 0}^\infty
\alpha (a, n, n', \tau)
\langle  a \, X \, n' | \psi_{\rm I\!I} (0) \rangle
,
\end{equation}
where 
\begin{eqnarray}
\label{alpha}
& &
\hspace*{-2mm}
\alpha (a, n, n', \tau)
=
\int_{-\infty}^{+\infty} d\xi \, 
\langle  n | \xi \rangle 
 \exp 
 \bigg(
 - i \tau \, \xi 
 \frac{v_{Ta}}{ v_c}
\bigg)
\langle  \xi | n'  \rangle 
\nonumber \\  
&  & 
=
\frac{1}{
( 2^{n+n'} \, n! \, n' ! \, \pi )^{1/2}
}
\int_{-\infty}^{+\infty} d\xi \,
H_n (\xi) H_{n'} (\xi ) 
\nonumber \\ & & 
\hspace*{5mm}
\mbox{} \times 
 \exp 
 \bigg( - \xi^2
 - i \tau \, \xi 
 \frac{v_{Ta}}{ v_c}
\bigg)
\nonumber \\  
&  & 
=
\frac{1}{
( 2^{n+n'} \, n! \, n' ! \, \pi )^{1/2}
}
e^{
-
( v_{Ta}^2/4 v_c^2 )
\tau^2
}
\int_{-\infty}^{+\infty} d\eta \,
e^{- \eta^2} 
\nonumber \\ & & 
\hspace*{5mm}
\mbox{} \times 
H_n 
\bigg( \eta
 - i \tau \, 
 \frac{v_{Ta}}{2 v_c}
\bigg)
H_{n'}
\bigg( \eta
 - i \tau \, 
 \frac{v_{Ta}}{2 v_c}
\bigg)
\nonumber \\ 
& & 
=
\bigg(
\frac{n! \, n'!}{2^{n+n'} }
\bigg)^{1/2}
e^{
-
( v_{Ta}^2/4 v_c^2 )
\tau^2
}
\sum_{k=0}^{\min(n,n')}
\frac{2^k}
{k!(n-k)!(n'-k)!}
\nonumber \\ & & 
\hspace*{5mm}
\mbox{} \times 
\bigg(
- i \tau \, 
 \frac{v_{Ta}}{v_c}
\bigg)^{n+n'-2k}
.
\end{eqnarray}
Thus, 
$\langle a \, X \, n  | \psi_{\rm I\!I} (\tau) \rangle$ 
damps rapidly due to 
the exponential factor 
$
 \exp 
[
 - 
 (v_{Ta}^2 / 4 v_c^2)
\tau^2
]
=
 \exp 
[
 - 
 (k_\parallel^2 v_{Ta}^2 / 4 )
t^2
]
$
appearing in Eq.~(\ref{alpha}). 

We note that the effects of the electromagnetic fields self-consistently produced by charged particles are included in $|\psi_{\rm I}(\tau)\rangle$, but not in the ballistic-mode component $|\psi_{\rm I\!I}(\tau)\rangle$.

\section{Conclusions}

This study presents a framework for applying the fluctuation theorem and the detailed fluctuation theorem to classical systems 
whose 
dynamics are governed by Schr\"{o}dinger-type equations and which 
possess either a unitary or an antiunitary time-reversal operator.
To apply these theorems, the initial state vector is treated as a random variable obeying a specified time-reversal-symmetric probability distribution, and the stochastic relative entropy is defined from the probability distribution of the random state vector.
The linear Vlasov-Poisson and linear gyrokinetic systems are investigated as examples of collisionless plasmas to which the fluctuation theorems can be applied.
In both systems, an invariant proportional to the squared perturbed distribution function exists, which serves as the basis for defining the state vector satisfying the Schr\"{o}dinger equation.
Unitary and antiunitary time-reversal operators are also identified for these systems.

In the linear Vlasov-Poisson system, the Hamiltonian eigenvectors correspond to the Case--Van Kampen (CVK) modes, and a discrete subset of these eigenvectors is used to construct the solutions for which the fluctuation theorem is formulated.
In this example, the stochastic relative entropy is interpreted as the entropy generated through Landau damping, which transfers the electric-field energy of the $n=0$ Hermite state to thermal reservoirs consisting of the $n \geq 1$ Hermite states.
In addition, a novel analytical expression for the probability density function of the stochastic relative entropy is derived under specific conditions, and its validity is verified numerically.

The governing equations of the linear gyrokinetic system are also transformed into Schr\"{o}dinger form, and the corresponding unitary and antiunitary time-reversal operators are defined.
Thus, the fluctuation theorems can also be applied to this system.
The tensor product of the three vector spaces representing particle species, perpendicular velocity space, and parallel velocity space is employed to define the state vectors.
The state vectors, Hamiltonian, and time-evolution operator decompose into two classes.
The first class is associated with perpendicular velocity-space structures generated by the zeroth- and first-order Bessel functions, whereas the second is orthogonal to the first.
An analytical expression is obtained for the solution in the second class, which represents a ballistic mode that decays rapidly and independently of electromagnetic fluctuations.
We also show how the solution in the first class can be represented using basis vectors in the tensor-product state-vector space.

The present study contributes to nonequilibrium statistical-mechanical formulations of collisionless plasma phenomena and to future applications of quantum-computing algorithms to plasma simulations.
Numerical investigations of the present framework for the linear gyrokinetic system, along with extensions to other systems, remain topics for future study.
\\[3mm]
{\bf ACKNOWLEDGMENTS}
%\begin{acknowledgments}

This work was supported in part by the JSPS Grants-in-Aid for Scientific Research (Grant No.\ 24K07000) 
and in part by the NINS program of Promoting Research by Networking among Institutions (Grant No.\ 01422301).
%\end{acknowledgments}
\\[3mm]

\section*{AUTHOR DECLARATIONS}

\subsection*{Conflict of Interest}

The author has no conflicts of interest to disclose.

\subsection*{Author Contributions}
\noindent
{\bf Hideo Sugama}: Conceptualization (lead); Data curation (lead); Formal analysis (lead); Funding acquisition (lead); Writing – original draft (lead).

\section*{DATA AVAILABILITY}
%Data sharing is not applicable to this article as no new data were created or analyzed in this study.

The data that support the findings of this study are available from
the corresponding author upon reasonable request.

\appendix

\section{Unitary and Antiunitary Time-Reversal Operators} 

We consider unitary and antiunitary time-reversal operators, 
which are denoted by $\widehat{T}$ and $\widehat{K}$, respectively. 
Here, $\widehat{T}$ and $\widehat{K}$ are the linear and antilinear 
operators~\cite{QM} 
which satisfy the following conditions
\begin{equation}
\label{A1}
\widehat{T} ( c_1 |\psi_1 \rangle + c_2 |\psi_2 \rangle )
=
 c_1 ( \widehat{T} |\psi_1 \rangle ) + c_2 ( \widehat{T} |\psi_2 \rangle )
 , 
\end{equation}
and
\begin{equation}
\label{A2}
\widehat{K} ( c_1 |\psi_1 \rangle + c_2 |\psi_2 \rangle )
=
 c_1^* ( \widehat{K} |\psi_1 \rangle ) + c_2^*  (\widehat{K} |\psi_2 \rangle )
 , 
\end{equation}
respectively, 
for arbitrary state vectors $|\psi_i \rangle$ $(i=1, 2)$ 
and arbitrary complex-valued coefficients $c_i$ $(i =1, 2)$. 
They also satisfy 
\begin{equation}
\label{TK}
\widehat{T} 
= \widehat{T}^{-1}
=  \widehat{T}^\dagger
\hspace*{3mm}
\widehat{K} 
= \widehat{K}^{-1}
=  \widehat{K}^\dagger
.
\end{equation}
In quantum mechanics,~\cite{QM} 
the antiunitary operator $\widehat{K}$ is generally employed as the time-reversal operator rather than the unitary operator $\widehat{T}$, and 
$\widehat{K}^{-1} = \widehat{K}^\dagger$ is equal to $-\widehat{K}$, 
not to $\widehat{K}$ for quantum mechanical systems including containing an odd number of spin-$1/2$ particles. 
However, we find that, at least in classical systems described by the Schr\"{o}dinger equation in the present work, 
there exist the unitary operator $\widehat{T}$ and the antiunitary operator $\widehat{K}$ 
which satisfy the conditions given in this appendix. 
In addition to the conditions in Eq.~(\ref{TK}), 
$\widehat{T}$ and 
$\widehat{K}$ are required to satisfy the anticommutation and 
commutation relations with the Hamiltonian operator $\widehat{H}$, 
\begin{equation}
\label{THK}
 \widehat{T} \widehat{H}
= 
-  \widehat{H} \widehat{T}
\hspace*{3mm}
\mbox{and}
\hspace*{3mm}
 \widehat{K} \widehat{H}
= 
 \widehat{H} \widehat{K}
 ,
\end{equation}
respectively. 
Noting that the unitary and antiunitary operators $\widehat{T}$ and $\widehat{K}$ 
satisfy
$\widehat{T} i =  i \widehat{T}$ and 
$\widehat{K} i = - i \widehat{K}$, respectively, 
we find that the conditions in Eq.~(\ref{THK}) are expressed as 
the anticommutation relations of $\widehat{T}$ and 
$\widehat{K}$ to the anti-Hermitian operator $i \widehat{H}$, 
\begin{equation}
 \widehat{T}  i \widehat{H}
= 
-  i \widehat{H} \widehat{T}
\hspace*{3mm}
\mbox{and}
\hspace*{3mm}
 \widehat{K} i \widehat{H}
= 
-i \widehat{H} \widehat{K}
, 
\end{equation}
from which we obtain 
\begin{equation}
 \widehat{T} \widehat{U} (\tau) 
 =
\widehat{U} (- \tau)  \widehat{T} 
\hspace*{3mm}
\mbox{and}
\hspace*{3mm}
 \widehat{K} \widehat{U} (\tau) 
 =
\widehat{U} (- \tau)  \widehat{K} 
, 
\end{equation}
where 
$
\widehat{U} (\tau) \equiv \exp ( - i \tau \widehat{H} )
$. 
Therefore, 
when $| \psi (\tau) \rangle = \widehat{U} (\tau) | \psi (0) \rangle$ is a solution to 
the Schr\"{o}dinger equation and there exists 
$\widehat{T}$ or $\widehat{K}$ satisfying 
the above-mentioned conditions, 
$ \widehat{T} | \psi (- \tau) \rangle = \widehat{U} (\tau) \widehat{T} | \psi (0) \rangle$ 
or $\widehat{K} | \psi (- \tau) \rangle = \widehat{U} (\tau) \widehat{K} | \psi (0) \rangle$ 
becomes a solution as well. 

Here, 
we assume that the basis vectors of the state vector space are given by 
$\{ | n  \rangle \}_{n = 0, 1, 2, \cdots, N_c -1}$. 
Then, the state vector $| \psi \rangle$ can be represented by 
the $N_c$-dimensional column vector 
$\boldsymbol{\psi} \equiv \mbox{}^t [ \psi_0, \psi_1, \cdots , \psi_{N_c-1} ] $
where $\psi_n \equiv \langle n | \psi \rangle$ 
$( n = 0, 1, \cdots, N_c -1)$. 
In addition, the $N_c \times N_c$ unitary matrices 
${\bf T} \equiv [ T_{n n'} ] $ and ${\bf K} \equiv [ K_{n n'} ]$ are defined 
from  $\widehat{T}$ and $\widehat{K}$ by 
$T_{n n'} \equiv \langle n | \widehat{T} | n' \rangle$ 
and 
$K_{n n'} \equiv \langle n | ( \widehat{K} | n' \rangle ) \equiv 
((\langle n | \widehat{K}) | n' \rangle )^*$, respectively, 
where the parentheses are required to specify the order of action of the antiunitary operator 
$\widehat{K}$ (see Ref.~\cite{QM}).
We also define the volume element in the state vector space by 
$
d \Gamma [\boldsymbol{\psi}] 
\equiv
\prod_{n = 0}^{N_c-1}
d\psi_{r,n} d\psi_{i,n}
$, 
where 
$\psi_{r,n} \equiv {\rm Re} \, \psi_n$ 
and 
$\psi_{i,n} \equiv {\rm Im} \, \psi_n$. 
Since unitary transformations and complex conjugation 
preserve the volume element, we obtain
\begin{equation}
\label{veconv}
d\Gamma [{\bf T} \boldsymbol{\psi}]
= 
d\Gamma [ \boldsymbol{\psi} ]
,
\hspace*{3mm}
d\Gamma [{\bf K} \boldsymbol{\psi}^*]
= 
d\Gamma [ \boldsymbol{\psi} ]
.
\end{equation}

\section{Proof of Fluctuation Theorem}

Using Eq.~(\ref{srentropy}), we have  
$
 P[ \boldsymbol{\psi}(0) ; 0 ] 
 = 
  P[ \boldsymbol{\psi}(\tau) ; 0 ] 
 \exp \Delta S[ \boldsymbol{\psi}(0); \tau] 
 ,
$
and substitute it into Eq.~(\ref{PDSdef}) to 
obtain 
\begin{eqnarray}
\label{B1}
P(\Delta S)
& = & 
\exp \Delta S
\int 
d \Gamma (\tau) \, 
 P[ \boldsymbol{\psi}(\tau) ; 0 ] 
 \nonumber \\ 
 & & 
 \mbox{}
 \times
   \delta [ \Delta S [ \boldsymbol{\psi}(0) ; \tau ] -  \Delta S  ]
,
\end{eqnarray}
where Eq.~(\ref{dGdG}) is also used. 
Here, from Eqs.~(\ref{TUUT}) and  (\ref{PT0}), 
$\Delta S [ \boldsymbol{\psi}(0) ; \tau ]$ defined in Eq.~(\ref{srentropy}) 
is expressed as 
\begin{eqnarray}
\label{B2}
\Delta S [ \boldsymbol{\psi}(0) ; \tau ]
& = &
- \log 
\left[ 
\frac{ 
P[\boldsymbol{\psi}(\tau), 0]
}{
P[\boldsymbol{\psi}(0), 0]
}
\right]
\nonumber \\
& = & 
- \log 
\left[ 
\frac{ 
P[{\bf T} \boldsymbol{\psi}(\tau), 0]
}{
P[{\bf T} {\bf U}(-\tau) \boldsymbol{\psi}(\tau), 0]
}
\right]
\nonumber \\
& = & 
- \log 
\left[ 
\frac{ 
P[{\bf T} \boldsymbol{\psi}(\tau), 0]
}{
P[{\bf U}(\tau) {\bf T}  \boldsymbol{\psi}(\tau), 0]
}
\right]
\nonumber \\
& = & 
- \Delta S [ {\bf T} \boldsymbol{\psi}(\tau) ; \tau ]
.
\end{eqnarray}
Substituting Eq.~(\ref{B2}) into Eq.~(\ref{B1}) and 
using Eq.~(\ref{PT0}) and 
$d  \Gamma (\tau) \equiv d \Gamma [\boldsymbol{\psi}(\tau)]
=  d \Gamma [{\bf T} \boldsymbol{\psi}(\tau)]$
[see Eq.~(\ref{veconv})], 
we obtain
\begin{eqnarray}
P(\Delta S)
& = & 
\exp \Delta S
\int 
d  \Gamma [{\bf T} \boldsymbol{\psi}(\tau)] \, 
 P[ {\bf T} \boldsymbol{\psi}(\tau) ; 0 ] 
 \nonumber \\ 
 & & 
 \mbox{}
 \times
   \delta [ \Delta S [  {\bf T} \boldsymbol{\psi}(\tau) ; \tau ] + \Delta S  ]
    \nonumber \\ 
 & =  & 
\exp \Delta S \cdot P(- \Delta S)
,
\end{eqnarray}
which gives the fluctuation theorem 
shown in Eq.~(\ref{FT}). 

Replacing
${\bf T} \boldsymbol{\psi}(\tau)$ 
in the procedures shown above 
with 
$\widehat{K} [\boldsymbol{\psi}(\tau)] \equiv {\bf K} \boldsymbol{\psi}^*(\tau)$, 
we can derive Eq.~(\ref{FT}) under the conditions 
given by Eqs.~(\ref{KUUK}) and (\ref{PK0}).

\section{Proof of the Detailed Fluctuation Theorem}

From Eqs.~(\ref{zzetayeta}) and (\ref{DSYdef}), we have 
\begin{eqnarray}
\label{C1}
P_Y [ \boldsymbol{\eta} [ 0 ; \boldsymbol{\psi}(0)]  ;  0]
& = & 
P_Y [ \boldsymbol{\eta} [ 0 ; \boldsymbol{\psi}(\tau)]  ;  0]
\nonumber \\ & & \mbox{}
\times
\exp 
\Delta S_Y [ \boldsymbol{\zeta} [ 0 ; \boldsymbol{\psi}(0)] , \boldsymbol{\eta} [ 0 ; \boldsymbol{\psi}(0)]   ; \tau ]
.
\nonumber \\ & &
\end{eqnarray}
Substituting Eq.~(\ref{C1})  into Eq.~(\ref{PBDSA}) and using Eq.~(\ref{PYTY0}), 
we obtain 
\begin{eqnarray}
\label{C2}
& & 
\hspace*{-5mm}
P({\bf z}_B, \Delta S' | {\bf z}_A)
\nonumber \\ 
& = & 
\exp \Delta S'
\int 
d \Gamma [{\bf T} \boldsymbol{\psi}(\tau)] 
\;    P_Y [ {\bf T}_y \boldsymbol{\eta}   [  0 ; \boldsymbol{\psi} (\tau) ]  ; 0 ]
\nonumber \\ 
& & 
\mbox{}
\times
\delta^{2N_z}
( \boldsymbol{\zeta}   [  0 ; \boldsymbol{\psi} (0) ]  - {\bf z}_A) 
\;
\delta^{2N_z}
( \boldsymbol{\zeta}   [  \tau ; \boldsymbol{\psi} (0) ] - {\bf z}_B) 
\nonumber \\ & & 
\mbox{} \times 
   \delta \bigl(
   \Delta S_Y [ \boldsymbol{\zeta}   [  0 ; \boldsymbol{\psi} (0) ] , 
   \boldsymbol{\eta}   [  0 ; \boldsymbol{\psi} (0) ]  ; \tau ]
    -  \Delta S' \bigr)
.
\end{eqnarray}
where
\begin{equation}
d \Gamma [\boldsymbol{\psi}(0)] 
=  
d \Gamma [\boldsymbol{\psi}(\tau)] 
= 
d \Gamma [{\bf T} \boldsymbol{\psi}(\tau)] 
\end{equation}
is used. 
From Eqs.~(\ref{zzetayeta}) and (\ref{TzTy}), 
we also have 
\begin{eqnarray}
\label{C5}
{\bf T}_z {\bf z} (\tau) 
& = & 
{\bf T}_z \boldsymbol{\zeta} [ 0 ; \boldsymbol{\psi} (\tau) ]
=
\boldsymbol{\zeta} [ 0;  {\bf T} \boldsymbol{\psi} (\tau) ]
, 
\nonumber \\ 
{\bf T}_y {\bf y} (\tau) 
& = & 
{\bf T}_y \boldsymbol{\eta} [ 0 ; \boldsymbol{\psi} (\tau) ]
=
\boldsymbol{\eta} [ 0;  {\bf T} \boldsymbol{\psi} (\tau) ]
.
\end{eqnarray}
Using Eqs.~(\ref{TUUT}) and (\ref{TzTy}) yields 
\begin{eqnarray}
\label{C6}
\left[
\begin{array}{c}
{\bf T}_z {\bf z} (0) 
\\
{\bf T}_y {\bf y} (0) 
\end{array}
\right]
& = & 
{\bf T} \boldsymbol{\psi} (0)
=
{\bf T}  {\bf U} ( - \tau)
\boldsymbol{\psi} (\tau)
\nonumber \\ 
& = &
 {\bf U} (\tau) {\bf T}
\boldsymbol{\psi} (\tau)
,
\end{eqnarray}
from which we obtain 
\begin{eqnarray}
\label{C7}
{\bf T}_z {\bf z} (0) 
& = & 
{\bf T}_z \boldsymbol{\zeta} [ 0; \boldsymbol{\psi} (0) ]
=
\boldsymbol{\zeta} [ \tau;  {\bf T} \boldsymbol{\psi} (\tau) ]
,
\nonumber \\ 
{\bf T}_y {\bf y} (0) 
& = & 
{\bf T}_y \boldsymbol{\eta} [ 0; \boldsymbol{\psi} (0) ]
=
\boldsymbol{\eta} [ \tau;  {\bf T} \boldsymbol{\psi} (\tau) ]
.
\end{eqnarray}
Then, from Eqs.~(\ref{C5}) and (\ref{C7}), we can derive   
\begin{eqnarray}
\label{C8}
\delta^{2N_z}
( \boldsymbol{\zeta}   [  0 ; \boldsymbol{\psi} (0) ]  - {\bf z}_A) 
& = & 
\delta^{2N_z}
( {\bf T}_z \boldsymbol{\zeta}   [  0 ; \boldsymbol{\psi} (0) ]  - {\bf T}_z  {\bf z}_A) 
\nonumber \\ 
& = & 
\delta^{2N_z}
( \boldsymbol{\zeta}   [  \tau ;  {\bf T} \boldsymbol{\psi} (\tau) ]  - {\bf T}_z {\bf z}_A) 
,
\nonumber \\ 
\delta^{2N_z}
( \boldsymbol{\zeta}   [  \tau ; \boldsymbol{\psi} (0) ] - {\bf z}_B) 
& = & 
\delta^{2N_z}
( \boldsymbol{\zeta}   [  0 ; \boldsymbol{\psi} (\tau) ] - {\bf z}_B) 
\nonumber \\ 
& = & 
\delta^{2N_z}
( {\bf T}_z \boldsymbol{\zeta}   [  0 ; \boldsymbol{\psi} (\tau) ] - {\bf T}_z {\bf z}_B) 
\nonumber \\ 
& = & 
\delta^{2N_z}
( \boldsymbol{\zeta}   [ 0 ; {\bf T} \boldsymbol{\psi} (\tau) ] - {\bf T}_z {\bf z}_B) 
.
\nonumber \\ & & 
\end{eqnarray}
Using Eqs.~(\ref{DSYdef}), (\ref{PYTY0}), (\ref{C5}), and (\ref{C7}), we have
\begin{eqnarray}
\label{C9}
& & 
 \Delta S_Y [ \boldsymbol{\zeta} [0; \boldsymbol{\psi}(0); 0] , 
   \boldsymbol{\eta} [0; \boldsymbol{\psi}(0); 0]  ; \tau ]
\nonumber \\ 
& & 
=
- \log 
\left[
\frac{
P_Y [ {\bf T}_y  \boldsymbol{\eta} [0; \boldsymbol{\psi}(\tau)]  ; 0 ] 
}{
P_Y [ {\bf T}_y   \boldsymbol{\eta} [0; \boldsymbol{\psi}(0)]; 0 ] 
}
\right]
\nonumber \\
& & 
=
- \log 
\left[
\frac{
P_Y [  \boldsymbol{\eta} [0; {\bf T} \boldsymbol{\psi}(\tau)] ; 0
}{
P_Y [  \boldsymbol{\eta} [\tau; {\bf T} \boldsymbol{\psi}(\tau)]  ; 0 ] 
}
\right]
\nonumber \\
& & 
=
- 
 \Delta S_Y [ \boldsymbol{\zeta} [0; {\bf T} \boldsymbol{\psi}(\tau); 0] , 
   \boldsymbol{\eta} [0; {\bf T} \boldsymbol{\psi}(\tau) ; \tau ]
. 
\end{eqnarray}
Then, substituting Eqs.~(\ref{C5}), (\ref{C8}), and (\ref{C9}) into Eq.~(\ref{C2}) 
leads to 
\begin{eqnarray}
\label{C10}
& & 
\hspace*{-5mm}
P({\bf z}_B, \Delta S' | {\bf z}_A)
\nonumber \\ 
& = & 
\exp \Delta S'
\int 
d \Gamma [{\bf T} \boldsymbol{\psi}(\tau)] 
\;    P_Y [ \boldsymbol{\eta}   [  0 ; {\bf T} \boldsymbol{\psi}(\tau) 
 ]  ; 0 ]
\nonumber \\ 
& & 
\mbox{}
\times
\delta^{2N_z}
( \boldsymbol{\zeta}   [  \tau ;  {\bf T} \boldsymbol{\psi} (\tau) ]  - {\bf T}  {\bf z}_A) 
\;
\delta^{2N_z}
( \boldsymbol{\zeta}   [ 0 ; {\bf T} \boldsymbol{\psi} (\tau) ] - {\bf T}_z {\bf z}_B) 
\nonumber \\ & & 
\mbox{} \times 
   \delta \bigl(
 \Delta S_Y [ \boldsymbol{\zeta} [0; {\bf T} \boldsymbol{\psi}(\tau); 0] , 
   \boldsymbol{\eta} [0; {\bf T} \boldsymbol{\psi}(\tau) ; \tau ]
    + \Delta S' \bigr)
.
\end{eqnarray}
Note that the integral in Eq.~(\ref{C10}) is obtained from the last integral in 
Eq.~(\ref{PBDSA}) by  
replacing the integration variable 
$\boldsymbol{\psi} (0)$, ${\bf z}_A$, ${\bf z}_B$, and $\Delta S'$  
with ${\bf T} \boldsymbol{\psi} (\tau)$,  ${\bf T}_z {\bf z}_B$, ${\bf T} {\bf z}_A$, and 
$- \Delta S'$, 
respectively. 
Thus, Eq.~(\ref{C10}) is rewritten as 
\begin{equation}
P({\bf z}_B, \Delta S' | {\bf z}_A)
=
\exp \Delta S' \; 
P({\bf T}_z {\bf z}_A, - \Delta S' | {\bf T}_z {\bf z}_B)
, 
\end{equation}
which completes the proof of the detailed fluctuation theorem, Eq.~(\ref{DFT}). 
The detailed fluctuation theorem can also be proved by replacing
${\bf T}(\cdot)$ with ${\bf K}(\cdot)^*$
throughout the above derivation.

\section{Relation between $\{ N \}$ and $\{ \rm{CVK} \}$ representations} 

The orthonormal basis vectors of the $\{ N \}$ and $\{ \rm{CVK} \}$ 
representations are related to each other by 
\begin{eqnarray}
& & 
| n \rangle = \int_{-\infty}^{+ \infty} 
| {\rm CVK}, \zeta \rangle d \zeta 
\langle {\rm CVK}, \zeta | n \rangle, 
\nonumber 
\\ & & 
| {\rm CVK}, \zeta \rangle 
= \sum_{n=0}^\infty
| n \rangle \langle n 
| {\rm CVK}, \zeta \rangle
.
\end{eqnarray}
Here, 
$\langle n | {\rm CVK}, \zeta \rangle
= \langle {\rm CVK}, \zeta | n \rangle^* $  
is regarded as the $(n, \zeta)$ component of the 
unitary matrix $[ \langle n | {\rm CVK}, \zeta \rangle ]$ 
which gives the transformation between the $\{ N \}$ and CVK representations. 
The $(n, \zeta)$ component of the unitary matrix is given by 
\begin{equation}
\label{D2}
\langle n | {\rm CVK}, \zeta \rangle
\equiv 
\frac{h_0(\zeta)} {|\epsilon(\zeta)|}
\langle n | \psi_{{\rm CVK}, \zeta} \rangle
, 
\end{equation}
where 
\begin{eqnarray}
\langle n | \psi_{{\rm CVK}, \zeta} \rangle 
& = & 
(1+ \kappa^{-2} \delta_{n0} )^{1/2}
\biggl[
\frac{h_n (\zeta)}{h_0 (\zeta)}
{\rm Re}[\epsilon (\zeta) ]
\nonumber 
\\ & & \mbox{}
- \int_{-\infty}^{+ \infty} d \xi \, 
\frac{1}{\pi} P
\left( \frac{1}{\xi - \zeta} \right)
\frac{h_n (\xi )}{h_0 (\xi)} {\rm Im}[\epsilon (\xi) ]
\biggr]
, 
\nonumber \\ & &
\end{eqnarray}
is written as a polynomial of order $n$ in $\zeta$. 
It is also found that $\langle n | \psi_{{\rm CVK}, \zeta} \rangle$ and 
$\langle n | {\rm CVK}, \zeta \rangle$ are even (odd) functions of $\zeta$ 
when $n$ is an even (odd) number. 
Here,  $| \psi_{{\rm CVK}, \zeta} \rangle$  is the eigenvector of the 
Hamiltonian $\widehat{H}$, 
and the eigenvector equation in Eq.~(\ref{Heveq}) 
is rewritten as 
\begin{equation}
\label{Heveq2}
\sum_{n' = 0}^\infty 
H_{n n'}
\langle n' | \psi_{{\rm CVK}, \zeta} \rangle
=
\zeta \, 
\langle n | \psi_{{\rm CVK}, \zeta} \rangle
,
\end{equation}
where $H_{n n'} \equiv \langle n | \widehat{H} | n' \rangle$ is given
by Eq.~(\ref{Hnn}). 
Using Eqs.~(\ref{Hnn}) and (\ref{Heveq2}), 
we obtain the recurrence relation, 
\begin{eqnarray*}
& & 
\langle n + 1 | \psi_{{\rm CVK}, \zeta} \rangle
\\ & & 
=
\sqrt{\frac{2}{n+1}} \zeta \, 
\langle n | \psi_{{\rm CVK}, \zeta} \rangle
-
\sqrt{\frac{n+ \kappa^{-2} \delta_{n1}}{n+1}}
\langle n - 1 | \psi_{{\rm CVK}, \zeta} \rangle
\\ & & 
\hspace*{60mm}
(n =1, 2, 3, \cdots)
\end{eqnarray*}
We can also express $\langle n | \psi_{{\rm CVK}, \zeta} \rangle$ 
using the Hermite polynomials as 
\begin{eqnarray}
\label{npsiH}
\langle n | \psi_{{\rm CVK}, \zeta} \rangle
 & = & 
\sqrt{ \frac{1+ \kappa^{-2} \delta_{n 0}}{2^n n!}  }
\biggl[ 
H_n (\zeta)
+
\kappa^{-2}
\sum_{k = 1}^{\lfloor n/2 \rfloor}
(-1)^k 2^k 
\nonumber 
\\ & & \mbox{} 
\hspace*{5mm} 
\times
\mbox{}_{n-k-1}P_{k-1} H_{n-2k} (\zeta)
\biggr]
\end{eqnarray}
where 
$\lfloor x \rfloor$ denotes the floor function of $x$ that is defined as 
the greatest integer less than or equal to $x$, and 
$
\mbox{}_n P_k = n! / (n-k)!
$
represents the number of $k$-permutations of $n$. 
For $n=0, 1, 2$, Eq.~(\ref{npsiH}) gives
\begin{eqnarray}
\langle 0 | \psi_{{\rm CVK}, \zeta} \rangle
 & = & 
 \sqrt{1+ \kappa^{-2} }
 ,
 \nonumber 
\\ 
\langle 1 | \psi_{{\rm CVK}, \zeta} \rangle
& = &
 \frac{1}{\sqrt{2}} H_1 (\zeta) 
 =
 \sqrt{2} \zeta
  ,
 \nonumber 
\\
\langle 2 | \psi_{{\rm CVK}, \zeta} \rangle
 & = &
 \frac{1}{2\sqrt{2}} \bigl( H_2 (\zeta) 
 - 2 \kappa^{-2} \bigr)
=
 \sqrt{2} 
 \Bigl(
 \zeta^2 - \frac{1+ \kappa^{2}}{2}
 \Bigr)
 .
 \nonumber 
 \\ & & \mbox{} 
\end{eqnarray}
It is also shown that, for $N \geq 1$, 
\begin{equation}
\langle N | \psi_{{\rm CVK}, \zeta} \rangle
= 
\sqrt{\frac{2^N}{N!}}
\det 
\bigl[
\zeta\, \delta_{n n'} -
 \langle n | \widehat{H} | n' \rangle 
 \bigr]_{n, n'=0, 1, \cdots, N-1}
 .
\end{equation}

We now consider the time evolution of the state vector 
$ | \psi  (\tau) \rangle $. 
Since
$\langle {\rm CVK}, \zeta | \psi  (\tau) \rangle 
= 
\langle {\rm CVK}, \zeta |  \exp (- i \tau \widehat{H} ) | \psi  (0) \rangle
= e^{-i \zeta \tau} \langle {\rm CVK}, \zeta |  \psi  (0) \rangle$
holds, 
there exist an infinite number of invariants given by 
\begin{eqnarray}
& & 
\hspace*{-3mm}
C(\{ \psi_n  (\tau) \}_{n=0, 1, 2, \cdots} ; \zeta) 
 \equiv 
 \Bigl|
\sum_{n=0}^\infty
\langle {\rm CVK}, \zeta | n \rangle 
\psi_n  (\tau) 
\Bigr|^2
\nonumber 
\\ &  & 
=
| \langle {\rm CVK}, \zeta | \psi  (\tau) \rangle |^2
\hspace*{3mm}
( -\infty < \zeta < + \infty )
, 
\end{eqnarray}
where 
$\psi_n (\tau)\equiv \langle n | \psi (\tau) \rangle$. 
Therefore, an arbitrary functional 
$F[C]$ of the invariants 
$C(\{ \psi_n  (\tau) \}_{n=0, 1, 2, \cdots} ; \zeta)$ 
$( -\infty < \zeta < + \infty )$ is an invariant as well. 
Then, we can consider that the probability density functional of 
$\{ \psi_n  (\tau) \}_{n=0, 1, 2, \cdots}$ in the statistically steady state 
takes the form of $F[C]$. 
The squared norm $\langle \psi  (\tau) | \psi  (\tau) \rangle$ is such an 
invariant as given by a functional of 
$C(\{ \psi_n  (\tau) \}_{n=0, 1, 2, \cdots} ; \zeta)$ $( -\infty < \zeta < + \infty )$
because it is written as 
\begin{eqnarray}
\langle \psi  (\tau) | \psi  (\tau) \rangle
& =  & 
\sum_{n=0}^\infty
| \psi_n (\tau) |^2
\nonumber \\ 
& = & 
 \int_{-\infty}^{+\infty} d \zeta \;
\Bigl|
\sum_{n=0}^\infty
\langle {\rm CVK}, \zeta | n \rangle 
\psi_n  (\tau) 
\Bigr|^2
\nonumber \\ 
& = & 
\int_{-\infty}^{+\infty} d \zeta \;
C(\{ \psi_n  (\tau) \}_{n=0, 1, 2, \cdots} ; \zeta)
.
\hspace*{5mm}
\end{eqnarray}

\section{Derivation of Eq.~(\ref{PDSaprox})}

Derivation of Eq.~(\ref{PDSaprox}) is shown in this appendix. 
It is recalled that the initial state-vector distribution 
$P[\boldsymbol{\psi} (0); 0]$ is given by Eq.~(\ref{initialP}) 
under the assumption that $\beta_n = \beta_0 / \rho$ 
for $n = 1, 2, \cdots, N_{\rm cvk}-1$. 
Then, using  
$d [ || \boldsymbol{\Psi}(\tau) ||^2 ] / d \tau = 
\sum_{n=0}^{N_{\rm cvk}-1} d [ | \Psi_n( \tau) |^2 ] / d \tau
=0$, 
the stochastic relative entropy 
$\Delta S [ \boldsymbol{\Psi}(0) ; \tau ]$ 
defined in Eq.~(\ref{srentropy}) is expressed as  
\begin{eqnarray}
\label{DSPsi1}
& & 
\hspace*{-6mm}
\Delta S [ \boldsymbol{\Psi}(0) ; \tau ]
=
\log
\left[
\frac{
P [ \boldsymbol{\Psi}(0) ; 0 ] 
}{
P [ \boldsymbol{\Psi}(\tau) ; 0 ] 
}
\right]
\nonumber \\
&  & 
=
\sum_{n=0}^{N_{\rm cvk} - 1}
\beta_n
\big( | \Psi_n (\tau) |^2 - | \Psi_n (0) |^2 \big)
\nonumber \\
& & 
=
\beta_0 ( \rho^{-1}  - 1 )
\big( | \Psi_0 (0) |^2 - | \Psi_0 (\tau) |^2 \big)
, 
\end{eqnarray}
where $\Psi_0 (\tau)$ is given by  
\begin{equation}
\label{Psi0}
\Psi_0(\tau) = U_{00}(\tau) \Psi_0(0)
+ \sum_{n=1}^{N_{\rm cvk} - 1}
U_{0n}(\tau) \Psi_n(0)
, 
\end{equation}
and 
$U_{00}(\tau)$ is a real-valued function of $\tau$. 
Substituting Eq.~(\ref{Psi0}) into Eq.(\ref{DSPsi1}) yields 
\begin{eqnarray}
\label{DSDS}
& &
\hspace*{-8mm}
\Delta S [ \boldsymbol{\Psi}(0) ; \tau ]
- 
\Delta S
\nonumber \\
& = & 
\frac{\beta_0 ( 1 - \rho )  }{ \rho }
\bigg[
\{ 1 - | U_{00} (\tau ) |^2 \} | \Psi_0(0) |^2
- 2  U_{00}  (\tau ) 
\nonumber \\ & & \mbox{}
\times
{\rm Re} 
 \bigg\{
\big( \Psi_0(0) \big)^*
\sum_{n=1}^{N_{\rm cvk} - 1}
U_{0n}(\tau) \Psi_n(0)
\bigg\}
\nonumber \\ & & \mbox{}
-
\bigg|
\sum_{n=1}^{N_{\rm cvk} - 1}
U_{0n}(\tau) \Psi_n(0)
\bigg|^2
\bigg] 
- \Delta S
\nonumber \\
& = & 
\frac{\beta_0 ( 1 - \rho )  }{ \rho }
\big[
a(\tau) | \Psi_0(0) |^2
-  2 b(\tau) |\Psi_0(0)|
- c(\tau) 
\big]
\nonumber \\ 
& & 
\mbox{}
- \Delta S
,
\end{eqnarray}
where 
\begin{eqnarray}
\label{abc}
a(\tau) 
& \equiv &
1 - | U_{00} (\tau ) |^2 
, 
\nonumber \\ 
b(\tau) 
& \equiv &
U_{00} (\tau ) 
\sum_{n=1}^{N_{\rm cvk} - 1}
 | U_{0n} (\tau ) |  |\Psi_n(0)|
\nonumber \\ & & \mbox{}
\hspace*{10mm}
\times
 \cos ( \phi_n (\tau) + \theta_n - \theta_0 )
, 
\nonumber \\ 
c(\tau) 
& \equiv &
\bigg| 
\sum_{n=1}^{N_{\rm cvk} - 1}
U_{0n} (\tau ) \Psi_n(0) 
\bigg|^2
,
\end{eqnarray}
$U_{0n}(\tau) = | U_{0n}(\tau) | \exp ( i \phi_n (\tau) )$, 
and 
$\Psi_n(0) = | \Psi_n(0) | \exp ( i \theta_n )$ 
$(n=0, 1, \cdots, N_{\rm cvk} - 1)$ 
are used. 
Since $[U_{nn'}(\tau)]$ is a unitary matrix, 
we have 
\begin{equation}
\label{U00U0n}
| U_{00}(\tau) |^2
+
\sum_{n=1}^{N_{\rm cvk} - 1}
| U_{0n}(\tau) |^2 
= 1
.
\end{equation}
Using Eqs.~(\ref{initialP}), (\ref{abc}), and (\ref{U00U0n}), 
we obtain 
\begin{eqnarray}
\langle
b(\tau)\}
\rangle_{\rm ens}
& =  & 0
,
\nonumber \\ 
\langle
\{b(\tau)\}^2
\rangle_{\rm ens}
& = & 
\frac{1}{2}
| U_{00}(\tau) |^2
( 1 - | U_{00} (\tau ) |^2 )
\frac{\rho}{\beta_0}
, 
\nonumber \\
\langle
c(\tau)
\rangle_{\rm ens}
& = & 
( 1 - | U_{00} (\tau ) |^2 )
\frac{\rho}{\beta_0}
. 
\end{eqnarray}

We see from Eq.~(\ref{DSDS}) 
that 
$\Delta S [ \boldsymbol{\Psi}(0) ; \tau ] -  \Delta S = 0$ 
is a second-order algebraic equation for $|\Psi_0(\tau)|$, 
the solutions of which are given by 
\begin{eqnarray}
\label{rpm}
r_\pm 
& \equiv &
\frac{1}{a(\tau)}
\Bigg[
\;
b (\tau )
\nonumber \\ & & \mbox{}
\pm
\sqrt{
\{b(\tau)\}^2
 + a(\tau) 
\bigg\{
c (\tau)
+ \frac{\rho \Delta S}{\beta_0 (1 - \rho )}
\bigg\}
}
\;
\Bigg]
.
\hspace*{8mm}
\end{eqnarray}
Then, 
when 
$
\Delta S \geq 0
$, 
we can use  
\begin{equation}
   \delta \big( \Delta S [ \boldsymbol{\Psi}(0) ; \tau ] -  \Delta S  \big)
   =
   \frac{
   \rho  \, \delta \big( | \Psi (0) | - r_+  \big)
   }{\beta_0 ( 1 - \rho ) a (\tau)  ( r_+ - r_- ) }
 ,
\end{equation}
and Eq.~(\ref{initialP}) 
to rewrite Eq.~(\ref{PDSdef}) as 
\begin{eqnarray}
& & 
\hspace*{-5mm}
\label{PDS2}
P(\Delta S)
 = 
  \frac{2 \rho}{ (1 - \rho )  a(\tau)} 
\bigg\langle
\frac{r_+
\exp 
\big(
- \beta_0
\,
{r_+}^2
\big)
}{r_+ - r_-}
\bigg\rangle_{\rm ens}
\nonumber \\ 
& & =
  \frac{\rho}{ (1 - \rho )  a(\tau)}
\exp 
\bigg[
-
  \frac{\rho \Delta S}{ (1 - \rho )  a(\tau)}
\bigg]
\nonumber \\ 
& & 
\mbox{}
\hspace*{1mm}
\times
\bigg\langle
\frac{2 r_+}{r_+ - r_-}
\exp
\bigg[
- \beta_0 {r_+}^2
+
  \frac{\rho \Delta S}{ (1 - \rho )  a(\tau)}
\bigg]
\bigg\rangle_{\rm ens}
. 
\nonumber \\ & & 
\end{eqnarray}
Here, Eq.~(\ref{rpm}) is used to obtain 
\begin{eqnarray}
\label{beta0rp2}
& & 
\hspace*{-3mm}
 \beta_0 {r_+}^2
-
  \frac{\rho \Delta S}{ (1 - \rho )  a(\tau)}
\nonumber \\
& &
= \frac{\rho}{a (\tau)} 
\bigg[
\frac{\beta_0}{\rho}
\bigg( c (\tau) + 2 \frac{\{ b(\tau) \}^2}{a (\tau)} \bigg)
+
2 b(\tau)
\sqrt{
\frac{\beta_0}{\rho \, a(\tau)}
}
\nonumber \\
& & \mbox{}
\hspace*{5mm}
\times 
\sqrt{
\frac{\Delta S}{1 - \rho}
+ 
\frac{\beta_0}{\rho}
\bigg( c (\tau) +  \frac{\{ b(\tau) \}^2}{a (\tau)} \bigg)
}
\; 
\bigg]
.
\hspace*{5mm}
\end{eqnarray}
Then, we see from Eqs.~(\ref{rpm}) and (\ref{beta0rp2}) that, 
when 
\begin{equation}
\frac{\rho}{1 - | U_{00} (\tau) |^2}  
\ll 
 1 
 \ll 
\frac{ \Delta S }{ 1 - \rho }
 \ll 
 \bigg(
 \frac{ 1 - | U_{00} (\tau) |^2 }{
 \rho | U_{00} (\tau) | }
 \bigg)^2
 ,
\end{equation}
we can use 
\begin{equation}
r_+ 
\simeq 
- r_-
\simeq
\sqrt{
 \frac{\rho \Delta S}{\beta_0 (1 - \rho ) a(\tau)}
}
,
\end{equation}
and 
\begin{equation}
\bigg|
 \beta_0 {r_+}^2
-
  \frac{\rho \Delta S}{ (1 - \rho )  a(\tau)}
\bigg|
\ll 1
, 
\end{equation}
in the last line of Eq.~(\ref{PDS2}), and 
rewrite Eq.~(\ref{PDS2}) as 
\begin{equation}
P(\Delta S)
=
  \frac{\rho}{ (1 - \rho )  a(\tau)}
\exp 
\bigg[
-
  \frac{\rho \Delta S}{ (1 - \rho )  a(\tau)}
\bigg]
, 
\end{equation}
which gives Eq.~(\ref{PDSaprox}).

\section{Orthonormal Basis Vectors in ${\cal E}_{ps} \otimes {\cal E}_{v_\perp}$ }

We here present two kinds of orthonormal basis vectors, $\{ | a \, l_j \rangle \}$ 
and $\{ | a \, j \rangle \}$, in the vector space 
 ${\cal E}_{ps} \otimes {\cal E}_{v_\perp}$. 
The Laguerre polynomials are used to define $\{ | a \, l_j \rangle \}$ 
while $\{ | a \, j \rangle \}$ are defined so as to efficiently describe  
the linear gyrokinetic system as shown in Sec.~IV. 

\subsection{Orthonormal basis vectors $\{ | a \, l_j \rangle \}$ }

The Laguerre polynomials $L_j (X)$ $(j = 0, 1, 2, \cdots)$ are defined by 
\begin{equation}
L_j (X) 
\equiv
\sum_{r=0}^j
(-1)^r
\left(
\begin{array}{c}
j \\
r
\end{array}
\right)
\frac{X^r}{r!}
\equiv 
\frac{e^X}{j!}
\frac{d^j}{d X^j} 
\Big( e^{-X} X^j \Big)
, 
\end{equation}
from which, for example, we have 
$
L_0 (X) 
= 
1
$,
$
L_1 (X) 
= 
1 - X
$, 
and 
$
L_2 (X) 
= 
1 - 2 X + \frac{1}{2} X^2
$. 
They satisfy the orthogonality relation, 
$
\int_0^{+\infty}
e^{-X}
L_j (X) 
L_{j'} (X) 
dX
= 
\delta_{j j'}
$.
Using the Laguerre polynomials, 
we define functions $l_j (X)$ and ket vectors $| l_j \rangle$
$(j = 0, 1, 2, \cdots)$ 
 in ${\cal E}_{v_\perp}$  by 
\begin{equation}
\label{ljX}
l_j (X)
\equiv
\langle X | l_j \rangle
\equiv 
e^{-X/2}
L_j (X)
, 
\end{equation}
which satisfy 
\begin{eqnarray}
\langle l_j | l_{j'} \rangle
& = & 
\int_0^{+\infty}
\langle l_j | X \rangle dX
\langle X | l_{j'} \rangle
\nonumber \\
& = & 
\int_0^{+\infty}
 l_j (X) l_{j'} (X) d X 
 =
 \delta_{j j'}
 .
 \end{eqnarray}
The basis vectors $\{ | a \, l_j \rangle \}$ 
in ${\cal E}^\times \equiv {\cal E}_{ps} \otimes 
{\cal E}_{v_\perp}$ are defined by 
\begin{equation}
| a \, l_j \rangle 
\equiv 
| a \rangle 
\otimes 
| l_j \rangle 
,  
\end{equation}
for which the orthonormality condition, 
\begin{equation}
\langle a \, l_j | a \, l_j \rangle 
=
\langle a | a' \rangle 
\langle l_j  | l_{j'} \rangle 
= 
\delta_{a a'} \delta_{j j'}
\end{equation}
holds. 

\subsection{Orthonormal basis vectors $\{ | a \, j \rangle \}$ }

We consider functions $l^a_j (X)$ and ket vectors 
$| l^a_j \rangle$ $(j = 0, 1, 2, \cdots)$ in ${\cal E}_{v_\perp}$ 
which are related to each other by 
\begin{equation}
l^a_j (X)
= 
\langle X |  l^a_j \rangle
.
\end{equation}
In the case of  $j=0$, 
we define $l^a_0 (X)$ by 
\begin{equation}
\label{la0X}
l^a_0 (X)
\equiv 
\langle X | l^a_0 \rangle
\equiv
C_{a0}
l_0 (X) J_0 (\sqrt{2 b_a X})
, 
\end{equation}
where  $b_a \equiv k_\perp^2 c^2 m_a T_a / (e_a B)^2$ and  
$l_0 (X) \equiv e^{-X/2}$ [see Eq.~(\ref{ljX})]. 
The coefficient $C_{a0}$ on the right-hand side of 
Eq.~(\ref{la0X}) is determined 
from the normalization condition, 
$
\langle l^a_0 | l^a_0 \rangle
=
\int_0^{+\infty}
\big[ l^a_0 (X) \big]^2
d X
=
1
$, 
as
\begin{equation}
C_a
=
1 / \sqrt{ \Gamma_0 (b_a) }
.
\end{equation}
We see that 
$C_a \rightarrow 1$ and $l^a_0 (X) \rightarrow l_0 (X)$ 
as  $b_a \rightarrow +0$.  

Next, the function $l^a_1(X)$ is defined such that 
$\langle l^a_1 | l^a_0 \rangle = 0$ and 
$\langle l^a_1 | l^a_1 \rangle = 1$. 
Following the Gram-Schmidt algorithm, 
$l^a_1(X)$ is written as
\begin{eqnarray}
\label{la1X}
l^a_1 (X)
& \equiv & 
C_{a1}
\bigg[
l_0 (X) \sqrt{X} J_1 (\sqrt{2 b_a X})
- l^a_0 (X)
\nonumber \\ & & 
\mbox{}
\times 
\int_0^{+\infty} l^a_0 (X')
l_0 (X') \sqrt{X'} J_1 (\sqrt{2 b_a X'})
d X'
\bigg]
,
\nonumber \\ & & 
\end{eqnarray}
where $C_{a1}$ is a positive coefficient determined from 
\begin{equation}
\label{Ca1}
1 = 
\langle l_1^a | l_1^a \rangle
=
\frac{C_{a1}^2 b_a}{2 \Gamma_0 (b_a)}
\Big(
\big[ \Gamma_0 (b_a) \big]^2
-
\big[ \Gamma_1 (b_a) \big]^2
\Big)
.
\end{equation}
In deriving Eq.~(\ref{Ca1}) from Eq.~(\ref{la1X}), we use the following formulas, 
\begin{eqnarray}
& & 
\int_0^{+\infty} e^{-X}
\sqrt{X} J_0 (\sqrt{2 b_a X})
J_1 (\sqrt{2 b_a X})
d X
\nonumber \\
& & 
=
\sqrt{\frac{b_a}{2}}
\big[
\Gamma_0 (b_a) 
-
\Gamma_1 (b_a) 
\big]
\end{eqnarray}
and
\begin{equation}
\int_0^{+\infty} e^{-X}
X 
\big[
J_1 (\sqrt{2 b_a X})
\big]^2
d X
=
b_a 
\big[
\Gamma_0 (b_a) 
-
\Gamma_1 (b_a) 
\big]
\end{equation}
It can be shown that 
$C_{a1} \sqrt{b_a/2} \rightarrow 1$ and 
$l^a_1(X) \rightarrow l_1(X)$ 
as $b_a \rightarrow + 0$. 

Subsequently, the functions $l^a_j (X)$ 
$(j=2, 3, \cdots)$ are defined such that 
$\{ | l^a_j \rangle \}_{j=0, 1, 2, \cdots}$ 
become the basis vectors in 
${\cal E}_{v_\perp}$, which satisfy the orthonormality 
condition,
\begin{equation}
\langle l^a_j | l^a_{j'} \rangle
=
\delta_{j j'}
. 
\end{equation}
For this purpose, we utilize 
 $l_j (X)$ $(j=2, 3, \cdots)$ given by Eq.~(\ref{ljX}) 
and apply the Gram-Schmidt algorithm to recursively define 
$l^a_j (X)$ $(j=2, 3, \cdots)$ by 
\begin{equation}
l^a_j (X)
\equiv 
C_{aj}
\bigg[
l_j (X) 
- 
\sum_{j'=0}^{j-1} l^a_{j'} (X)
\int_0^{+\infty} 
l^a_{j'}(X')
l_{j'} (X')
d X'
\bigg]
,
\end{equation}
where $C_{aj}$ is a positive coefficient determined from 
$\langle l^a_j | l^a_j \rangle = 1$. 
Again, we see that 
$l^a_j (X) \rightarrow l_j (X)$ 
as $b_a \rightarrow +0$. 

Finally, we define the basis vectors $\{ | a \, j \rangle \}$ 
in ${\cal E}^\times \equiv {\cal E}_{ps} \otimes 
{\cal E}_{v_\perp}$ by
\begin{equation}
| a \, j \rangle 
\equiv 
| a \, l^a_j \rangle 
\equiv
| a \rangle
\otimes
| l^a_j \rangle 
,
\end{equation}
for which the orthonormality condition 
\begin{equation}
\langle a \, j
| a'  j' \rangle 
=
\langle a | a' \rangle
\langle l^a_j | l^{a'}_{j'} \rangle
=
\delta_{a a'}
\delta_{j j'}
\end{equation}
holds. 
In Sec.~IV.C, 
 $| a \, j \rangle \equiv 
| a \, l^a_j \rangle $ $(j=2, 3, \cdots)$  
are used as the basis vectors in 
${\cal E}^\times_{\rm I\!I}$, 
and they are associated with the ballistic-mode part of the solution of 
the linearized gyrokinetic system of equations. 
Actually, as explained after Eq.~(\ref{psiII}), 
the detailed expressions of $l^a_j (X)$ 
$(j=2, 3, \cdots)$ are not necessary to express that linear solution. 
However, in the nonlinear gyrokinetic system, 
fine structures develop in the perpendicular velocity space 
as energy is transferred from low- to high-$j$ modes.
In that case, appropriate basis functions for high-$j$ orders 
are required for the velocity-space spectral analysis, for which 
$\{ l^a_j (X) \}$ [or $\{ l_j (X) \}$] given above can be 
practically useful.

% Create the reference section using BibTeX:
%\bibliography{basename of .bib file}

%\nocite{*}
%\bibliography{aipsamp}% Produces the bibliography via BibTeX.

\end{document}